\documentstyle[twoside,fleqn,espcrc2,epsf]{article}

\catcode`\@=11
\def\mref#1{\ifx\und@fined#1
{need to supply reference \string#1.}
\else #1 \fi}
\catcode`\@=12 %
\let\lref=\def
\newif\ifhypertex
\ifx\hyperdef\UnDeFiNeD
    \hypertexfalse
    \def\hyperdef#1#2#3#4{#4}
    \def\hfoot#1{}%
    
    \def\e@tf@ur#1{}
    \def\hth/#1#2#3#4#5#6#7{{\tt hep-th/#1#2#3#4#5#6#7}}
    
    \def\CERN{CERN, Geneva, Switzerland}
    \def\WLerche{W.\ Lerche}
    \def\email{{e-mail:{\tt lerche@nxth04.cern.ch}}}
\else
    \hypertextrue
\def\hth/#1#2#3#4#5#6#7{
  {\tt hep-th/#1#2#3#4#5#6#7}}

\def\CERN{
 CERN, Geneva, Switzerland}
\def\hfoot#1{\foot{#1}}%
\def\WLerche{

 W.\ Lerche}
    \def\email{{e-mail:

{\tt lerche@nxth04.cern.ch}}}
\fi
    \hypertextrue
\def\hth/#1#2#3#4#5#6#7{
  {\tt hep-th/#1#2#3#4#5#6#7}}

\def\CERN{
 CERN, Geneva, Switzerland}
\def\hfoot#1{\foot{#1}}%
\def\WLerche{

 W.\ Lerche}
    \def\email{{e-mail:

{\tt lerche@nxth04.cern.ch}}}
\makeatletter
\@addtoreset{equation}{section}
\makeatother

\newcommand{\foot}{\footnote}
\newcommand\al{\alpha} \newcommand\be{\beta} 
 \newcommand\g{\gamma}
 \newcommand\la{\lambda}
\newcommand\La{\Lambda} 
 
\newcommand\cC{{\cal C}} 
\newcommand\cF{{\cal F}}

\newcommand\cL{{\cal L}} \newcommand\cM{{\cal M}}
 \newcommand\cO{{\cal O}}

\newcommand\inbar{\vrule height1.5ex width.4pt depth0pt}
\newcommand\IC{\relax\,\hbox{$\inbar\kern-.3em{\rm C}$}}
\newcommand\IQ{\relax\,\hbox{$\inbar\kern-.3em{\rm Q}$}}
\newcommand\IR{\relax{\rm I\kern-.18em R}}
\newcommand\IP{\relax{\rm I\kern-.18em P}}
\newcommand\ZZ{\relax{\hbox{\rm Z\kern-.42em Z}}}
\newcommand{\beq}{\begin{equation}}
\newcommand{\eeq}{\end{equation}}
\newcommand{\eel}[1]{\label{#1}\end{equation}}
\newcommand{\bea}{\begin{eqnarray}}
\newcommand{\eea}{\end{eqnarray}}
\newcommand{\eeal}[1]{\label{#1}\end{eqnarray}}
\newcommand{\beac}{\begin{equation}\begin{array}{rcl}}
\newcommand{\eeacn}[1]{\end{array}\label{#1}\end{equation}}
\newcommand{\non}{\nonumber}
\newcommand{\equ}[1]{(\ref{#1})}
\newcommand{\eq}[1]{(\ref{#1})}
\newcommand{\nup}[3]{Nucl.\ Phys.\ {\bf B#1#2#3}\ }
\newcommand{\plt}[3]{Phys.\ Lett.\ {\bf B#1#2#3}\ }

\newcommand{\del}{\partial}
\newcommand\nid{\noindent}
\newcommand\hyp{\vrule height 2.3pt width 2.5pt depth -1.5pt}

\newcommand{\nex}[1]{$N\!=\!#1$\ }
\newcommand\Coeff[2]{{#1\over #2}}
\newcommand\coeff[2]{\relax{\textstyle {#1 \over #2}}\displaystyle}
\newcommand\shalf{\relax{\textstyle {1 \over 2}}\displaystyle}
\newcommand\half{\shalf}

\newcommand\ds{\displaystyle}

\newcommand\bifset{\Sigma}
\newcommand\CM{{\cal M}_0}
\newcommand\simpA{W_{\!A_{n-1}}}

\newcommand\vp{\varpi}
\newcommand\Om{\Omega}

\newcommand\figinsert[4]
{\begin{figure}[htb]
\newcommand\figsize{#3}
\epsfysize\figsize
\vskip -.5cm
\centerline{\epsffile{#4}}
\vskip -.5cm
\caption{\leftskip 1pc \rightskip 1pc
\baselineskip=10pt plus 2pt minus 0pt {\sl {#2}}}
\label{fig:#1}
\vskip -.3cm\end{figure}}

\newcommand\figref[1]{Fig.\ref{fig:#1}}

\newcommand\abs{
{We give an elementary introduction to the recent solution
of $N=2$ supersymmetric Yang-Mills theory.
In addition, we review how
it can be re-derived from string duality.}}
\newcommand\proc{{Contribution to the Proceedings of the {\it Spring
School and Workshop on String Theory, Gauge Theory and Quantum
Gravity}, I.C.T.P., Trieste, Italy, March 18-29, 1996. Based, in
part, on lectures given at {\it ``Gauge Theories, Applied
Supersymmetry and Quantum Gravity'',} Leuven, Summer 1995, and at Les
Houches, Summer 1995.}}

\begin{document}
\begin{titlepage}
\topskip0.5cm
\hfill\hbox{CERN-TH/96-332}\\[-1.cm]
\flushright{\hfill\hbox{hep-th/9611190}}\\[3.3cm]
\begin{center}{\Large\bf
{Introduction to Seiberg-Witten Theory
and its Stringy Origin\foot{\proc}}\\[2cm]}{
\Large \WLerche\foot{\email}\\[1.2cm]}
{\large\CERN}\\[3.5cm]
\end{center}
\flushleft{\large{\abs}}
\vskip3cm
\vfill
\hbox{CERN-TH/96-332}\hfill\\[-.4cm]
\hbox{November 1996}\hfill\\
\end{titlepage}
%
\title{Introduction to Seiberg-Witten Theory and its Stringy Origin}

\author{\WLerche\address{\CERN}}

\begin{abstract}\abs\end{abstract}

\maketitle

\section{Introduction}

In the last two years, there has been a remarkable progress in
understanding non-perturbative properties of supersymmetric field and
string theories. This dramatic development was initiated by the work
of Seiberg and Witten on \nex2 supersymmetric Yang-Mills theory
\cite{SW}, and by Hull and Townsend on heterotic-type II string
equivalence \cite{HT}. By now, many non-perturbatively exact
statements can be made about various types of supersymmetric
Yang-Mills theories with and without matter, and even more drastic
statements about superstring theories in various dimensions.

It has become evident that the main insight is of conceptional nature
and goes far beyond original expectations. The picture that seems to
emerge is that the various known, perturbatively defined string
theories represent non-perturbatively equivalent, or dual,
descriptions of one and the same fundamental theory. Moreover,
strings do not appear to play a very privileged role in this theory,
besides higher dimensional $p$-branes \cite{pbranes}. It may well
turn out, ultimately, that there is just one theory that is fully
consistent at the non-perturbative level, or a just small number of
such theories. Though the number of free parameters (``moduli'') may
a priori be very large --which would hamper predictive power-- it is
clear that investigating this kind of issues is important and will
shape our understanding of the very nature of grand unification.

A full treatment of these matters is surely outside the scope of
these lecture notes, and would be premature anyway. We therefore
limit ourselves to discussing some of the basic concepts, and since
some of these arise already in supersymmetric Yang-Mills theory, we
think it is a good idea to start in section 2 with a very basic
introduction to the original work of Seiberg and Witten (for gauge
group $SU(2)$). We will emphasize the idea of analytic continuation
and the underlying monodromy problem, and show how the
effective action can be explicitly computed.
For other reviews on this subject, see \cite{reviews}.

In section 3 we will then explain the generalization to other gauge
groups, emphasizing the role of the ``simple singularities'' that are
canonically associated with the simply laced Lie groups of type $ADE$
\cite{Arn}. The simple singularities will turn out to be the key to
understand how the SW theory arises in string theory. Indeed, as we
will explain in section 4, the SW theory can actually be {\it
derived} from Hull-Townsend string duality. This string duality will
also allow to interpret the geometrical structure of the SW theory,
in particular the ``auxiliary'' Riemann surface, in concrete physical
terms. In section 5, it will turn out \cite{KLMVW} that the SW
geometry has indeed a natural interpretation in terms of a very
peculiar string theory~!


\section{$N=2$ Yang-Mills Theory}

\subsection{Overview}

So, in a nutshell, what is all the excitement about that has made
furor even in the mass media ? As one of the main results one may
state the exact non-perturbative low energy effective Lagrangian of
\nex2 supersymmetric Yang-Mills theory with gauge group $SU(2)$; it
contains, in particular, the effective, renormalized gauge coupling,
$g_{{\rm eff}}$, and theta-angle, $\theta_{{\rm eff}}$:
\bea
\Big({\theta_{{\rm eff}}(a)\over\pi}+
{8 \pi i\over {g_{{\rm eff}}}^2(a)}\Big)\ =&
\label{geff}\\
\underbrace{{8 \pi i\over {g_0}^2}}_{{\rm bare}}+
\underbrace{{2i\over\pi}\,\log\Big[{a^2\over\Lambda^2}\Big]}_{{\rm
one-loop}} &-\
\underbrace{{i\over \pi}\sum_{\ell=1}^\infty c_\ell
\Big({\Lambda\over a}\Big)^{4\ell}}_{{\rm instanton\ corrections}}
\non\eeal{geff1}
Here, $\Lambda$ is the dynamically generated
scale at which the gauge coupling becomes
strong, and $a$ is the Higgs field. This effective, field dependent
coupling arises by setting the renormalization scale, $\mu$, equal to
the characteristic scale of the theory, which is given by the Higgs
VEV: $g_{{\rm eff}}(\mu)\to g_{{\rm eff}}(a)$. The
running of the perturbative coupling constant thus looks as
follows \cite{GKMMM}:
\figinsert{running}
{At scales above the Higgs VEV $a$, the masses of the non-abelian
gauge bosons, $W^\pm$, are negligible, and we can see the ordinary
running of the coupling constant of an asymptotically free theory. At
scales below $a$, $W^\pm$ freeze out, and we are left with just an
effective $U(1)$ gauge theory with vanishing $\beta$-function.}
{1.6truein}{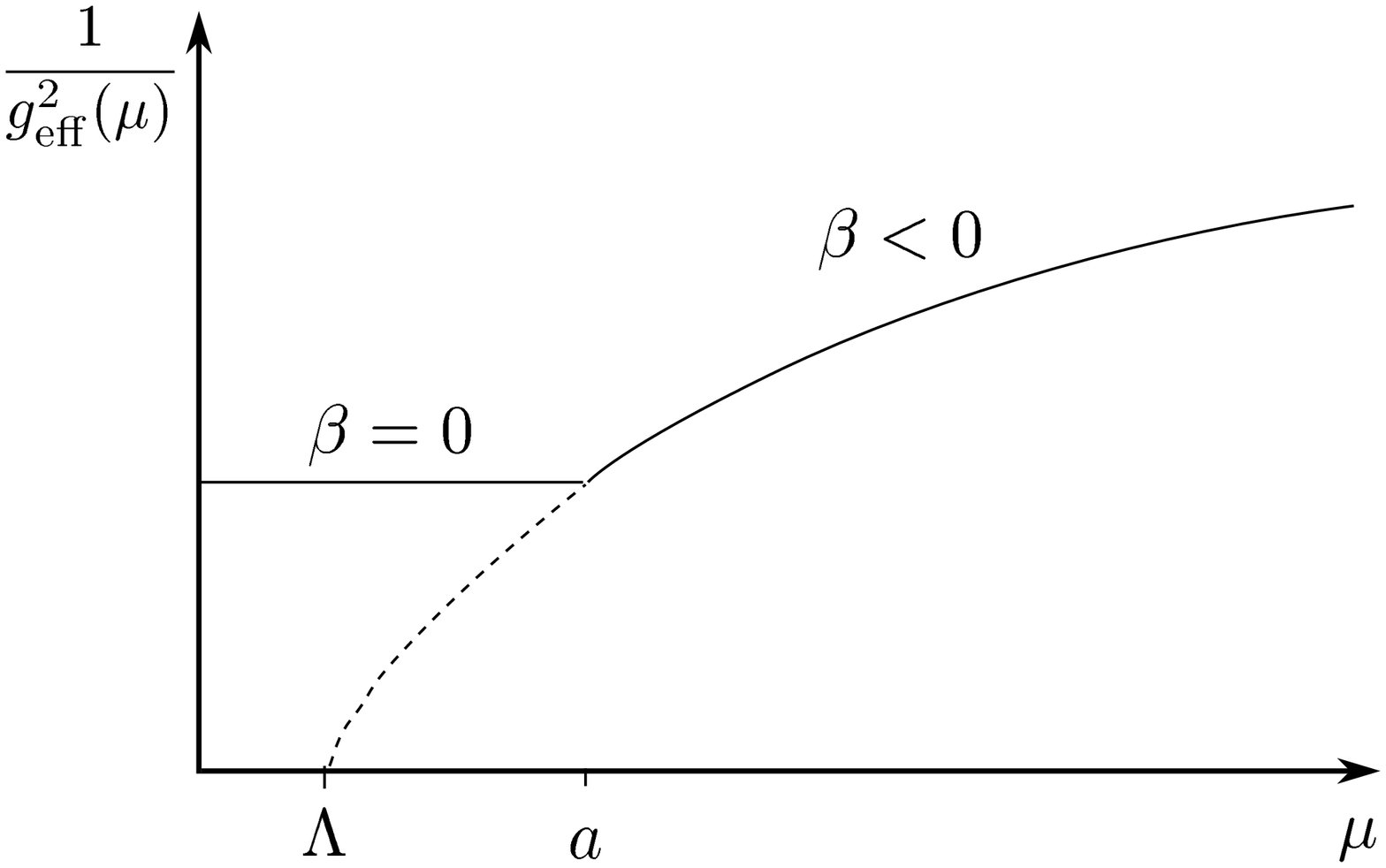}

The general form of the full non-perturbatively corrected coupling
\eq{geff} has been known for some time \cite{NSbeta}. One knows in
particular that all what can come from perturbation theory arises up
to one loop order only \cite{SCHIF}, and the amount of $R$-charge
violation (given by $8\ell$) of the $\ell$-instanton process; the
latter gives rise to the powers $4\ell$ in \eq{geff}. What is a
priori not known in \eq{geff} are the precise values of the instanton
coefficients $c_\ell$, and it is the achievement of Seiberg and
Witten to determine all of these coefficients explicitly. These
coefficients give infinitely many predictions for zero momentum
correlators involving $a$ and gauginos in non-trivial instanton
backgrounds. Such correlators are topological and also have an
interpretation \cite{WDo} in terms of Donaldson theory, which deals
with topological invariants of four-manifolds. It is the ease of
determination of such topological quantities that has been one of the
main reasons for excitement on the mathematician's side. The fact
that highly non-trivial mathematical results can be reproduced gives
striking evidence that S\&W's approach for solving the Yang-Mills
theory is indeed correct, even though some details, like a rigorous
field theoretic definition of the theory, may not yet be completely
settled. Furthermore, explicit computations \cite{instaComp} of some
of the instanton coefficients by more conventional field theoretical
methods have shown complete agreement with the predicted $c_\ell$.

It is, however, presently not clear what lessons can ultimately be
drawn for non-supersymmetric theories, like ordinary QCD. The hope
is, of course, that even though supersymmetry is an essential
ingredient in the construction, it is only a technical device that
facilitates computations,\foot{This may also apply to the role of
space-time supersymmetry in string theory; there is no intrinsic
relation between string theory and (low scale) space-time
supersymmetry.} and that nevertheless the supersymmetric toy model
displays the physically relevant features. See \cite{luisetal} for an
analysis in this direction.

Let us list some typical features of supersymmetric field
theories:\\

\nid {$\bullet$}{\it Non-renormalization properties:} perturbative
quantum corrections are less violent; this is related to a\\

\nid {$\bullet$}{\it Holomorphic structure,} which leads to vacuum
degeneracies, and allows to use powerful methods of complex
analysis.\\

\nid {$\bullet$}{\it Duality symmetries} between electric and
magnetic, or weak and strong coupling sectors, are more or less
manifest, depending on the number of supersymmetries.\\

\nid The maximum number of supersymmetries is four in a globally
supersymmetric theory:\\

\nid {$\bullet$}{\it N=4} supersymmetric Yang-Mills theory is
conjectured to be self-dual \cite{OW}, i.e., completely invariant
under the exchange of electric and magnetic sectors. However, though
interesting, this theory is too simple for the present purpose of
investigating non-trivial quantum corrections, since there aren't any
in this theory.\\

\nid {$\bullet$}{\it N=1} supersymmetric Yang-Mills theory, on the
other hand, is presumably not exactly solvable, since the quantum
corrections are not under full control; only certain sub-sectors of
the theory are governed by holomorphic objects (like the chiral
superpotential), and thus are protected from perturbative quantum
corrections. Indeed many interesting results on exact effective
superpotentials have been obtained recently \cite{Neqone}. \\

\nid {$\bullet$}{\it N=2} supersymmetric Yang-Mills theory is at the
border between ``trivial'' and ``not fully solvable'', in that it is
in the low-energy limit exactly solvable. It is governed by a
holomorphic function, the ``prepotential'' $\cal F$, for which the
perturbative quantum corrections are under complete control, ie.,
occur just to one loop order.\\

Having motivated why it is particularly fruitful to study \nex2
Yang-Mills theory, we now turn to discuss it in more detail.
\goodbreak

\subsection{The Semi-Classical Theory for $G=SU(2)$}

The fields of pure \nex2 Yang-Mills theory are vector supermultiplets
in the adjoint representation of the gauge group. For convenience,
one often rewrites such multiplets in terms of \nex1 chiral
multiplets, $W^i_\al, \Phi^i$, as follows:
\def\addots{\mathinner{
\mkern1mu\raise1pt
\vbox{\kern7pt\hbox{.}}
\mkern2mu\raise4pt\hbox{.}
\mkern2mu\raise7pt\hbox{.}
\mkern1mu}}
$$
\vbox{\offinterlineskip \tabskip=0pt \halign{\strut $#$ & $#$ & $#$ &
$#$ & $#$ & $#$ & $#$ & $#$ & $#$ & $#$ & $#$ & $#$ & $#$ & $#$ \cr
&&&&&&&&&&&{\rm spin}\hfil \cr &&&& A_\mu^i &&&&&&&1 \cr &&& \addots
&&&&&&&&& \cr && \lambda_\al^i &&&& \psi^{\be i} &&&&&\shalf \cr &
\swarrow &&&& \addots &&&&&& \cr W_\al^i &&&& \phi^i &&&&&&&0 \cr &&&
\swarrow &&&&&&&& \cr && \Phi^i &&&&&&&&& \cr }}
$$
The bottom component, the scalar field $\phi\equiv \phi_i \sigma_i$,
has the following potential:
\beq
V(\phi)\ =\ {\rm Tr}[\phi,\phi^\dagger]^2\ .
\eeq
This potential displays a typical feature of supersymmetric theories,
namely flat directions along which $V(\phi)\equiv0$. That is,
field configurations
\beq
\phi\ =\ \,a\,\sigma_3
\eeq
do not cost any energy. Of course, if $a\not=0$,
there is a spontaneous symmetry breakdown: $SU(2)
\hookrightarrow U(1)$. A more suitable ``order'' parameter
is given by the gauge invariant Casimir
\beq
u(a)\ =\ {\rm Tr}\,\phi^2\ =\ 2 a^2\ .
\eeq
It is in particular invariant under the Weyl group of $SU(2)$, which
acts as $a\to-a$ and is, physically, the discrete remnant of the
gauge transformations. that act within the Cartan subalgebra. The
quantity $u$ represents a good coordinate of the manifold $\cM_c$ of
inequivalent vacua, which one usually calls ``moduli space''. Since
$u$ can be any complex number, the moduli space is given by the
complex plane, which may be compactified to the Riemann sphere by
adding a point at infinity.



In the bulk of $\cM_c$ one has an unbroken $U(1)$ gauge
symmetry, which is enhanced to $SU(2)$ just at the origin. What we
are after is a ``Wilsonian'' effective lagrangian description of the
theory, for any given value of $u$. Such an effective lagrangian can
in principle be obtained by integrating out all fluctuations above
some scale $\mu$ (that, as we have indicated earlier, is chosen to be
equal to $a$). In particular, we would integrate out the massive
non-abelian gauge bosons $W^\pm$, to obtain an effective action that
involves only the neutral gauge multiplet, $W^0=(A\equiv
\Phi^0,W^0_\la)$. It is clear that, semi-classically, this theory can
possibly be meaningful only outside a neighborhood of $u=0$, since at
$u=0$ the non-abelian gauge bosons $W^\pm$ become massless, and the
effective description in terms of only $W^0$ cannot be accurate --
actually, it would become meaningless. This tells that $u=0$
will be a singular point on $\cM_c$ (besides the point of infinity).
In order to have a well-defined theory near $u=0$, one would need to
include the charged $W$-bosons in the effective theory;
one then says that the gauge bosons $W^\pm$ ``resolve'' the
 singularity.

It is clear from \figref{running} that, because of
asymptotic freedom, the region near $u=\infty$ will correspond to
weak coupling, so that only in this ``semi-classical'' region
reliable computations can be done in perturbation theory. On the
other hand, the theory will be strongly coupled near the classical
$SU(2)$-enhancement point $u=0$, so that a priori no reliable quantum
statements about the theory can be made here.

It is known (just from supersymmetry) that the low energy effective
lagrangian\footnote {By this we mean the piece of the effective
lagrangian that is leading for vanishing momenta, i.e., that contains
at most two derivatives. There are of course infinitely many higher
derivative terms in the full effective action. These are not governed
by holomorphic quantities, and thus we do not have much control of
them. See \cite{higherDer} for some results in this direction.} is
completely determined by a holomorphic prepotential $\cF$ and must be
of the form:
\bea
\cL\ &=&\ {1\over4\pi}{\rm Im}\,\Big[\, \int \!d^4\theta\,K(A,\bar A)
\cr &&\ +\, \int
\!d^2\theta\, \big(\Coeff12\sum
\tau(A)W^{\alpha }W_{\alpha }\big)\Big].
\eeal{effL}
Here, $\Phi\equiv:A\,\sigma_3$, and
\beq
K(A,\bar A)\ =\ {\partial \cF(A)\over\partial A}\bar A
\eeq
is the ``K\"ahler potential'' which gives a supersymmetric non-linear
$\sigma$-model for the field $A$, and
\beq
\tau(A)\ =\ {\partial^2 \cF(A)\over\partial^2A}\ .
\eeq
That is, the bosonic piece of \eq{effL} is, schematically,
\beq
\cL = {\rm Im}(\tau)\Big\{\partial a\,\partial \bar a+ F\cdot F\Big\}
+ {\rm Re}(\tau)\, F\cdot \tilde F\ + \dots,
\eel{bospart}
from which we see that
\beq
\tau(a)\ \equiv\ {\theta(a)\over\pi}+{8 \pi i\over {g}^2(a)}
\eel{taudef}
represents the complexified effective gauge coupling, and
Im($\tau$) is the $\sigma$-model metric on $\cM_c$. Classically,
$\cF(A)=\shalf \tau_0A^2$, where $\tau_0$ is the bare coupling
constant. However, the full quantum prepotential will receive
\cite{SCHIF} perturbative (one-loop) and non-perturbative
corrections, and must be of the form \cite{NSbeta}:
\bea
&&\!\!\!\!\!\!\cF(A) =\\ &&\!\!\!\!\!\!{1\over 2}\tau_0A^2 +
{i\over\pi}\,A^2\log\Big[{A^2\over\Lambda^2}\Big]+
{1\over 2\pi i}A^2\sum_{\ell=1}^\infty c_\ell
\Big({\La\over A}\Big)^{4\ell}.\non
\eeal{Feff}
By taking two derivatives, $\cF$ gives rise to the effective coupling
\eq{geff} mentioned in the introduction. Note that indeed for large
$a\equiv A|_{\theta=0}$, the instanton sum converges well, and the
theory is dominated by semi-classical, one-loop physics.

A crucial insight \cite{SW} is that the global properties of the
effective gauge coupling $\tau(a)$ are very important.
Specifically,
we know that near $u=\infty$:
\beq
\tau\ =\ {\rm const}+{2i\over\pi}\,\log\Big[{u\over\Lambda^2}\Big]+
{\rm single\hyp valued}.
\eel{multival}
This implies that if we loop around $u=\infty$ in the moduli space,
the logarithm will produce an extra shift of $2\pi i$ because of its
branch cut, and thus:
\beq
\tau\ \longrightarrow\ \tau-4\ .
\eeq
{}From \eq{taudef} it is clear that this monodromy just corresponds
to
an irrelevant shift of the $\theta$-angle, but what we learn is that
$\tau$, as well as $\cF$, are not functions but rather multi-valued
sections. Actually, the full story is more complicated than that, in
that also the imaginary part, Im$\tau={8\pi\over g^2}$, will be
globally non-trivial.

More specifically, we see from \eq{bospart} that Im($\tau$)
represents a metric on the moduli space, and the physical
requirement of unitarity implies that it must be positive
throughout the moduli space:
\beq
{\rm Im}(\tau(u))\ >\ 0\,.
\eel{taupos}
It is now a simple mathematical fact that since Im($\tau$) is a
harmonic function (ie., $\partial\overline{\partial}{\rm
Im}(\tau)=0$),
it cannot have a minimum if it is globally defined. Thus, in order
not to conflict with unitarity, we learn that Im$(\tau)$ can only
be locally defined -- a priori, it is defined only in the
semi-classical coordinate patch near infinity, cf., \eq{multival}. We
thus conclude that the global structure of the true "quantum" moduli
space, $\cM_q$, must be very different as compared to the classical
moduli space, $\cM_c$. In particular, any situation with just two
singularities must be excluded.

\subsection{The exact quantum moduli space}

The question thus arises, how many and what kind of singularities the
exact quantum moduli space should have, and what the physical
significance of these singularities might be. Seiberg and Witten
proposed that there should be two singularities at $u=\pm \La^2$,
where $\La$ is the dynamically generated quantum scale, and that the
classical singularity at the origin disappears -- see \figref{Mq}.

\figinsert{Mq}
{The transition from the classical to the
exact quantum theory involves splitting and shifting of
the strong coupling singularity away from $u=0$ to
$u=\pm \La^2$.}
{1.3in}{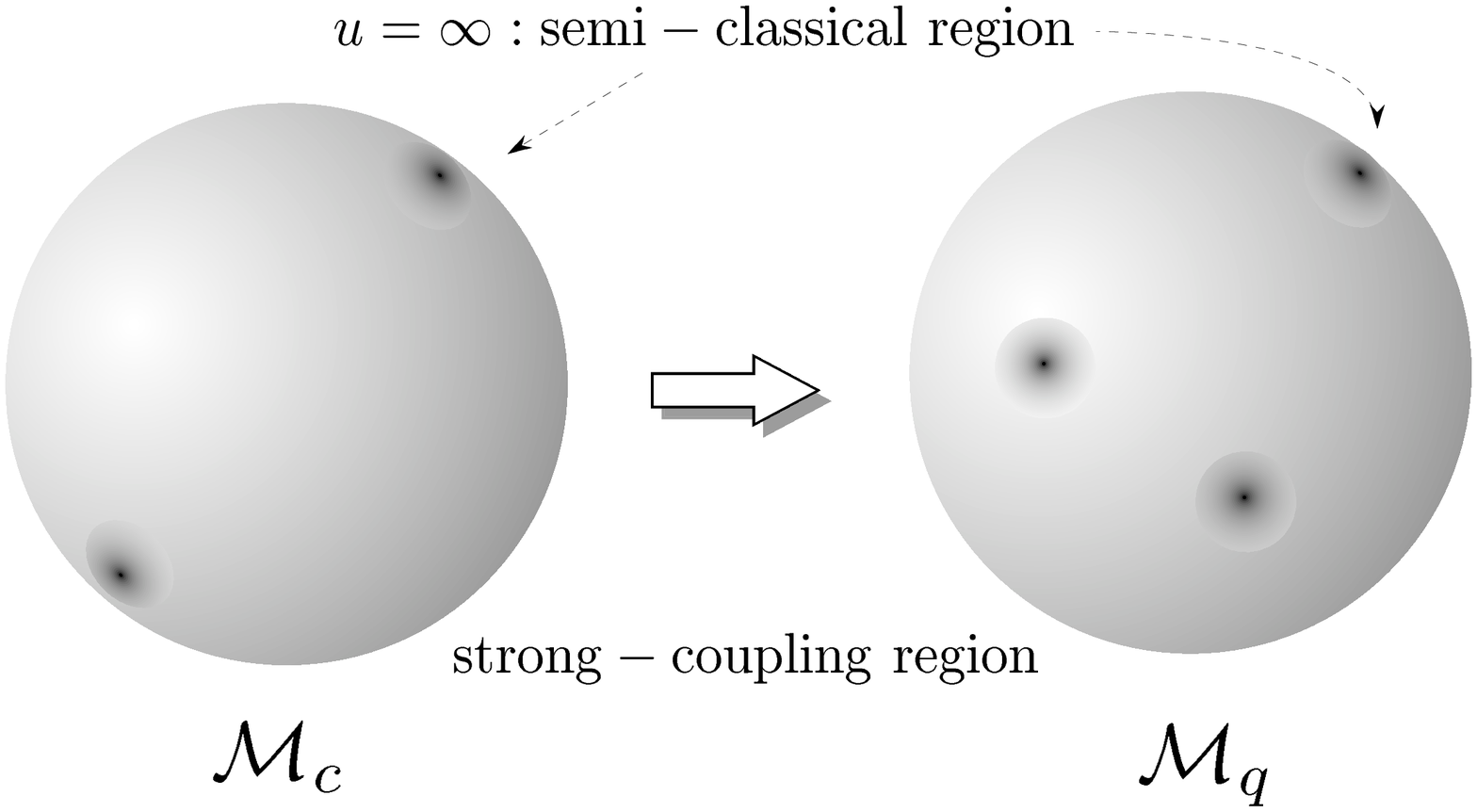}
 Though this proposal will prove to be a physically motivated and
self-consistent assumption about the strong coupling behavior, it is
very difficult, at least for for now, to derive it rigorously. But
there is a whole bunch of arguments, with varying degree of rigor,
why precisely the situation depicted in \figref{Mq} must be the
correct one. For example, the absence of a singularity at $u=0$
(which implies that there are, in the full quantum theory, no extra
massless gauge fields $W^\pm$) is motivated by the absence of an
$R$-current that a superconformal theory with massless gauge bosons
would otherwise have \cite{SW}. Furthermore, the appearance of just
two, and not\footnote{The number of singularities must be consistent
with global $R$-symmetry, which acts as $u\to-u$.} $2n$ strong
coupling singularities reflects that the corresponding \nex1 theory
(obtained by explicitly breaking the \nex2 theory by a mass term for
$\Phi$) has precisely two vacua (from Witten's index, Tr$(-1)^F=n$
for $SU(n)$). More mathematically speaking, the singularity structure
poses, as will be explained later, a particular non-abelian monodromy
problem, and it can be shown that there is no solution for this
problem for any other arrangement of singularities (under mild
assumptions about the form of these singularities) \cite{unique}.

The most interesting question is clearly what the physical
significance of the extra strong coupling singularities is. One
expects in analogy to the classical theory, where the singularity at
$u=0$ is due to the extra massless gauge bosons $W^\pm$, that the
strong coupling singularities in the quantum moduli space should be
attributed to certain excitations becoming massless as well. Guided
by the early ideas of 't Hooft about confinement \cite{thooft},
Seiberg and Witten postulated that near these singularities certain
't Hooft-Polyakov monopoles must become arbitrarily light.

There is a powerful tool to get a handle on
soliton masses in theories with extended supersymmetry, namely
the BPS-formula \cite{OW}:
\beq
m^2\ \ge\ |Z|^2\ ,
\eel{BPS}
where $Z$ is the central charge of the superalgebra in question.
For \nex2 supersymmetry, this formula immediately follows
from unitarity ($\bar Q\,Q>0$), in combination with
the anti-commutator
\bea
\big\{Q_{\alpha i},\overline Q_{\beta j}\big\}\ =\
\delta_{ij}\gamma^\mu_{\al\beta}P_\mu +
\delta_{\al\beta}\epsilon_{ij}U+
(\gamma_5)_{\al\beta}\epsilon_{ij}V,\non
\eeal{anti}
where $|Z|^2\equiv U^2+V^2$. The important point is that the BPS
bound \eq{BPS} is saturated by a certain class of excitations, namely
by the ``BPS-states'' that obey $Q|\psi\rangle=0$. The idea is that
if a state obeys this condition semi-classically, it obeys it also in
the exact quantum theory. This is because the number of degrees of
freedom of a ``short'' (or ``chiral'') multiplet that obeys
$Q|\psi\rangle=0$ is smaller as compared to those of a generic
supersymmetry multiplet, and the number of degrees of freedom is
supposed not to jump when switching on quantum corrections. In
particular, since 't~Hooft-Polyakov monopoles do satisfy the BPS
bound semi-classically, they must obey it in the exact theory as
well. From semi-classical considerations we can also learn that the
monopoles lie in \nex2 hypermultiplets, which have maximum
spin~$\shalf$.

\nid For \nex2 supersymmetric Yang-Mills theories,
the central charge takes the form
\beq
Z\ =\  q\,a + g\,a_D\ ,
\eel{zdef}
where $(g,q)$ are the (magnetic,electric) quantum numbers of the BPS
state under consideration. Above, $a_D$ is the ``magnetic dual'' of
the electric Higgs field $a$ and belongs to the $N=2$ vector
multiplet $(A_D,W_{\al,D})$ that contains the dual, magnetic photon,
$A^\mu_D$. By studying the electric-magnetic duality transformation,
under which the ordinary electric gauge potential $A^\mu$ transforms
into $A^\mu_D$, it turns out \cite{SW} that in the \nex2 Yang-Mills
theory the dual variable $a_D$ is simply given by:
\beq
a_D\ =\ {\partial\over\partial a}\,\cF(a)\ .
\eel{aDdef}
The general idea is that at the singularity at $u=\La^2$, one would
have $a\not=0$ but $a_D=0$, such that (by \eq{zdef}) a monopole
hypermultiplet with charges $(g,q)=(\pm 1,0)$ would be massless. On
the other hand, one would have that in the exact theory $u=0$ does
not imply $a=0$, so that in contrast to the classical theory, no
gauge bosons (with charges $(0,\pm 2)$) become massless. This in
particular would imply that the classical relation $u=2 a^2$ can hold
only asymptotically in the weak-coupling region.

The point is to view $a_D(u)$ as a variable
that is on a equal footing as $a(u)$; it just belongs to a dual
gauge multiplet that couples locally to magnetically charged
excitations, in the same way that $a$ couples locally to electric
excitations (such as $W^\pm$). A priori, it would not matter
which variable we use to describe the theory, and which variable
we actually use will rather depend on the region of
$\cM_q$ that we are looking at. More specifically,
in the original semi-classical, ``electric'' region near
$u=\infty$, the preferred local variable is $a$, and an
appropriate lagrangian is given by \eq{Feff}.
As mentioned above, the instanton sum converges
well for large $a\simeq \sqrt {u/2}$.

However, if we try to extend $\cF(a)$ to a region far enough away
from $u=\infty$, we will leave the domain of convergence of the
instanton sum, and we cannot really make any more much sense of
$\cF$. That is, in attempting to {\it globally extend} the effective
lagrangian description outside the semi-classical coordinate patch,
we face the problem of suitably analytically continuing $\cF$. The
point is that even though we cannot have a choice of $\cF$ that would
be globally valid anywhere on $\cM_q$ (it would be in conflict with
positivity, cf., \eq{taupos}), we can resum the instanton terms in
$\cF$ in terms of other variables, to yield another form of the
lagrangian that converges well in another region of $\cM_q$.

The reader might already have guessed that while $a$ is the preferred
variable near $u=\infty$, it is $a_D$ that is the preferred variable
in the ``magnetic'' strong coupling coordinate patch centered at
$u=\La^2$. More precisely, near $u=\La^2$ we expect to have the
following, dual form of the effective lagrangian:
\goodbreak
\bea
\cF_D(a_D)\ &=&\ \Coeff12\tau^D_0{a_D}^2-
{i\over4\pi}\,{a_D}^2\log\Big[{a_D\over\Lambda}\Big]
\nobreak\cr\nobreak &&\ - {1\over 2\pi i}\La^2\sum_{\ell=1}^\infty
c^D_\ell \Big({i a_D\over \La}\Big)^\ell\ .
\eeal{FDeff}
\goodbreak
The infinite sum indeed converges well, because at this singularity
$a_D\to0$.

{}From the coefficient of the logarithm we see that the theory is
non-asymptotically free (positive $\beta$-function), and thus weakly
coupled for $a_D\to0$ (though strongly coupled in terms of the
original variable, $a$). Indeed the dual theory is simply given by an
abelian $U(1)$ gauge theory (contributing zero to the
$\beta$-function), coupled to charged matter that is integrated out
(and that would be massless at $a_D=0$). The magnitude of the
coefficient shows that there should be a single matter field with
unit charge coupling to the (dual) photon, which belongs to a \nex2
hypermultiplet. This extra matter hypermultiplet is just the dual
representative of the massless magnetic monopole. To the dual
magnetic photon related to $a_D$, the monopole looks like an
ordinary, elementary (local) field, in spite of that it couples to
the original electric photon in a non-local way. It is this dual,
abelian reformulation of the original non-abelian instanton problem
what leads to substantial simplifications, especially to the
mathematician's profit.

Note that the infinite sum of correction terms in \eq{FDeff} reflects
the effect of integrating out infinitely many massive BPS states, and
though its physical meaning is completely different, has the same
information content as the instanton sum in the original lagrangian,
\eq{Feff}. Note also that the situation at the other singularity,
$u=-\La^2$, does not present anything new, in that (by $u\to-u$
symmetry) it is isomorphic to the the situation at $u=\La^2$ and
related to it by simply replacing $a_D$ in $\cF_D(a_D)$ by $a_D-2a$.
The whole scheme can therefore be depicted as in \figref{bigpicture}.

\figinsert{bigpicture}
{The exact quantum moduli space is covered by three distinct regions,
in the center of each of which the theory is weakly coupled when
choosing suitable local variables. A local effective lagrangian
exists in each coordinate patch, representing a particular
perturbative approximation. None of such lagrangians is more
fundamental than the other ones, and no local lagrangian exists that
would be globally valid throughout the moduli space. }
{1.5in}{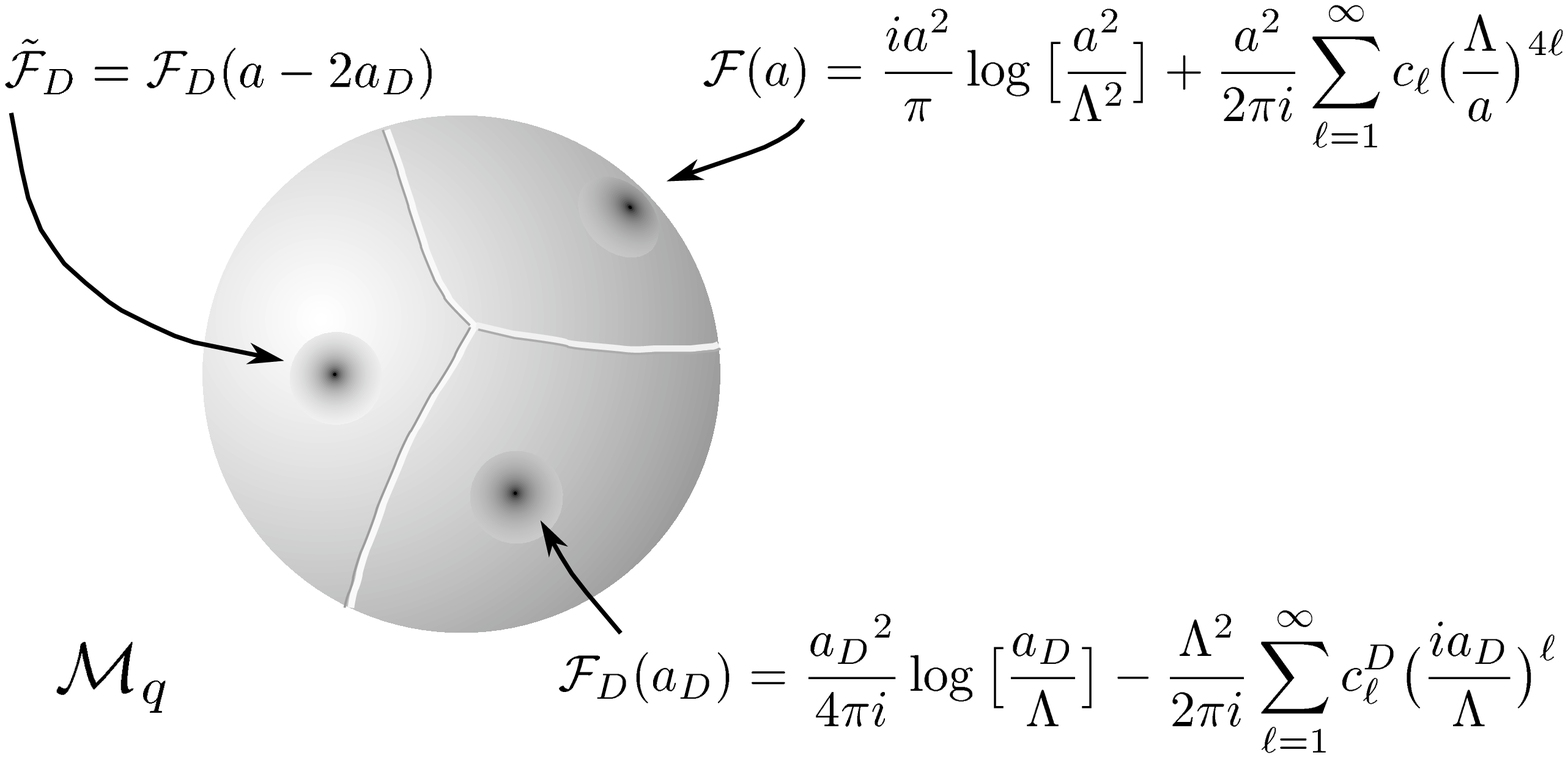}

The alert reader might have noticed that so far nothing
concrete was achieved yet -- instead, we have introduced another
set of infinitely many unknowns, $ c^D_\ell$-- and also that we have
just guessed the coefficient of the logarithm in \eq{FDeff}. Indeed,
this specific coefficient cannot be derived at this point, but rather
is part of the assumption that a single monopole with unit charge
becomes massless at $u=\La^2$.

The issue is now to determine the values of all the unknown
coefficients in $\cal F$, ${\cal F}_D$ \eq{Feff},\eq{FDeff} from the
assumptions that govern the {\it local}, i.e., perturbative behavior
of the theory in each of the three coordinate patches in
\figref{bigpicture}. The local behavior is determined by the
coefficients of the logarithms, which can reliably be computed in
one-loop perturbation theory and directly reflect the charge quantum
numbers of the fields that are supposed to be light near a given
singularity.

The key idea is that it is the {\it patching together of the known
local data in a globally consistent way} that will completely fix the
theory (up to irrelevant ambiguities like $\theta$-shifts). More
precisely, the one-loop term determines the local monodromy $M$
around a given singularity, and this acts on the section $({a_D\atop
a})$ as follows:
\beq
\Big({a_D(u)\atop a(u)}\Big)\ \longrightarrow\
M\,\Big({a_D(u)\atop a(u)}\Big)\ .
\eel{monotrans}
In particular, from our knowledge of the asymptotic behavior of
$a_D(u),a(u)$ at semi-classical infinity,
\beq
\Big({a_D(u)\atop a(u)}\Big)\ \simeq\ \Big({{i\over\pi}\sqrt
{2u}\log(u/\La^2))\atop \sqrt {u/2}}\Big)
\eeq
we infer that for a loop around $u=\infty$:
\beq
M_\infty\ =\ \pmatrix{-1&4\cr0&-1\cr}\ .
\eel{Minf}
As for the strong coupling singularities at $u=\pm\La^2$,
we choose a different strategy: we know on general grounds that
the monodromy of a dyon with charges $(g,q)$ that becomes massless
at a given singularity is given by:
\beq
{M}^{(g, q)}= \pmatrix{1 +  q g&
q^2\cr - g^2&1-  g q}
\eel{monmatrix}
This can be seen in various ways, one of which will be explained
further below.

The global consistency condition on how to patch together
the local, perturbative data is then simply
\beq
{M}_{+\La^2}\cdot {M}_{-\La^2}\ =\ M_\infty\ ,
\eel{globalcond}
since we can smoothly pull the monodromy paths $\gamma$ around the
Riemann sphere ($u_0$ is an arbitrary base point):

\figinsert{monpath}
{Monodromy paths in the $u$-plane.}
{1.1in}{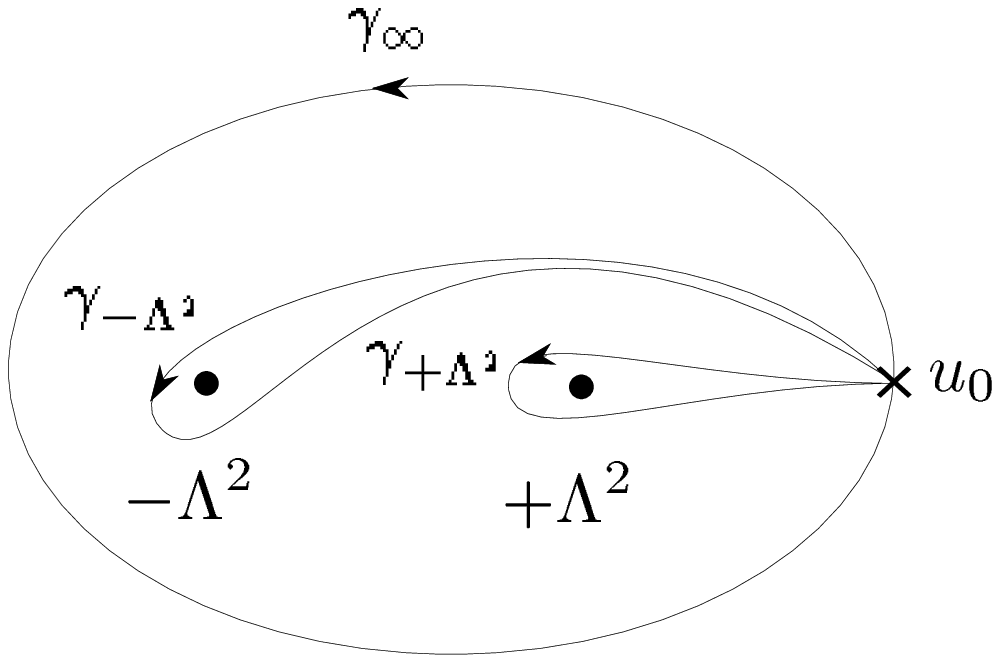}

One may view equation \eq{globalcond} as a condition on the possible
massless spectra at $u=\pm\La^2$. For matrices of the restricted form
\eq{monmatrix},  its solution is:
\bea
{M}_{+\La^2}\ &=&\ M^{(1, 0)}\ \cr
{M}_{-\La^2}\ &=&\ M^{(1, -2)}\ ,
\eeal{diosol}
which is unique up to irrelevant conjugacy. From this we can read off
the allowed (magnetic, electric) quantum numbers of the massless
monopoles/dyons. They indeed give back the coefficient of the
logarithmic term of $\cF_D$ that we had anticipated in
eq.\ \eq{FDeff}.

If we would consider a situation with more than two strong coupling
singularities, we would have to solve an equation like
\eq{globalcond}\ with the corresponding product of matrices
\eq{monmatrix}. However, it can be deduced \cite{unique} that such
equations for more than two such matrices do not have any solution.

\subsection{Solving the monodromy problem}

The physics problem has now become a mathematical one, namely simply
to find multi-valued functions $a(u),a_D(u)$ that
display the required monodromies ${M}_{\pm\La^2,\infty}$ around the
singularities (and that in addition
lead to a coupling $\tau\equiv\partial_a a_D$
with Im$\tau>0$). This is a classical mathematical problem, the
``Riemann-Hilbert'' problem, which is known to have a
unique\footnote{Unique up to multiplication of $\big({a_D\atop
a}\big)(u)$ by an entire function; this can however be fixed by
imposing the correct, semi-classical asymptotic behavior.} solution.

The RH problem can be accessed from two complementary point of views:
either by considering $a,a_D$ as solutions of a differential equation
with regular singular points, or from considering $a,a_D$ as certain
period integrals related to some auxiliary ``spectral surface'' $X$.
The latter approach, to be discussed momentarily, allows an easy
geometric implementation of the right monodromy properties, while the
differential equation approach, to be considered later, is more
useful for obtaining explicit expressions for $a(u)$ and $a_D(u)$.

Any two of the monodromy matrices ${M}_{\pm\La^2,\infty}$ generate
the monodromy group $\Gamma_M$, which constitutes the subgroup
$\Gamma_0(4)$ of the modular group $SL(2,\ZZ)$ and consists of
matrices of the form
\bea
\Gamma_0(4)\ =\ \Big\{\pmatrix{a & b\cr c & d \cr}\in SL(2,\ZZ),
\ b=0\ {\rm mod}\ 4
\Big\}.\non
\eeal{gammfour}
Mathematically speaking, the quantum moduli space can thus be viewed
as the upper half-plane modulo the monodromy group:
\beq \cM_q\
\cong\ H^+\Big/\Gamma_0(4)\ .
\eeq
This group represents the quantum symmetries of the theory, and acts
(because of \eq{monotrans}) on the gauge coupling $\tau={\del
a_D(u)\over \del a(u)}$ via $\tau\to {a\tau+b\over c\tau+d}$. Is
particular, we see that $S:\ \tau\to-{1\over\tau}$ is {\it not} part
of $\Gamma_M$, and this means that the theory is not weak-string
coupling duality invariant (in contrast to \nex4 Yang-Mills theory).

Now, motivated by the appearance of a subgroup of the
modular group (which is the group of the discontinuous
reparametrizations of a torus), the basic idea is that the monodromy
problem can be formulated in terms of a toroidal Riemann surface,
whose moduli space is precisely $\cM_q$ \cite{SW}. Such an elliptic
curve indeed exists and can be algebraically characterized by:
\foot{There are various physically equivalent forms of this curve.}
\bea
X_1:\ \ y^2(x,u)\ &=&\ (x^2-u)^2 - \La^4\cr
&\equiv&:\ \prod_{i=1}^4(x-e_i(u,\La))\ .
\eeal{thecu}
The point is to interpret the gauge coupling $\tau(a)$ as the period
``matrix'' of this torus, and this has the added bonus that
manifestly Im$(\tau)>0$ is guaranteed, by virtue of a mathematical
theorem called ``Riemann's second relation''. As such is $\tau$
defined by a ratio of period integrals:
\beq
\tau(u)\ =\ {\varpi_D(u)\over\varpi(u)}\ ,
\eel{tauw}
where
\beq
\varpi_D(u)\ =\ \oint_{\beta}\omega \ ,\qquad\
\varpi(u)\ =\ \oint_{\al}\omega
\eel{perdef}
with the holomorphic differential
$\omega\ \equiv {1\over \sqrt2\pi} {dx\over y(x,u)}$.
Here, $\al, \beta$ are canonical basis homology cycles of the
torus, like shown as in \figref{torus}.

\figinsert{torus}
{Basis of one-cycles on the torus.}
{.8in}{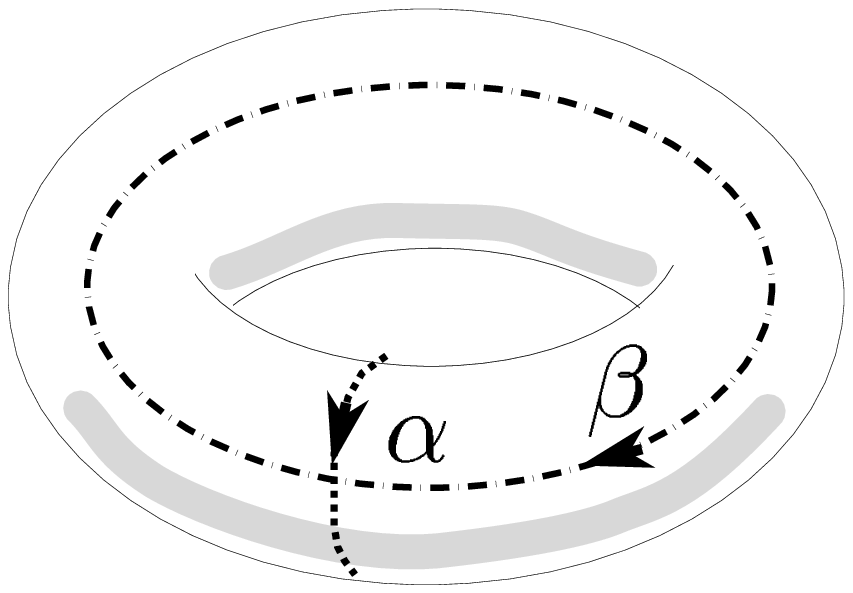}

\nid From the relation $\tau=\partial_a a_D$ we thus infer that
\beq
\varpi_D(u)\ =\ {\partial a_D(u)\over \partial u} \ ,\qquad\
\varpi(u)\ =\ {\partial a(u)\over \partial u}\ .
\eel{adelu}
That is, the yet unknown functions $a_D(u),a(u)$, and consequently
the prepotential $\cF=\int_a a_D(a)$, are supposed to be obtained by
integrations of torus periods. Note that \eq{adelu} implies that we
can
also write
\beq
a_D(u)\ =\ \oint_{\be}\la_{SW} \ ,\qquad\
a(u)\ =\ \oint_{\al}\la_{SW}\ ,
\eel{aaDdef}
where
\beq
\la_{SW}\ =\ {1\over\sqrt2\pi}x^2{dx\over y(x,u)}
\eel{lamdef}
(up to exact pieces) is a particular
{\it mero}morphic one-form (for $z\equiv {1\over x}\to0$
it has a second oder pole: $\lambda_{SW}\sim {dz\over z^2}$).

What needs to be shown is that the periods, derived from the specific
choice of elliptic curve given in \eq{thecu}, indeed enjoy the
correct monodromy properties. The periods \eq{perdef} and \eq{aaDdef}
are actually largely fixed by their monodromy properties around the
singularities of $\cM_q$, and obviously just reflect the monodromy
properties of the basis homology cycles, $\al$ and $\be$. It
therefore suffices to study how the basis cycles $\al,\be$ of the
torus transform when we loop around a given singularity.

\figinsert{bascyc}
{Representation of the auxiliary elliptic curve $X_1$ in
terms of a two-sheeted covering of the branched $x$-plane. The two
sheets are meant to be glued together along the cuts that run between
the branch points $e_i(u)$. Shown is our choice of homology basis,
given by the cycles $\al,\be$. This picture corresponds to the choice
of the basepoint $u_0> \Lambda^2$ real.}
{0.9in}{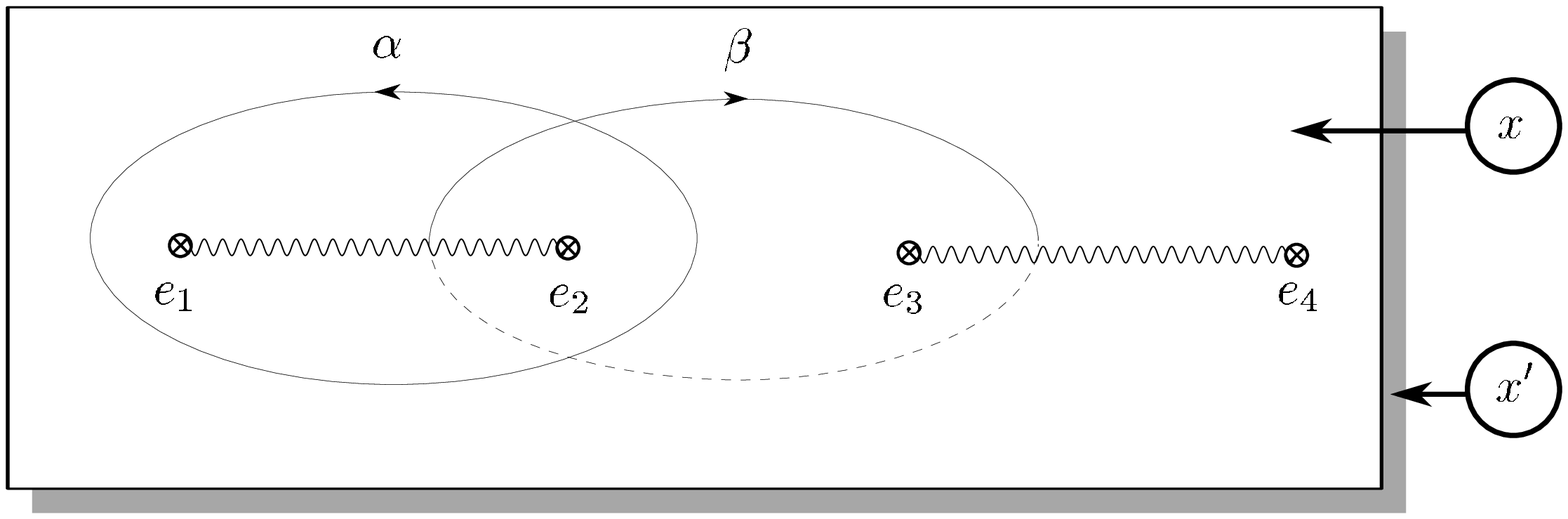}

For this, we represent the above torus in a convenient way
that is well-known in the mathematical literature:
we will represent it in terms of a two-sheeted cover of the
branched $x$-plane. More precisely,
denoting the four zeroes of $y^2(x,u)=0$ by
\bea
e_1\ &=& -\sqrt{u+\Lambda^2}\ ,\ \ e_2\ = -\sqrt{u-\Lambda^2}\cr
e_3\    &=&\ \sqrt{u-\Lambda^2}\ , \ \ \,e_4\ =
\sqrt{u+\Lambda^2}\ ,
\eeal{zeros}
we can specify the torus in the way depicted in \figref{bascyc}.

The singularities in the quantum moduli space arise when the torus
degenerates, and this obviously happens when any two of
the zeros $e_i$ coincide. This can be expressed as the
vanishing of the ``discriminant''
\beq
\Delta_\La\ =\ \prod_{i<j}^4(e_i-e_j)^2\ =\
(2\Lambda)^8(u^2-\Lambda^4)\ .
\eel{discdef}
The zeroes of $\Delta_\La$ describe the following degenerations of
the elliptic curve:

\noindent
$i_+$) $u\rightarrow +\Lambda^2$, for which $(e_2\rightarrow
e_3)$, i.e., the cycle $\nu_{+\Lambda^2}\equiv\be$ degenerates,

\noindent
$i_-$) $u\rightarrow -\Lambda^2$, for which $(e_1\rightarrow
e_4)$, i.e., the cycle $\nu_{-\Lambda^2}\equiv\be- 2\, \al$
degenerates,

\noindent
 $ii$) ${\Lambda^2/ u}\rightarrow 0$, for which
$(e_1\rightarrow e_2)$ {\sl and } $(e_3\rightarrow e_4)$.

\figinsert{vancyc}
{Vanishing cycles on the torus that shrink to zero
as one moves towards a degeneration point.}
{1.5in}{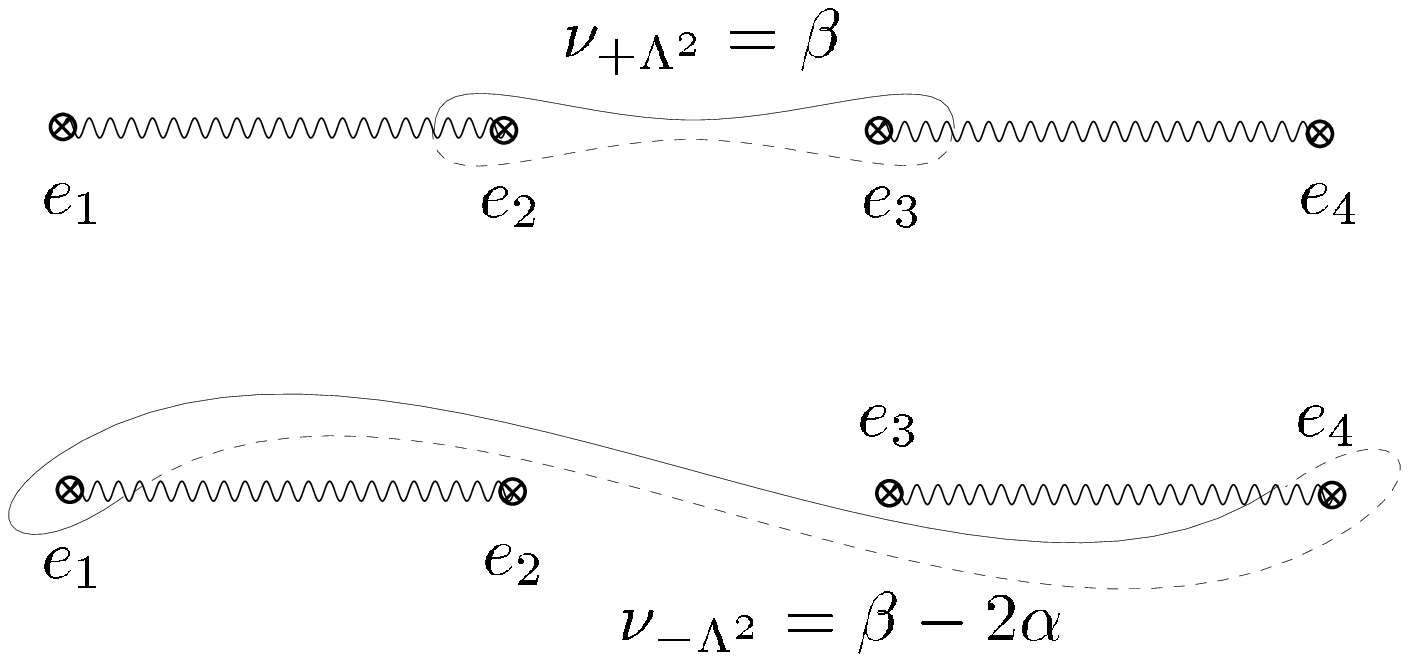}

Is is now easy to see that a loop $\gamma_{+\La^2}$ around
the singularity at $u=\La^2$ makes $e_2$ and $e_3$ rotate around
each other, so that the cycle $\al$ gets transformed into $\al-\be$,
as can be seen from \figref{vancyc}. This means that on the basis
vector
$({\be\atop\al})$, the monodromy action looks
\beq
\left(\matrix{1& 0\cr -1& 1}\right)\ \equiv\
M^{(1,0)}\ =\ M_{+\Lambda^2}\ .
\eel{Mone}
Similarly, from \figref{vancyc} one can see that the monodromy around
$u=-\La^2$ is given by
\beq
\left(\matrix{-1& 4\cr -1& 3}\right)\ \equiv\
M^{(1,-2)}\ =\ M_{-\Lambda^2}\ .
\eeq
To obtain the monodromy around ${\Lambda^2/ u}\rightarrow 0$, one
can compactify the $u-$plane to $\IP^1$, as we did before, and get
the monodromy at infinity from the global relation
$M_\infty=M_{+\Lambda^2} M_{-\Lambda^2}$ (cf., \figref{monpath}.).

We thus have reproduced the monodromy matrices associated with the
exact quantum moduli space directly from the the elliptic curve
\eq{thecu}, and what this means is that the integrated torus periods
$a_D(u),a(u)$ defined by \eq{adelu} must indeed have the requisite
monodromy properties. However, before we are going to explicitly
determine these functions in the next section, let us say some more
words on the general logic of what we have just been doing.

We have seen in \figref{vancyc} that when we loop around a
singularity in $\cM_q$, the branch points $e_i(u)$ exchange along
certain paths, $\nu$, which shrink to zero as $e_i\to e_j$. Such
paths are called ``vanishing cycles'' and play, as we will see, an
important role for the properties of BPS states. Indeed, in a quite
general context, many features of a BPS spectrum can be encoded in
the singular homology of an appropriate auxiliary surface $X$.

Concretely, assume that a path vanishes at a singularity
that has the following expansion in terms of given basis cycles:
\beq
\nu\ =\ g\,\be + q\,\al\ .
\eel{nudef}
Then obviously, assuming that $\lambda$ does not blow up, we have
\bea
0\ =\ \oint_\nu\la\ =\ g\oint_\be\la+q\oint_\al\la =
g \,a_D+q\,a\ \equiv\ Z,\non
\eeal{Znull}
so that we have at the singularity a massless BPS state with
(magnetic,electric) charges equal to $(g,q)$.
That is, we can simply read off the quantum numbers of massless
states from the coordinates of the vanishing cycle !
Obviously, under a change of homology basis, the charges
change as well, but this is nothing but a duality rotation. What
remains invariant is the intersection number
\beq
\nu_i\circ\nu_j\ =\ \nu^t\cdot\Omega\cdot\nu\
=  g_i q_j-g_jq_i\ \in\ \ZZ\ ,
\eel{dzw}
where $\circ$ is the intersection product of one-cycles and $\Omega$
is the symplectic (skew-symmetric) intersection metric for the basis
cycles. Note that this represents the well-known Dirac-Zwanziger
quantization condition for the possible electric and magnetic
charges, and we see that it is satisfied by construction. The
vanishing of the r.h.s.\ of \eq{dzw} is required for two states to be
local with respect to each other \cite{thooft,AD}. This means that
only states that are related to non-intersecting vanishing cycles are
mutually local. In our case, the monopole with charges $(1,0)$, the
dyon with charges $(1,-2)$ and the (massive) gauge boson $W^+$ with
charges $(0,2)$ are all mutually non-local, and thus cannot be
simultaneously represented in a local lagrangian.

Furthermore, there is a closed formula for the monodromy around a
given singularity associated with a vanishing cycle: the monodromy
action on any given cycle, $\gamma\in H_1(X,\ZZ)$, is directly
determined in terms of this vanishing cycle $\nu$ by means of the
``Picard-Lefshetz'' formula \cite{Arn}:
\beq
M_\nu:\ \ \ \gamma\ \longrightarrow\ \gamma - (\gamma\circ\nu)\,\nu\
{}.
\eel{PicLef}
This implies that for a vanishing cycle of the form \eq{nudef},
the monodromy matrix is precisely the one given in \eq{monmatrix}, as
promised.

\subsection{The BPS Spectrum}

We noted above that the global consistency relation \eq{globalcond}
is solved by monodromy matrices that correspond to a monopole with
charges $(g,q)=\pm(1,0)$ and to a dyon with charges $\pm(1,-2)$.
These excitations are massless at $u=\La^2$ and $u=-\La^2$,
respectively. We now like to ask about other BPS states that may
exist, though these must be massive throughout the moduli space.

For this, remember that the charge labels $(g,q)$ are highly
ambiguous, because they are defined only up to symplectic
transformations; this reflects the choice of homology basis. The
charges can thus be changed by conjugation by any monodromy
transformation belonging to $\Gamma_0(4)$. In particular, looping
around $u=\infty$ acts as
\beq
M_\infty\cdot M^{(g,q)}\cdot{M_\infty}^{-1}\ =\ M^{(-g,-q-4g)}\ ,
\eeq
and thus will shift the electric charge, $q\to-q-4g$. This
corresponds to $\tau\to\tau-4$ and to $\theta\to\theta-4\pi$, and
hence is a manifestation of the fact \cite{witdyon} that the
electric charge of a dyon changes if the $\theta$-angle is changed --
there is no absolute definition of the electric charge of a dyon.

It also means that the weak coupling spectrum of the theory must be
invariant under shifts $\theta\to\theta-4\pi n$, $n\in \ZZ$. That is,
under ``spectral flow'' induced by smoothly changing $\theta$ by
$4\pi$, the BPS spectrum must map back to itself, though individual
states need not map back to themselves. More precisely, since the
above monodromy conjugation can be induced by arbitrarily small loops
around $u=\infty$, we know that the BPS spectrum should consist in
the weak coupling patch at least of dyons with charges
$\pm(1,2\ell)$,
$\ell\in\ZZ$, besides the massive gauge bosons $W^\pm\sim(0,\pm2)$.

A very important point made in \cite{SW} is that the stable BPS
spectrum in the strong coupling region is, in fact, different and
consists only of a subset of the above semi-classical BPS spectrum.
This is because the moduli space $\cM_q$ decomposes into two regions,
$\cM_q^{{\rm weak}}$ and $\cM_q^{{\rm strong}}$, with different
physics. They are separated by a line $\cC$, on which most of the
semi-classical BPS states decay. This line is defined by
\beq
\cC\ =\  \Big\{u:\, {a_D(u)\over a(u)}\in\IR\Big\}\ ,
\eel{Cdef}
and turns out the be almost an ellipse passing through
the singular points at $u=\pm\La^2$; see \figref{margin}.
Indeed all possible singularities associated with massless BPS
states must lie on $\cC$, since if $Z=ga_D+qa=0$ for
$g,q\in \ZZ$, then $a_D/a\in\IR$.

\figinsert{margin}
{The line $\cC$ of marginal stability separates the strong coupling
BPS spectrum from the semi-classical BPS spectrum. Both spectra are
indicated here by the charges of the stable states. The dashed line
represents the logarithmic branch cut.}
{1.6in}{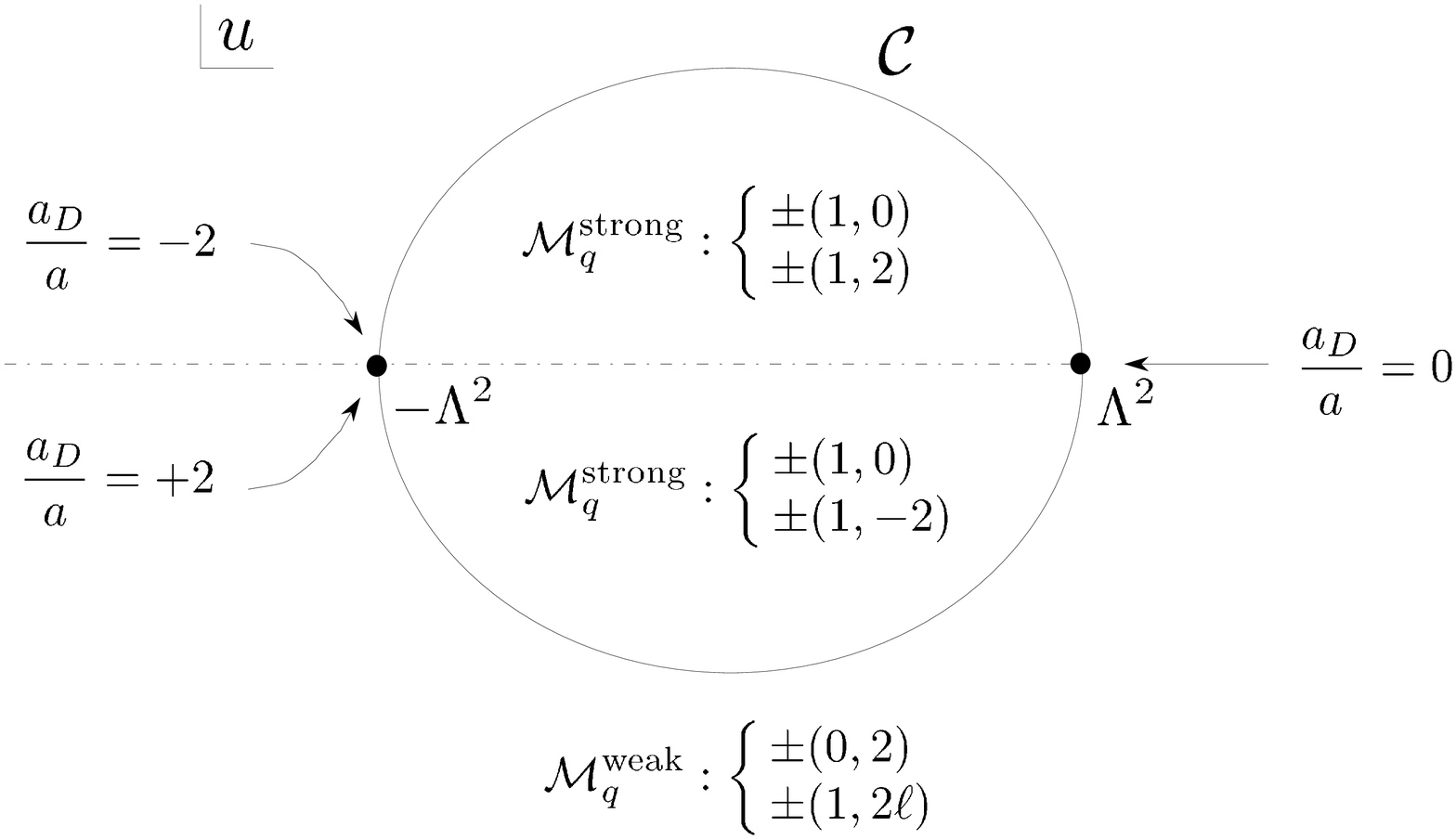}

One can check \cite{BiFe} that if one traces $\cC$ clockwise starting
from $u=-\La^2$, $(a_D/a)(u)$ varies monotonically from $-2$ to $+2$,
with $(a_D/a)(\La^2)=0$. That we do not map back to $(a_D/a)=-2$ is
due to the branch cut of the logarithm in $a(u)$. Thus there is
really an ambiguity in the electric charge of the dyon: if we
approach $u=-\La^2$ from the upper-half $u$-plane, the dyon has
charges $\pm(1,2)$, which is $M_\infty$-conjugate to $\pm(1,-2)$ that
we had before.

The physical significance of the marginal line of stability $\cC$ is
that when $(a_D/a)(u)\in\IR$, the lattice (or ``Jacobian'') of the
central charges $Z=ga_D+qa$ degenerates to a line. Then mass and
charge conservation do not any more prohibit BPS states to decay into
monopoles and dyons, because the triangle inequality
$|Z_{g_1+g_2,q_1+q_2}|\leq |Z_{g_1,q_1}| + |Z_{g_2,q_2}|$ becomes
saturated. For example, if $a_D=\xi\, a$, $\xi\!\in\![0,2]$, then the
gauge field with $(g,q)=(0,2)$ and $m_{(0,2)}=2|a|$ is unstable
against decay into a monopole-dyon pair, with $m_{(-1,2)}=(2-\xi)|a|$
and $m_{(1,0)}=\xi|a|$.

These purely kinematical considerations do not, a priori, prove that
such decays actually take place, but we will see later in section 5,
from an entirely different perspective, that the quantum BPS
states indeed do decay (or rather degenerate) precisely in this
manner.

With a more detailed analysis \cite{BiFe}, employing the global
symmetry $u\to-u$, one can show that the only stable BPS states in
$\cM_q^{{\rm strong}}$ are indeed precisely the monopole and the
dyon, and no other states. Furthermore, one can show that the
semi-classical, stable BPS spectrum in $\cM_q^{{\rm weak}}$ consists
precisely of the above-mentioned states $\pm(0,2)$ and
$\pm(1,2\ell)$, $\ell\in \ZZ$, and of no other states.

\subsection{Picard-Fuchs equations}

In order to obtain the effective action explicitly, one needs to
evaluate the period integrals \eq{perdef}. However, instead of
directly computing the integrals, one may use the fact that the
periods form a system of solutions of the Picard-Fuchs equation
associated with the curve \eq{thecu}. One then has to evaluate the
integrals only in leading order, just to determine the correct linear
combinations of the solutions.

Concretely, in order to derive the PF equations (see also refs.\
\cite{PF}), let us first write the defining relation of the
curve \eq{thecu} in homogenous form, by introducing another
coordinate $z$:
\beq
W(x,y,z,u)\ \equiv\ (x^2-u\, z^2)^2-z^4-y^2\ =\ 0\
\eel{homogW}
(here we have set $\La=1$). We also introduce the following integrals
over certain globally defined one-forms:
\beq
\Om_1\ =\ \oint_\gamma\Coeff1W\,\bar\omega\ ,\quad
\Om_2\ =\ \oint_\gamma\Coeff{x^2z^2}{W^2}\,\bar\omega\ ,
\eel{Omforms}
where $\gamma$ is a one-cycle that winds around the surface $W=0$,
and $\bar\omega$ is an appropriate volume form on $\IP^3$. The point
is that we do not need to evaluate these periods by explicitly
performing the integrations. Rather, the integrands should be
considered here as dummy variables, introduced only to conveniently
derive the PF equations that will then be solved by other means.
By elementary algebra one easily finds:
\beq
\!\!{\partial\over\partial u}\Om_1\ =\
\oint_\g {2z^2(x^2-u\,z^2)\over W^2}\omega
\eeq\vskip-.4truecm
\beq
= \-{2\over (u^2-1)}\Om_2 -
\oint_\gamma {u\,z\over2(u^2-1)}{\partial_z W\over W^2}\omega ,
\eel{firsteq}
where we have used in the second line the following expansion into
``ring elements and vanishing relations'':
$$
2z^2(x^2-u\,z^2) \equiv -{2\over (u^2-1)}x^2z^2-{u\over
2(u^2-1)}z\partial_zW .
$$
Integrating by parts, we can cancel $W$ in the second term to
get
$$
{\partial\over\partial u}\Om_1\ =\ -{2\over (u^2-1)}\Om_2-{u\over
2(u^2-1)}\Om_1\ .
$$
We can repeat a similar game for $\Om_2$, and obtain, after multiple
partial integrations, the following differential identity:
\bea
{\partial\over\partial u}\Om_2\
&=&\ \oint_\g x\,z^4{\partial_x W\over W^3}\omega\ \cr
&=&\ {1\over 8(u^2-1)}\Om_1+{u\over 2(u^2-1)}\Om_2\ .
\eeal{secondeq}
We now can eliminate $\Om_2$ from \equ{firsteq} and \equ{secondeq} to
obtain a differential equation for the fundamental period:
${\cal L}\Om_1=0$, with
$
{\cal L}=(\Lambda^4-u^2)\partial_u^2-2 u\partial_u-{1\over4}
$.
This Picard-Fuchs equation is supposed to be satisfied by all the
periods, in particular by $(\vp_D(u),\vp(u)) \equiv (\partial_u
a_D,\partial_u a)$. In terms of the variable $\alpha={u^2\over
\Lambda^4}$, the PF differential operator turns into
$(\theta_\alpha=\alpha\partial_\alpha)$
\beq
{\cal L}=
\theta_\alpha (\theta_\alpha-{1\over2})-
\alpha(\theta_\alpha+{1\over4})^2\ ,
\eel{PFone}
which constitutes a hypergeometric system of type $(a,b,c)=
({1\over4},{1\over4};{1\over2})$.

It is also possible to derive a second order differential equation
for the section $(a_D,a)$ directly \cite{KLT}. In fact, one easily
verifies that ${\cal L}\partial_u=\partial_u\tilde{\cal L}$ with
\beq
\tilde{\cal L}=\theta_\alpha (\theta_\alpha -{1\over 2}) -\alpha
(\theta_\alpha-{1\over 4})^2\ ,
\eel{PFtwo}
and this forms a hypergeometric system of type
$(-{1\over4},-{1\over4};{1\over2})$. One may also check
directly that $\tilde{\cal L}\cdot\oint\lambda=0$.

The solutions of $\tilde\cL\,(a_D(u),a(u))=0$ in terms of
hypergeometric functions, and their analytic continuation over the
complex plane, are of course well known. For $|u| > |\Lambda|$ a
system of solutions to the Picard-Fuchs equations is given by $w_0$
and $w_1$ with
$$
w_0(u)={\sqrt{u}\over\La}\sum c(n)({\Lambda^4\over u^2})^n\,,\ \
c(n)={({1\over4})_n (-{1\over4})_n\over(1)_n^2}
$$
$$
{\rm and\ }w_1(u)=w_0(u) \log({\Lambda^4\over u^2})+
{\sqrt{u}\over\La}\sum d(n) ({\Lambda^4\over u^2})^n,
$$
\bea
{\rm where\ }&&d(n) \equiv c(n)\bigl[2(\psi(1)-\psi(n+1))\non\\
&&\!\!\!\!\!\!\!+\psi(n+{1\over4})-\psi({1\over4})+
\psi(n-{1\over4})-\psi(-{1\over4})\bigr]\non
\eeal{dncn}
and where $(a)_m\equiv \Gamma(a+m)/\Gamma(a)$ is the Pochhammer
symbol and $\psi$ the digamma function.
Matching the asymptotic expansions of the period integrals
one finds
\bea
a(u)&=&{\Lambda\over \sqrt{2}}w_0(u)\\
a_D(u)&=&-{i\Lambda\over \sqrt{2}\pi}
\Big[w_1(u)+(4-6\log(2)) w_0(u)\Big]\ ,\non
\eeal{auaDu}
which transform under counter-clockwise continuation of $u$ along
$\gamma_\infty$ (c.f., \figref{monpath}) precisely as in \eq{Minf}.
These expansions correspond to particular linear combinations of
hypergeometric functions, the most concise form of which are
\bea
a_D(\al)&=&\!\!{i\over 4}\La (\al-1)\,
_2F_1\Big(\Coeff34,\Coeff34, 2 ; 1-\al\Big)\non\\
!\!\!\!\!\!a(\al) &=&\!\!  {1\over\sqrt2}\La {\al}^{1/4}\,
_2F_1\Big(\!-\!\Coeff14,\Coeff14,1;\Coeff1{\al}\Big).\non
\eeal{aaD}
{}From these expressions the prepotential in the semi-classical
regime near infinity in the moduli space can readily be computed to
any given order. Inverting $a(u)$ as series for large $a/\Lambda$
yields for the first few terms ${u(a)\over \Lambda^2}=2 \left(a\over
\Lambda\right)^2 + {1\over 16} \left(\Lambda \over a\right)^2+{5
\over 4096} \left(\Lambda\over a\right)^6 + O(\left(\Lambda\over
a\right)^{10})$. After inserting this into $a_D(u)$, one obtains
${\cal F}$ by integration as follows:
\bea
{\cal F}(a)\!=\!{i\, a^2\over 2 \pi}
 \!\left( 2\log {a^2\over \Lambda^2}\!-\! 6
+8 \log 2\!-\!\!
\sum_{\ell=1}^\infty\! c_\ell\! \left(\Lambda\over a\right)^{4
\ell}\right).\non
\eeal{Fa}
It has indeed the form advertised in \eq{Feff}.
Specifically, the first few terms of the instanton expansion are:
$$
\vbox{\offinterlineskip\tabskip=0pt
\halign{\strut\vrule#&
\hfil~~$#$~~&
\hfil~~$#$~~&
\hfil~~$#$~~&
\hfil~~$#$~~&
\hfil~~$#$~~&
\hfil~~$#$~~&
\hfil~~$#$~~&
\vrule#\cr
\noalign{\hrule}
&\ell  &1 & 2 &  3& 4   & 5   & 6     & \cr
\noalign{\hrule}
&c_\ell&
\ds{1\over 2^5}&
\ds{5\over 2^{14}}&
\ds{3\over 2^{18}}&
\ds{1469\over 2^{31}}&
\ds{4471\over 2^{34} \cdot 5} &
\ds{40397\over 2^{43}}  &\cr
\noalign{\hrule}}
\hrule}
$$

One can treat the dual magnetic semi-classical regime is an analogous
way. Near the point $u=\Lambda^2$ where the monopole becomes
massless, we introduce $z=(u-\Lambda^2)/(2 \Lambda^2)$ and rewrite
the Picard-Fuchs operator as
\beq
{\cal L}=z (\theta_z-{1\over 2})^2+\theta_z (\theta_z-1)\ .
\eel{PFdual}
At $z=0$, the indices are $0$ and $1$, and we have again one power
series
$$
w_0(z)=\Lambda^2\sum c(n) z^{n+1},
\ \ c(n)=(-1)^n{({1\over2})_n^2\over(1)_n (2)_n}\non
$$
and a logarithmic solution
$$
w_1(z)=w_0(z) \log(z)+
\sum d(n) z^{n+1} -4\ ,\non
$$
with
\bea
&&d(n) \equiv c(n)\Bigl[2(\psi(n+{1\over2})-\psi({1\over2}))
+\psi(n+{1\over4})\non\\
&&\!\!\!\!\!\!\!-\psi({1\over4})+
+\psi(1)-\psi(n+1)+\psi(2)-\psi(n+2)\Bigr]\ .\non
\eeal{dncn1}

For small $z$ one can easily evaluate
the lowest order expansion of the period integrals and
thereby determine the analytic continuation of the
solutions from the weak coupling to the strong coupling domain:
\bea
a_D &=&\!2\! \int_{e_2}^{e_3}\!\!\lambda =
i \Lambda w_0(z)\non\\
a &=&\! 2\! \int_{e_1}^{e_2}\!\!\lambda\
\!=\! -\! {\Lambda\over 2 \pi}(w_1(z)\!-\!(1\!+\!\log (2))
w_0(z)).\non
\eeal{aDa}
This exhibits the monodromy of \eq{Mone} along the path
$\gamma_{+\Lambda^2}$. Inverting $a_D(z)$ yields $z(a_D)= -2{\tilde
a}_D+\coeff14{{\tilde a}_D}^2+\coeff1{32} {{\tilde a}_D}^3+{\cal
O}({{\tilde a}_D}^4)$, with ${\tilde a}_D\equiv i a_D/\La$. After
inserting this into $a(z)$ we integrate w.r.t. $a_D$ and obtain the
dual prepotential ${\cal F}_D$ as follows:
\bea
{\cal F}_D(a_D)={i\La^2\over 2 \pi} \left(
{\tilde a}_D^2\log \Big[-{i\over4}\sqrt{{\tilde a}_D}\Big]
+ \sum_{\ell=1}^\infty c^D_{\,\ell}\, {\tilde a}_D^\ell
\right),\non
\eeal{FaD}
where the lowest threshold correction coefficients $c^D_{\,\ell}$ are
$$
\vbox{\offinterlineskip\tabskip=0pt
\halign{\strut\vrule#&
\hfil~~$#$~~&
\hfil~~$#$~~&
\hfil~~$#$~~&
\hfil~~$#$~~&
\hfil~~$#$~~&
\hfil~~$#$~~&
\hfil~~$#$~~&
\vrule#\cr
\noalign{\hrule}
&\ell  &1 & 2 &  3& 4   & 5   & 6     &\cr
\noalign{\hrule}
&c^D_{\,\ell}&
\ds{4}&
-\ds{3\over 4}&
\ds{1\over 2^4}&
\ds{5\over 2^9}&
\ds{11\over 2^{12}} &
\ds{63\over 2^{16}}  &\cr
\noalign{\hrule}}
\hrule}$$
They reflect the effect of integrating out
the massive BPS spectrum near $u=\La^2$.

\section{Generalization to other Gauge Groups}

 The above construction for $SU(2)$ Yang-Mills theory can be
generalized in many ways; for example, one may add extra matter
fields \cite{withmatter,massprep}, and/or consider other
gauge groups \cite{suN,othergrps,MW}. For lack
of space, we will confine ourselves in the lectures to the extension
of pure Yang-Mills theory to simply laced gauge groups of type $ADE$
(though interesting phenomena can arise when matter is added
\cite{withmatter}).

\subsection{Simple Singularities}

We will first outline the group theoretical aspects for
$G=SU(n)$, and present the discussion in a particular way that
follows \cite{KLT}: namely by starting with the classical theory.
Indeed, interesting features appear in a simplified fashion
already at the classical level, and some of them will play an
important role in the generalization to string theory.

Just like as for $G=SU(2)$, the scalar superfield component $\phi$
labels a continuous family of inequivalent ground states that
constitutes the classical moduli space, $\cM_c$. One can always
rotate $\phi$ into the Cartan sub-algebra,
$
\phi=\sum_{k=1}^{n-1}a_k H_k
$,
with $H_k=E_{k,k}-E_{k+1,k+1},\,
(E_{k,l})_{i,j}=\delta_{ik}\delta_{jl}$. For generic eigenvalues of
$\phi$, the $SU(n)$ gauge symmetry is broken to the maximal torus
$U(1)^{n-1}$. However, if some eigenvalues coincide, then some
larger, non-abelian group $H\subseteq G$ remains unbroken. Precisely
which gauge bosons are massless for a given background $a=\{a_k\}$,
can easily be read off from the central charge formula. For an
arbitrary charge vector $q$, this formula reads:
\beq
Z\ =\ e_{q}(a)\  =\  q\cdot  a\ ,\ \ \ \
{\rm with\ }\ \  m^2\ =\ |Z|^2\ ,
\eel{edef}
and in the present context we take for the charge vectors $q$ of the
gauge bosons the roots $\alpha\in\La_R(G)$ in Dynkin basis.

The Cartan sub-algebra variables $a_k$ are not gauge invariant and in
particular not invariant under discrete Weyl transformations.
Therefore, one introduces other variables for parametrizing the
classical moduli space, which are given by the Weyl invariant
Casimirs $u_k(a)$, $k=2,...,n$. These variables parametrize the
(complexified) Cartan sub-algebra modulo the Weyl group, ie,
$\{u_k\}\cong \IC^{n-1}/S(n)$, and can formally be generated as
follows:
\bea
P^{{n}}_{A_{n-1}} &\equiv& \det_{n\times n}\big[\,x{\bf1}-
\phi\,\big]\ =\
\prod_{i=1}^n\big(x-e_{\la_i}(a)\big)\non\\ &=&\
x^n-\sum_{l=0}^{n-2}u_{l+2}(a)\,x^{n-2-l}\non\\
&\equiv& \simpA(x,u)\ .
\eeal{WAnSing}
Here, $\la_i$ are the weights of the $n$-dimensional fundamental
representation, and $\simpA(x,u)$ is nothing but the ``simple
singularity''\foot {More precisely, $W(x,0)=0$ has a singularity of
type $A_{n-1}$ at the origin, which is resolved by switching on the
$u_k$.} \cite{Arn,Slodo} associated with $SU(n)$, with
\bea
u_k(a)\ =\ (-1)^{k+1}\sum_{j_1\not=...
\not=j_k}e_{\la_{j_1}}e_{\la_{j_2}}\dots e_{\la_{j_k}}(a)\ .\non
\eeal{symmp}
These symmetric polynomials are manifestly invariant under the Weyl
group $S(n)$, which acts by permutation of the weights $\la_i$.

{}From the above we know that whenever $e_{\la_i}(a)=e_{\la_j}(a)$
for some $i$ and $j$, there are, classically, extra massless
non-abelian gauge bosons, since the central charge vanishes:
$e_{\al}=0$ for some root $\al$. For such backgrounds the effective
action becomes singular. The classical moduli space is thus given by
the space of Weyl invariant deformations, except for such singular
regions: $\CM=\{u_k\}\backslash\bifset_0$. Here, $\bifset_0 \equiv
\{u_k: \Delta_0(u_k)=0\}$ is the zero locus of the discriminant
\bea
\Delta_0(u) =
\!\!\prod_{i<j}^n(e_{\la_i}(u)-e_{\la_j}(u))^2 =\!\!\!\!
\prod_{{{\rm positive}\atop{\rm roots}\
\alpha}}\!\!\!\!(e_{\alpha}(u))^2\ \
\eeal{cdisc}
of the simple singularity \eq{WAnSing}. We schematically depicted
(the real slices of) the singular loci $\bifset_0$ for $n=2,3,4$
in \figref{clasdisc}.

\figinsert{clasdisc}
{Singular loci $\bifset_0$ in the classical moduli
spaces $\cM_c$ of pure $SU(n)$ \nex2 Yang-Mills theory. They are
nothing but the bifurcation sets of the type $A_{n-1}$ simple
singularities, and reflect all possible symmetry breaking patterns in
a gauge invariant way (for $SU(3)$ and $SU(4)$ we show only the real
parts). The picture for $SU(4)$ is known in singularity theory as the
``swallowtail''. }
{1.0in}{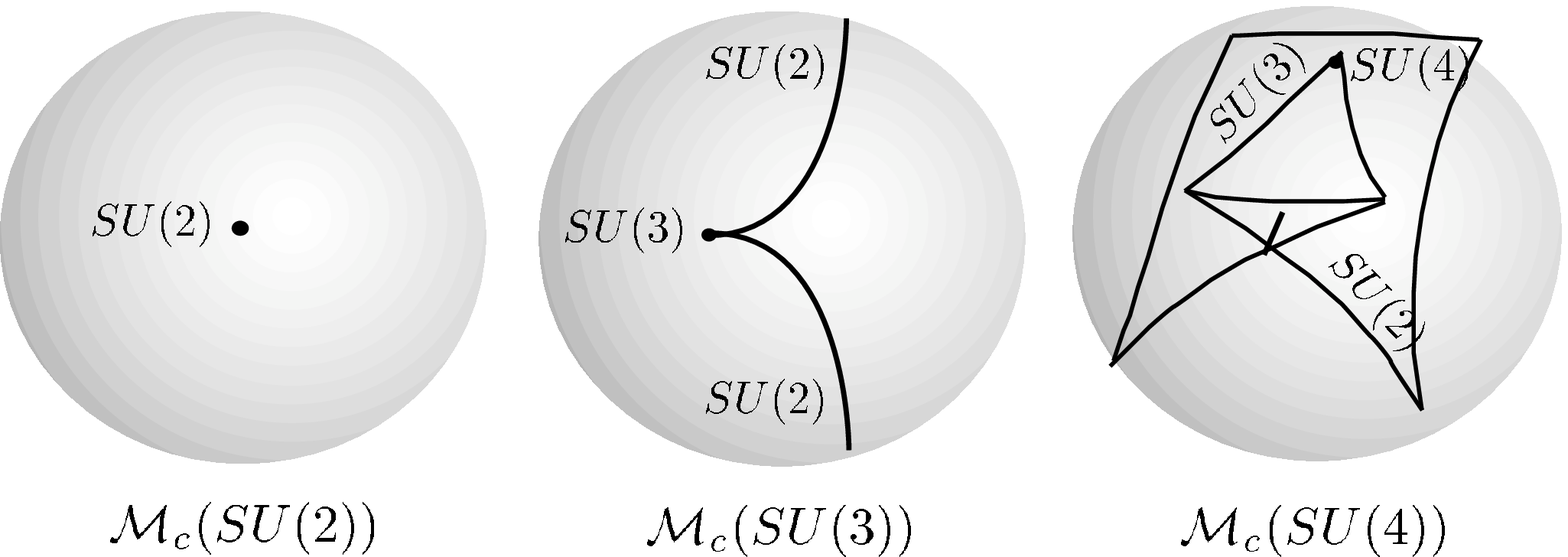}

The discriminant loci $\bifset_0$ are generally given by intersecting
hypersurfaces of complex codimension one. On each such surface one
has $e_{\al_i}=0$ for some pair of roots $\pm\al_i$, so that there is
an unbroken $SU(2)$. In total, there are ${1\over2}n(n-1)$ of such
branches $\Sigma_0^{\al_i}$. On the intersections of these branches
one has, correspondingly, larger unbroken gauge groups. All surfaces
intersect together in just one point, namely in the origin, where the
gauge group $SU(n)$ is fully restored. Thus, what we learn is that
all possible classical symmetry breaking patterns are encoded in the
discriminant loci of the simple singularities, $\simpA(x,u)$.

In previous sections we have seen that $SU(2)$ quantum Yang-Mills
theory is characterized by an auxiliary elliptic curve. In a more
general context, one may view it as a ``spectral'', or ``level''
manifold. The relationship between BPS states and cycles on an
auxiliary manifold $X$ seems in fact to be quite generic. As we will
see, there is a whole variety of such manifolds, describing various
different physical systems. In these notes, we will denote generic
spectral manifolds of complex dimension $d$ by $X_d$.

Indeed one may introduce here this concept to describe {\it
classical} YM theory as well, and characterize BPS states (the
non-abelian gauge bosons) by an auxiliary manifold $X=X_0$. This
level manifold is zero dimensional and simply given by the following
set of points:
\beq
X_0\, =\ \big\{\,x\,:\,\simpA(x,u)=0\,\big\}\ =\
\big\{\,e_{\la_i}(u)\,\big\}.
\eel{levsurf}
It is singular if any two of the $e_{\la_i}(u)$ coincide, and the
vanishing cycles are simply given by the formal differences: $\nu_\al
=e_{\la_i}-e_{\la_j}=e_{\al}$, i.e., by the central charges \eq{edef}
associated with the non-abelian gauge bosons. Obviously, massless
gauge bosons are associated with vanishing 0-cycles of the spectral
set $X_0$. It is indeed well-known \cite{Arn} that such 0-cycles
$\nu_\al$ generate the root lattice:
\beq
H_0(X_0,\ZZ)  \cong\ \Gamma_R^{SU(n)}\,.
\eel{H0rel}
We depicted the level surface for $G=SU(3)$ in \figref{claslevel}.
Such kind of pictures has a concrete group theoretical meaning
-- given the locations of the dots $\{x_i=\la_i\cdot a\}$, they just
represent projections of weight diagrams.

\figinsert{claslevel}
{Level manifold $X_0$ for classical $SU(3)$ Yang-Mills theory,
given by points in the $x$-plane that form a weight diagram. The
dashed lines are the vanishing cycles associated with non-abelian
gauge bosons (having corresponding quantum numbers, here in Dynkin
basis). The masses are proportional to the lengths of the lines and
thus vanish if the cycles collapse.}
{1.1in}{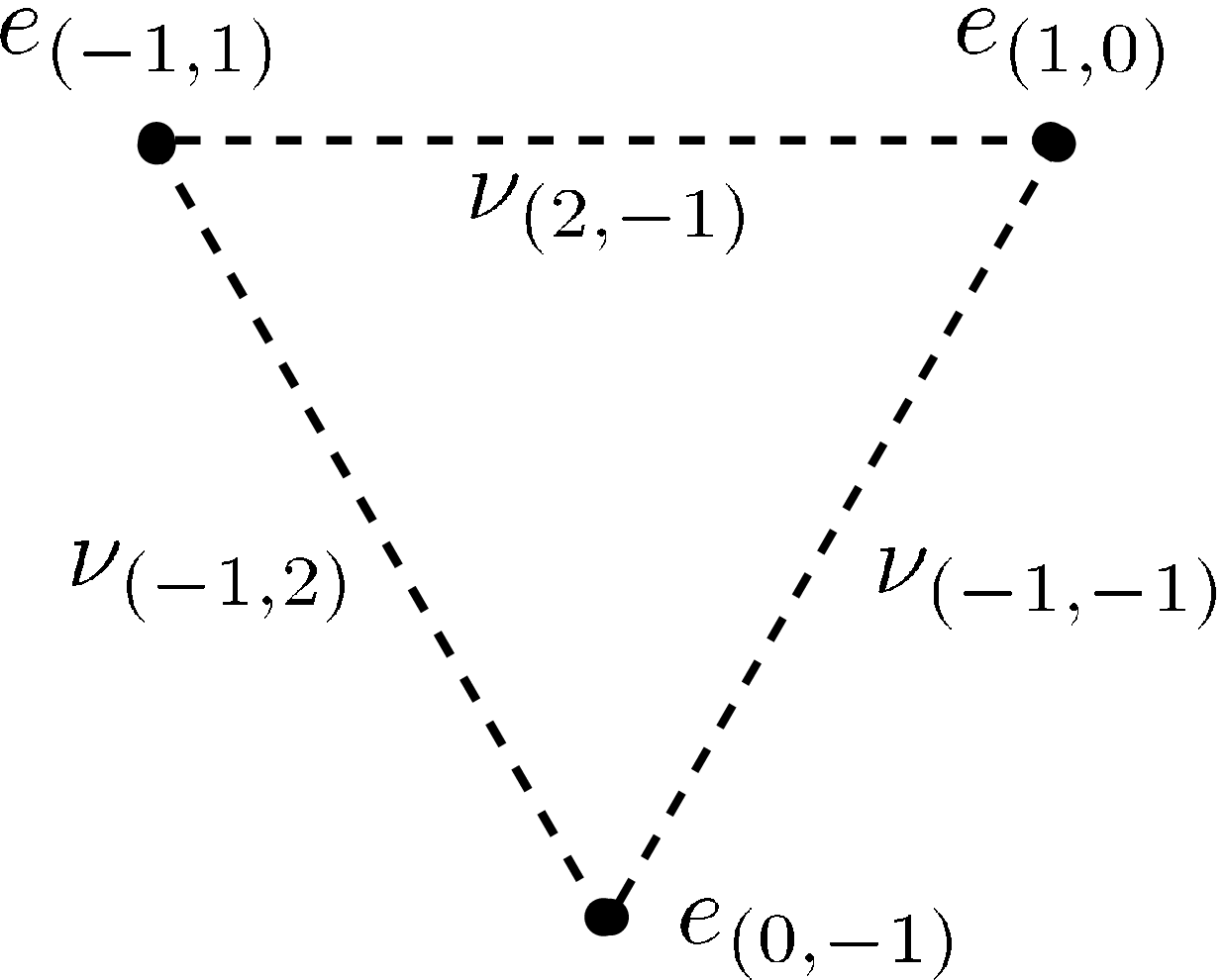}

We thus see a close connection between the vanishing homology of
$X_0$ and $SU(n)$ weight space. Indeed, the intersection numbers of
the vanishing cycles are just given by the inner products between
root vectors, $\nu_{\al_i}\circ\nu_{\al_j} =
\langle\al_i,\al_j\rangle$ (self-intersections counting $+2$), and
the Picard-Lefshetz formula \eq{PicLef} coincides in this case with
the well-known formula for Weyl reflections, with matrix
representation:
$
M_{\al_i} = {\bf 1} - \al_i\otimes w_i
$
(where $w_i$ are the fundamental weights).

\subsection{Classical Theory: Other Gauge Groups}

The above considerations apply \cite{Arn,othergrps,MW} more or less
directly to the other simply laced Lie groups of types $D$ and
$E$.\foot {And essentially to non-simply laced groups as well, for
which ``boundary singularities'' are relevant. See \cite{Arn} for
details.} However, there are marked differences in that the
corresponding simple singularities
\bea
W_{D_n} &=& {x_1}^{n-1}\!+\half x_1{x_2}^2-\!\sum_{k=1}^{n-1}\!
x^{n-1-k} u_{2k} \!-\! \tilde u_{n-1} x_2 \non\\
W_{E_6} &=& {x_1}^3+{x_2}^4 - u_2 x_1 x_2^2-u_5 x_1 x_2-u_6
x_2^2 \non\\ &&-u_8 x_1-u_9 x_2-u_{12} \label{WDE}\\
W_{E_7} &=& {x_1}^3+x_1{x_2}^3 - u_2 x_1^2 x_2-u_6 x_1^2-u_8
x_1 x_2  \non\\ &&-u_{10} x_2^2 -u_{12} x_1-u_{14} x_2-u_{18}\non\\
W_{E_8} &=& {x_1}^3+{x_2}^5 - u_2 x_1 x_2^3-u_8 x_1 x_2^2-
u_{12} x_2^3 \non\\ &&-u_{14} x_1 x_2 -u_{18} x_2^2-u_{24}
x_2-u_{30}\non
\eeal{WDE1}
involve an extra variable, $x_2$ ($u_k$ denote the $r\equiv rank(G)$
independent Casimirs, which have the degrees $k$ as indicated). And
in contrast to $A_{n-1}$ (cf., \equ{WAnSing}), these have a priori no
simple relationship to the characteristic polynomials
\bea
P^{{\cal R}}_{ADE} &\equiv& \det_{{\cal R}} [x {\bf1}-\phi]
\ = \prod_{\la_i\in{\cal R}}\big(x-\la_i\cdot a\big)\\
&=& x^{{\rm dim}{\cal R}}+\big[{\rm lower\ order\ terms\
in\ } x,u_k\big]\,, \non
\eeal{charP}
where $\cal R$ is some, say fundamental, representation of the gauge
group $G$. For example, for $G=E_6$, $W_{E_6}(x_1,x_2)$ is of degree
12, while $P^{{27}}_{E_6}(x)$ is of order 27 -- so these polynomials
are really quite different from each other. The point is that the
equations $W_{ADE}=0$ and $P^{{\cal R}}_{ADE}=0$ have the same
relevant information content (in fact, for arbitrary representations
$\cal R$); the simple singularities \equ{WDE} are in a sense more
efficient in encoding this information, in that the overall scaling
degree is minimized (given by the dual Coxeter number $h$), at the
expense of introducing another variable, $x_2$.

In effect, both equations $W=0$ and $P=0$ can be taken to define an
auxiliary spectral surface $X$. However, for $D,E$ gauge groups the
surfaces $W(x_1,x_2)=0$ happen to be no longer zero dimensional.
\foot{Actually, as we will see later, the best way to think about
this is to add a third variable $x_3$ and promote $W=0$ to an
``ALE space''.} For $D_n$ one can ``integrate out'' the variable
$x_2$ and thereby relate $W=0$ to $P=0$. That is, we can simply
eliminate $x_2$ via the ``equation of motion''
$\partial_{x_2}W_{D_n}(x_1,x_2)=0$. Multiplying $W_{D_n}$ by $x_1$
and substituting $x_1=x^2$, we then indeed get
$$
x^{2n}-\sum_{k=1}^{n-1}x^{2n-2k}u_{2k}-\half {\tilde u_{n-1}}^2
\ \equiv\ P^{{2n}}_{D_n}(x,u)=0.
$$
The relation between $W=0$ and $P=0$
is however much more complicated for the
exceptional groups; see \cite{ALESW} for $E_6$.

\subsection{Quantum $SU(n)$ Gauge Theory}

We now turn to the quantum version of the \nex2 Yang-Mills theories,
where the issue is to construct curves $X_1$ whose moduli spaces
give the supposed quantum moduli spaces, $\cM_q$. We have seen that
the classical theories are characterized by simple singularities,
so we may expect that the quantum versions should also have something
to do with them. Indeed, for $G=SU(n)$ the appropriate manifolds
were found in \cite{suN} and can be represented by
\beq
X_1:\ \ y^2 \ =\ \left(W_{A_{n-1}}(x,u_i)\right)^2-\Lambda^{2n}\
\ ,
\eel{aaa}
which corresponds to special genus $g=n-1$ hyperelliptic curves.
Above, $\La$ is the dynamically generated quantum scale.

 Since $y^2$ factors into $W_{A_{n-1}}\pm\La^n$, the situation is in
some respect like two copies of the classical theory, with the top
Casimir $u_n$ shifted by $\pm\La^n$. Specifically,
the points $e_{\la_i}$ of the
classical level surface \eq{levsurf} split as follows,
$$
e_{\la_i}(u)\ \to\ e_{\la_i}^\pm(u,\La) \equiv\
e_{\la_i}(u_2,,...,u_{n-1},u_n\pm\La^n)\ ,
$$
and become the $2n$ branch points of the Riemann surface \eq{aaa}.
The curve can accordingly be represented by the two-sheeted $x$-plane
with cuts running between pairs $e_{\la_i}^+$ and $e_{\la_i}^-$. See
\figref{xplane} for an example.

\figinsert{xplane}
{The spectral curve of quantum $SU(3)$ Yang-Mills theory is given by
a genus two Riemann surface, which is represented here as a
two-sheeted cover of the $x$-plane. It may be viewed as the quantum
version of the classical, zero dimensional surface $X_0$ of Fig.10,
whose points transmute into branch cuts. The dashed lines represent
the vanishing 1-cycles (on the upper sheet) that are associated with
the six branches $\Sigma_{\pm}^{i}$ of the singular locus
$\Sigma_\La(SU(3))$. The quantum numbers refer to $(g;q)$, where
$g,q$ are root vectors in Dynkin basis.} {1.75in}{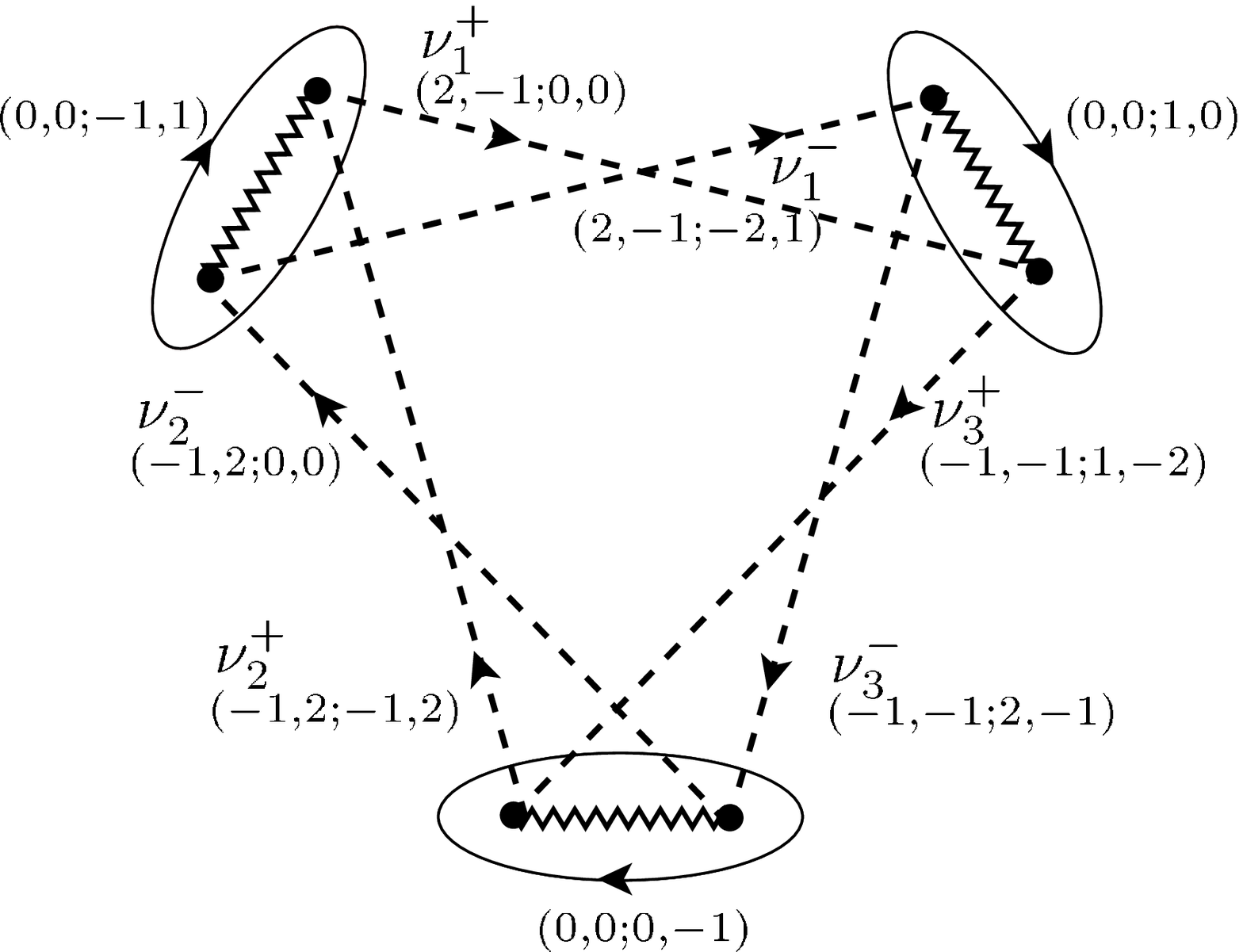}

Moreover, the ``quantum'' discriminant, whose zero locus
$\bifset_\La$ gives the singularities in the quantum moduli space
$\cM_q$, is easily seen to factorize as follows:
\bea
\Delta_\Lambda(u_k,\La)\ &=&\
\prod_{i<j}(e_{\la_i}^+-e_{\la_j}^+)^2(e_{\la_i}^--e_{\la_j}^-)^2
\non\\ &=&
{\rm const.}\,\La^{2n^2}
\delta_+\,\delta_-\ ,\label{DLdef}\\ \ \delta_\pm(u_k,\Lambda)
&\equiv&
\Delta_0(u_2,...,u_{n-1},u_n\pm\La^n)\ ,\non
\eeal{DLdef1}
is the shifted classical discriminant, \eq{cdisc}. Thus,
$\bifset_\La$ consists of two copies of the classical singular locus
$\Sigma_0$, shifted by $\pm\La^n$ in the $u_n$ direction. Obviously,
for $\La\to0$, the classical moduli space is recovered:
$\bifset_\La\to \bifset_0$. That is, when the quantum corrections are
switched on, a single isolated branch $\Sigma_{0}^{\al_i}$ of
$\bifset_0$ (associated with massless gauge bosons of a particular
$SU(2)$ subgroup) splits into two branches $\Sigma_{\pm}^{i}$
of $\bifset_\La$ (reflecting two massless dyons related to this
$SU(2)$). For $G=SU(3)$, this is depicted in \figref{qudisc}.

\figinsert{qudisc}
{When switching to the exact quantum theory, the classical singular
locus splits into two quantum loci that are associated with
massless dyons; this is completely analogous to Fig.2. The distance
is governed by the quantum scale $\La$. Shown are here the six
branches $\Sigma_{\pm}^{i}$ for $G=SU(3)$.
}{2.8in}{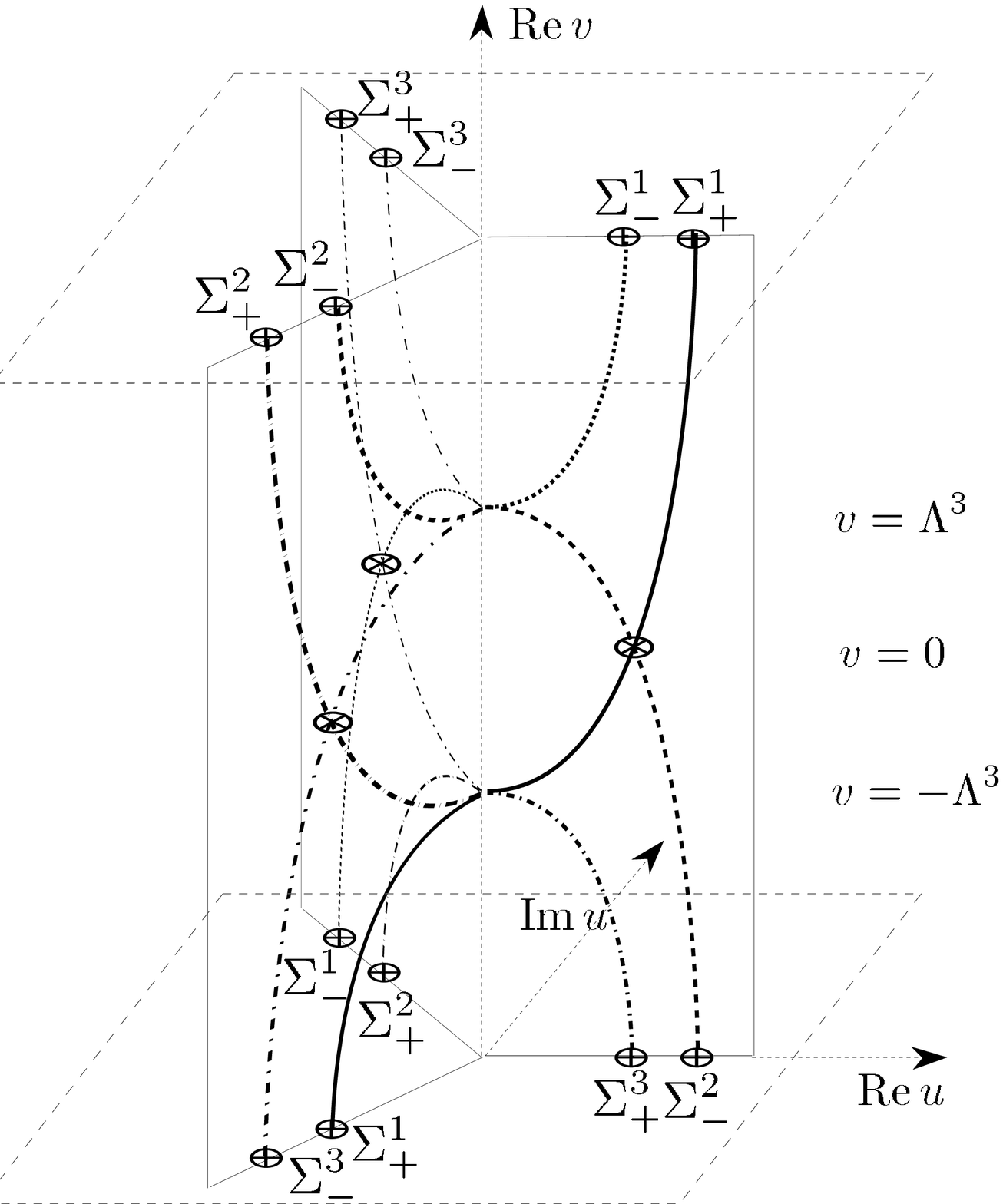}

\newcommand\here{
{\tt here}} According to the line of arguments
given below equ.~\equ{nudef}, all what it takes to determine the dyon
spectrum associated with the $n(n-1)$ singular branches, is to
determine the set of one-cycles $\nu^{i}_\pm$ that vanish on the
$\Sigma^{i}_\pm$, with respect to some appropriate symplectic basis
of $\al$ and $\be$ cycles. This can be done by tracing the exchange
paths of the branch points $e_{\la_i}^\pm$ when we encircle the
components of the discriminant $\Sigma^{i}_{\pm}$ in the moduli space
(starting from and ending at an arbitrary, but fixed base
point).\hfoot {Hypertex capable on-line readers may click \here\ to
obtain a Mathematica notebook that shows how this can be done in
practice.}

The result can be characterized in a very simple way: the punctured
$x$-plane (cf., \figref{xplane}) can be thought of as a ``quantum
deformation'' of the classical level surface $X_0$ (cf.,
\figref{claslevel}), and thus inherits its group theoretical
properties. We already mentioned above that the points $e_{\la_i}$ of
$X_0$, associated with the projected weight vectors $\la_i$, turn
into branch cuts, whose length is governed by the quantum scale,
$\Lambda$ (in fact, one obtains two, slightly rotated copies of the
weight diagram).

Now, a basis of cycles can be chosen such that the coordinates of the
``electric'' $\al$-type of cycles around the cuts are given precisely
by the corresponding weight vectors $\la_i$. That is, we can
associate charges $(g;q)=(0;\la_i)$ with the $\al$-cycles. Moreover,
the classical cycles of $X_0$ (related to the roots $\al_i$), turn
into pairs of ``magnetic'' $\be$-type of cycles. By consistently
assigning charge vectors to all vanishing cycles, we can then
immediately read off the electric and magnetic quantum numbers of the
massless dyons: they are given by specific combinations of root
vectors. For $G=SU(3)$, this is indicated in \figref{xplane}.

At this point a feature that is novel for $SU(n)$, $n>2$, becomes
evident: there are regions in $\cM_q$ where {\it mutually non-local}
dyons become simultaneously massless \cite{AD}. Indeed, as can be
inferred from Figs.10,11 for $G=SU(3)$, at $u\equiv u_2=0$, $v\equiv
u_3=\pm\La^n$, dyons are massless whose vanishing cycles have
non-zero intersection numbers, $\nu_i\circ\nu_j\not=0$; this
phenomenon persists for higher $n$ as well. In other words, their
Dirac-Zwanziger charge product \eq{dzw} does not vanish, and this
means, as mentioned before, that they are not local with respect to
each other.

Whenever this happens, then by general arguments \cite{conformal} the
theory becomes conformally invariant. From \eq{aaa} it is clear that
near such an ``Argyres-Douglas'' point the curve looks locally like
$y^2=W_{A_{n-1}}=x^n+...$, and thus effectively behaves like a genus
$g=(n-1)/2$ (for $n=$odd, $g=n/2-1$ for even $n$) curve that has a
singularity of type $A_{n-1}$. This is the same singularity type that
the classical level set $X_0$ \eq{levsurf} has at the conformally
invariant point, $u_l=0$. Indeed one may view the AD points as
arising from ``splitting and shifting'' the classical $A_{n-1}$
singularities,\foot {This generalized to a whole series of $d=4$,
\nex2 superconformal theories, classified by the $ADE$ Lie algebras
\cite{conformal}.} analogously to what saw in \figref{Mq} for
$SU(2)$. However, whereas the classical theory has a gauge symmetry
at the singularity, the SW theory appears not to have massless gauge
bosons at the AD points \cite{AD}. Rather, the SW theory may have
some sort of novel symmetry, but this is not yet completely settled.

To obtain the effective action (i.e., prepotential
\cite{prepot,massprep}), one must first determine the sections
$a_i(u_k), a_{D,i}(u_k)\equiv \partial_{a_i}{\cal F}(a)$,
appropriately defined as period integrals. For theories with more
than one modulus, the existence of a prepotential poses an
integrability condition, which can be solved by finding a suitable
meromorphic one-form $\la_{SW}$.

More specifically, the genus of the hyperelliptic curve $X_1$
\equ{aaa} is equal to $g=n-1$, so that its $2n-2$ periods can
naturally be associated with
\beq
\vec \pi\ \equiv\ \left({\vec a_{D}\atop\vec a}\right)\ .
\eeq
On such a curve there are $n-1$ holomorphic
differentials (abelian differentials of the first kind)
$\omega_{n-i}={x^{i-1}\,dx\over y},\, i=1,\dots,g$, out of which one
can construct $n-1$ sets of periods $\int_{\gamma_j}\omega_i$. (Here
$\gamma_j,\, j=1,\dots,2g$, is any basis of $H_1(X_1,\ZZ)$.)
All periods together can be combined in the
$(g,2g)$-dimensional period matrix
\beq
\Pi_{ij}=\int_{\gamma_j}\omega_i\ .
\eeq
If we chose a symplectic homology basis, i.e. $\al_i=\gamma_i,\,
\be_i=\gamma_{g+i},\,i=1,\dots,g$, with intersection pairing\foot{We
use the convention that a crossing between the cycles $\al$, $\be$
counts positively to the intersection $(\al\circ\be)$, if looking in
the direction of the arrow of $\al$ the arrow of $\be$ points to the
right.} $(\al_i\circ\be_i)=\delta_{ij},\,
(\al_i\circ\al_j)=(\be_i\circ\be_j)=0$, and if we write $\Pi=(A,B)$,
then $\tau\equiv A^{-1}B$ is the metric on the quantum moduli space.
By Riemann's second relation, Im$(\tau)\equiv 8\pi^2/g_{eff}^2$ is
manifestly positive, which is important for unitarity of the
effective $N=2$ supersymmetric gauge theory.

The precise relation between the periods and the components of the
section $\vec\pi$ is given by:
\bea
A_{ij} &=& \int_{\al_j}\!\omega_i =
{\partial\over\partial u_{i+1}}\, a_j\non\\
B_{ij} &=& \int_{\be_j}\!\omega_i =
{\partial\over\partial u_{i+1}}\, a_{D_j}
\eeal{qqq}
(where $i,j=1,\dots,n-1$). From the explicit expression \equ{aaa} for
the family of hyperelliptic curves, one immediately verifies that the
integrability conditions $\partial_{i+1} A_{jk}=\partial_{j+1}
A_{ik},\,\partial_{i+1} B_{jk}=\partial_{j+1} B_{ik}$ are satisfied.
It also follows that $\tau_{ij}\equiv \del_{a_i}\del_{a_j}\cF(a)$.
This reflects the special geometry of the quantum moduli space, and
implies that the components of $\vec \pi$ can directly be expressed
as integrals
\beq
a_{D_i}=\int_{\be_i}\lambda_{SW}\,,\qquad a_i=\int_{\al_i}
\lambda_{SW}\ ,
\eeq
over a suitably chosen meromorphic differential. One may take,
for example \cite{KLTY,AF}:
\bea
\!\!\!\lambda_{SW}\!\!\!&=&\!\!\!\!\! {dx\over4\sqrt2\pi}
\log\Big[{
W_{A_{n-1}}\!\!+\!\sqrt{W_{A_{n-1}}^2\!\!-\!\La^{2n}}
\over
W_{A_{n-1}}\!\!-\!\sqrt{W_{A_{n-1}}^2\!\!-\!\La^{2n}}
}\Big]\non\\ \!\! &=&\!\!\!\!\!
{1\over2\sqrt2\pi}{\Big({\partial \over\partial
x}W_{\!A_{n-1}}(x,u_i)\!\Big)}{x\,dx\over y}\! +\!\partial[*].
\eeal{Lamdef}

Explicit expressions for the prepotentials \cite{PF,prepot,massprep}
can be obtained by first solving Picard-Fuchs equations, and
consequently matching the solutions with the asymptotic expansions of
the period integrals (in analogy to what we discussed in section 2.5;
recently, a more efficient method has been developed in ref.\
\cite{HKP}.) In fact, for a given group one can write down a whole
variety of effective actions that are valid in appropriate coordinate
patches in the moduli space; this is similar to what was shown in
\figref{bigpicture} for $G=SU(2)$.

Specifically, in the semi-classical coordinate patch, where by
definition the classical central charges are large, $e_{\al_i}\equiv
\al_i\cdot a \gg \La$, the prepotential has the form:\foot {This form
is valid for all $ADE$ groups; $C$ denotes the Cartan matrix,
$\tau_0$ the bare coupling and $h$ the dual Coxeter number ($h\equiv
n$ for $SU(n)$).}
\bea
\cF(a_i)\ =\ \cF_{\rm class}+\cF_{\rm 1-loop}+\cF_{\rm non-pert}\ ,
\eeal{www}
where
\bea
\cF_{\rm class}&=&\Coeff 12 \tau_0\, (a^t\cdot C \cdot a)
\phantom{\sum_{{{\rm positive}\atop{\rm roots}\ \alpha}}}\non\\
\cF_{\rm 1-loop}&=&\Coeff i{4\pi}\!\sum_{{{\rm positive}\atop{\rm
roots}\ \alpha}} {e_\al}^2 \log\,[{e_\al}^2/\La^2]\label{Fdef}\\
\cF_{\rm non-pert}&=&\!\!- \Coeff{i}{2\pi}\!\big(\!\!\!\sum_{{{\rm
positive}\atop{\rm roots}\ \alpha}}\!\! {e_\al}^{\!2}\big)
\sum_{\ell=1}^\infty\!\cF_{2h\ell}({e_\al}^{\!\!-1})\La^{2h\ell}.\non
\eeal{Fdef1}
Here $\cF_{2h\ell}({e_\al}^{-1})$ are Weyl invariant
Laurent polynomials in the $e_\al$ of degree $-2h\ell$. For example,
for $G=SU(3)$, $\cF_{6}={1\over4}\prod_\al {e_\al}^{-2}$; see refs.\
\cite{KLT,HKP}. for some further explicit expressions for
$\cF_{2h\ell}$. The one-loop term, $\cF_{\rm 1-loop}$, here obtained
from solving a differential equation, indeed coincides exactly with
what one obtains by a standard perturbative quantum field theory
computation~!

\subsection{Fibrations of Weight Diagrams}

There is an alternative representation of the SW curves $X_1$, which
is not manifestly hypergeometric and thus perhaps slightly less
convenient to deal with, but which can easily be generalized to
arbitrary gauge groups. As we will see, it is also precisely this
geometrically more natural form of the curves that arises in
string theory \cite{KLMVW}.

\figinsert{SWfib}
{The Seiberg-Witten curve can be understood as fibration of a weight
diagram (the classical surface $X_0$ of Fig.10 over $\IP^1$. Pairs of
singular points in the base are associated with vanishing $0$-cycles
in the fiber, i.e., to root vectors $a_i-a_j$. In string theory, the
local fibers will be replaced by appropriate ALE spaces with
corresponding vanishing two-cycles.}{1.7truein}{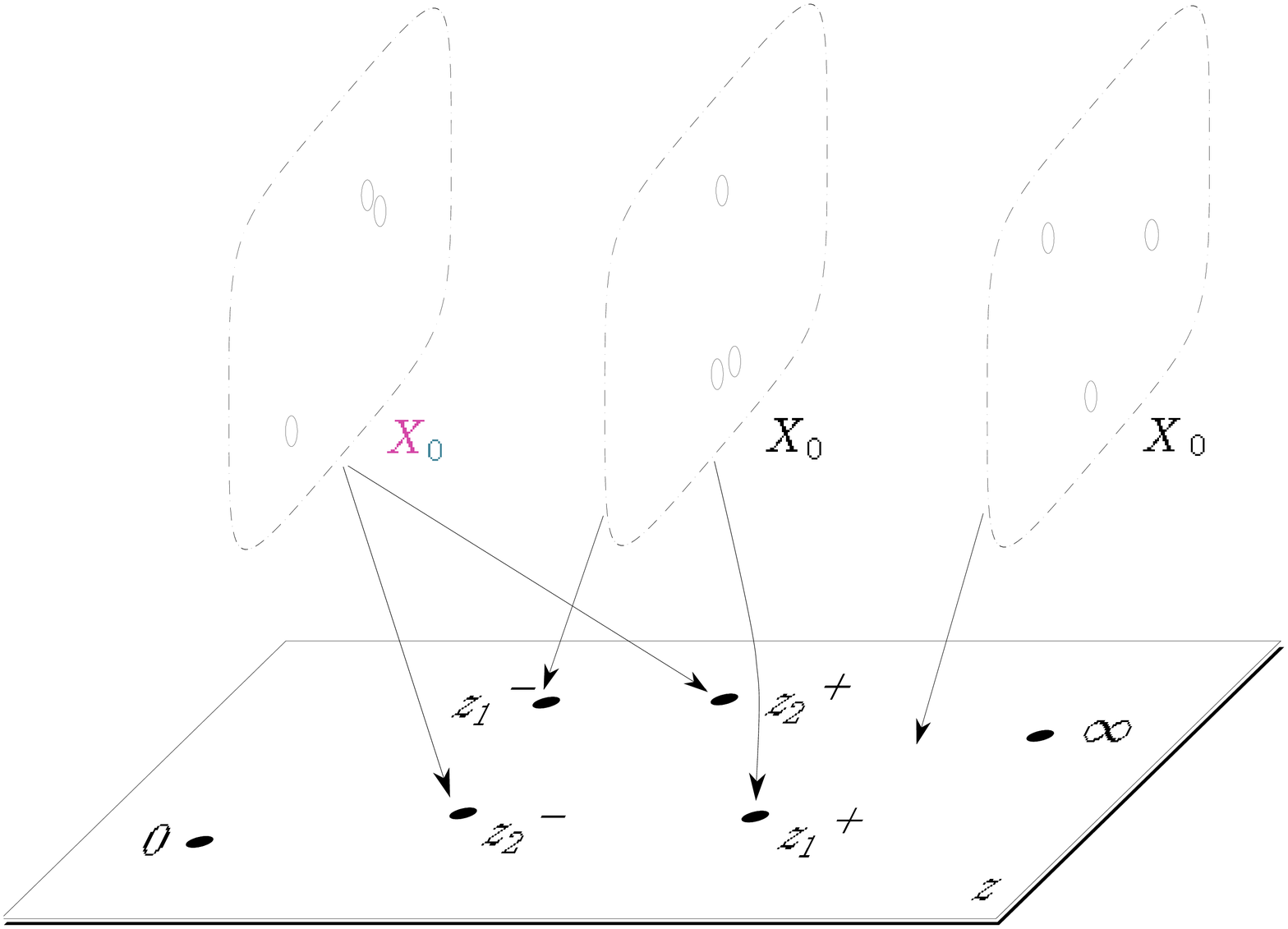}

Inspired by the role of spectral curves in
integrable systems \cite{GKMMM,MW,integ}, one is lead to consider
SW curves \equ{aaa} of the form \cite{GKMMM,MW}:
\beq
X_1:\ \
z+{\La^n\over z}+2\,P^{{n}}_{A_{n-1}}(x,u_k)\ =\ 0\ ,
\eel{nuCu}
where the characteristic polynomial \eq{charP} for $SU(n)$ obeys
``by accident'' : $P^{{n}}_{A_{n-1}}\equiv W_{A_{n-1}}$. These
curves are related to the hyperelliptic curves \equ{aaa} by a simple
reparametrization, $z\to y-P$, and thus are completely
equivalent to them. Moreover, note also that the classical limit
$\La\to 0$ gives $X_1: z+P(x)=0$, which is an (equivalent)
alternative to the classical level surfaces, $X_0$: $P(x)=0$.

A curve of the form \eq{nuCu} can be thought as fibration of the
classical level set $X_0$ \equ{levsurf} over $\IP^1$, coordinatized
by $z$ and whose size is measured by $1/\La$. There are $(n-1)$ pairs
of branch points in the $z$-plane, $z_{i}^\pm$, which are associated
with the basic degenerations of $X_0$, ie., with the simple roots
$\al_i$. There are two additional branch points $z_0, z_\infty$, and
cuts run between, say $z_{i}^-$ and $z_0$, and between $z_{\al_i}^+$
and $z_\infty$. See \figref{SWfib} for $G=SU(3)$, where $X_1:\,
z+\La^3/z+
2(x^3-u x -v)=0$ and
\bea
z_1^\pm &=& \ {{2{u^{{3\over 2}}} + 3{\sqrt{3}}v  \pm
     {\sqrt{ {{\Big( 2{u^{{3\over 2}}} +
               3{\sqrt{3}}v \Big) }^2} - \La^6} }}} \non\\
z_2^\pm &=& {{-2{u^{{3\over 2}}} + 3{\sqrt{3}}v \pm
     {\sqrt{ {{\Big( 2 {u^{{3\over 2}}} -
               3{\sqrt{3}}v \Big) }^2} - \La^6} }}}\ .\non
\eeal{nothing} The curve may also be viewed as a foliation, or
$n$-sheeted covering of the $z$-plane, the sheets being associated to
the points of $X_0$, ie., to the weights $\la_i$, see
\figref{genustwo}.\foot{One may draw the cuts also in different ways;
we have chosen them here such that the massless monopoles are related
to vanishing $\be$-type cycles.}

\figinsert{genustwo}
{The genus two curve resulting from the fibration shown in Fig.13 can
be viewed as a foliation with three leaves that are glued together
over the cuts. The sheets are one-to-one to the weights of the
fundamental representation of $SU(3)$.} {1.6in}{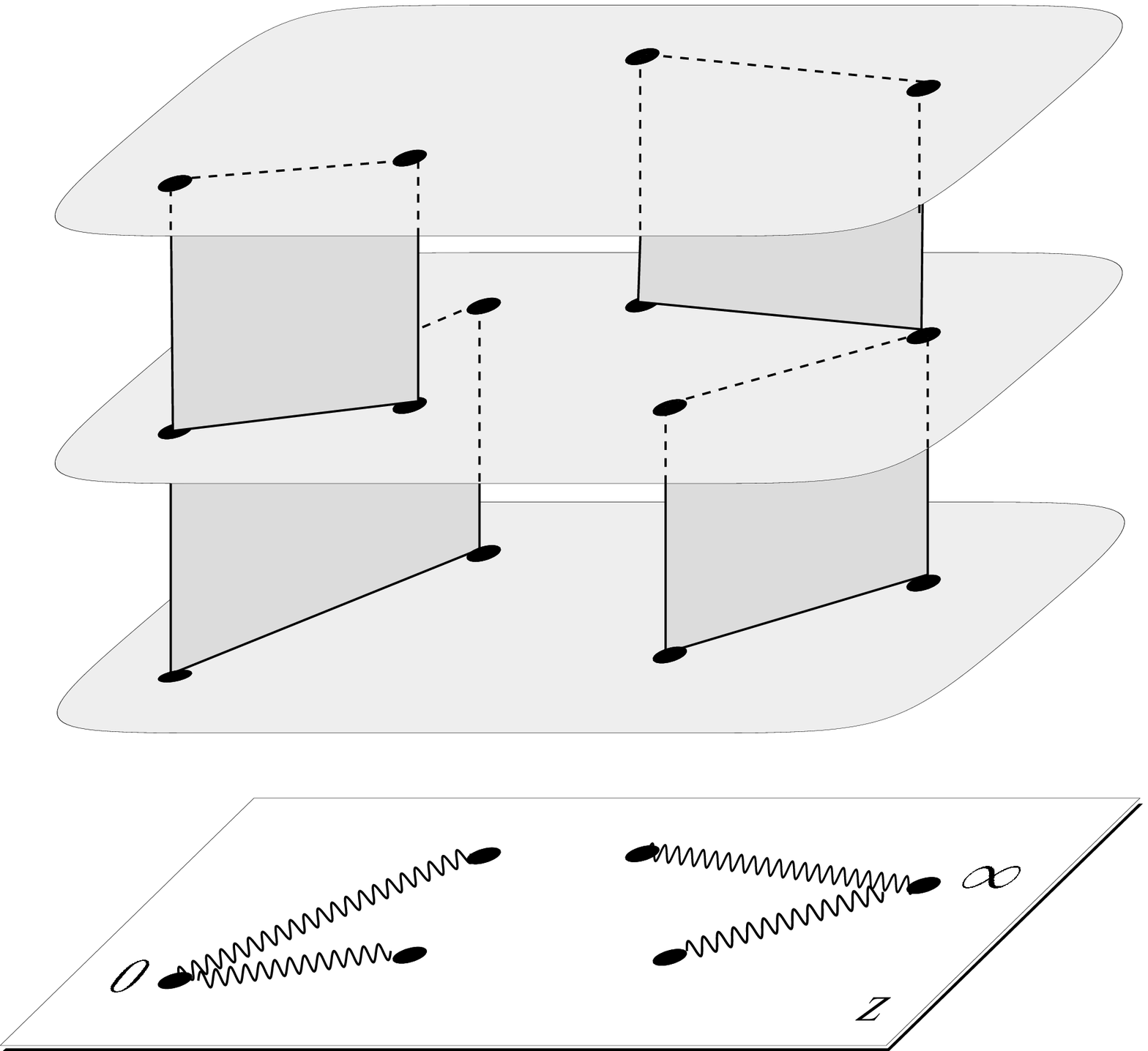}

The meromorphic differential takes the
following particularly simple form,
\beq
\la_{SW}\ =\ {1\over 2\sqrt 2\pi}x(z,u)\,{dz\over z}\ .
\eel{lswII}
which, via partial integration, is equivalent to \equ{Lamdef}.
Note that there are actually $n$ versions of $\la_{SW}$,
since $x(z)\sim z^{1/n}$, each associated to one of the sheets
of the foliation.

Having represented the $SU(n)$ curves in the form \equ{nuCu}, the
generalization to other simply laced groups is now easy to state
\cite{MW}: one just takes the characteristic polynomial $P^{{\cal
R}}_{ADE}$ \equ{charP} (for an arbitrary representation $\cal R$),
and shifts the top Casimir $u_h$ by $z+{\La^h\over z}$, i.e.,
\beq
X_1:\ \
P^{{\cal R}}_{ADE}(x,u_k,u_h+z+{\La^h\over z})\ =\ 0\ ,
\eel{nuCu1}
where $h$ is the dual Coxeter number. As explained in \cite{MW}, the
choice of the representation is irrelevant. That is, there is for
each gauge group an infinity of possible curves, with arbitrarily
high genera; however, one can restrict attention to a particular
subset of $r\equiv rank(G)$ $\al$- and $r$ dual $\be$-periods, which
carry all the relevant information.

On can in fact write down curves for the non-simply laced groups as
well. These are obtained by fibering $ADE$ level surfaces $P^{{\cal
R}}_{ADE}=0$ in a ``non-split'' fashion, i.e., in a way that
encircling the singular points in the $z$-plane gives rise not only
to Weyl transformations acting on the fiber $X_0$, but also to outer
automorphisms. This is similar to the considerations of \cite{PAMG},
and gives an orbifold prescription leading to a ``folding'' of the
$ADE$ Dynkin diagram into the corresponding non-simply laced one; it
also appropriately modifies the curves \eq{nuCu1}\cite{MW}.

\section{SW Geometry from String Duality}
\subsection{General Picture}

So far, the auxiliary ``spectral curves''
\equ{aaa},\equ{nuCu},\equ{nuCu1} have been introduced in a somewhat
{\it ad hoc} fashion, originally just in order to deal with the
monodromy problem in a systematic way. One may in fact approach the
SW theory without directly referring to a Riemann surface, for
example along the lines of \cite{nahm}, but this gets pretty quickly
out of hand for larger gauge groups. Thus the question comes up
whether the SW curves have a more concrete physical significance --
it would be rather absurd if all the geometrical richness of Riemann
surfaces would be nothing more than a technical convenience, and
would not have any deeper meaning.

This attitude is supported by
growing recent experience that whenever we meet a geometrical object
like a spectral manifold, {\it it represents a concrete physical
object
in some appropriate dual formulation of the theory}. For example, the
classical level set $X_0$ \equ{levsurf}, consisting of a discrete set
of points, has been associated with the locations of $D$-branes in a
dual formulation of gauge symmetry enhancement \cite{BSV} --
the dashed lines in \figref{claslevel} then are nothing but open
strings
linking $D$-branes.

Indeed it turns out \cite{KLMVW} that, in this sense, the natural
home of SW geometry is string theory, but of a
rather peculiar, ``non-critical'' kind \cite{witcom,selfd}.

For ease of discussion it is most convenient to start with ordinary,
``critical'' superstrings that naturally live in ten dimensions.
Remember that for these theories, complex manifolds play an
ubiquitous role as compactification manifolds -- and this already
hints at our aim to ultimately view the SW curves as some kind of
compactification manifolds as well.

\figinsert{CYfibr}
{Complex manifolds that are relevant in
heterotic-type II duality. The vertical direction is a local scaling
limit
that includes switching off $\al'$. The horizontal
step corresponds to switching on space-time quantum corrections. }
{1.4in}{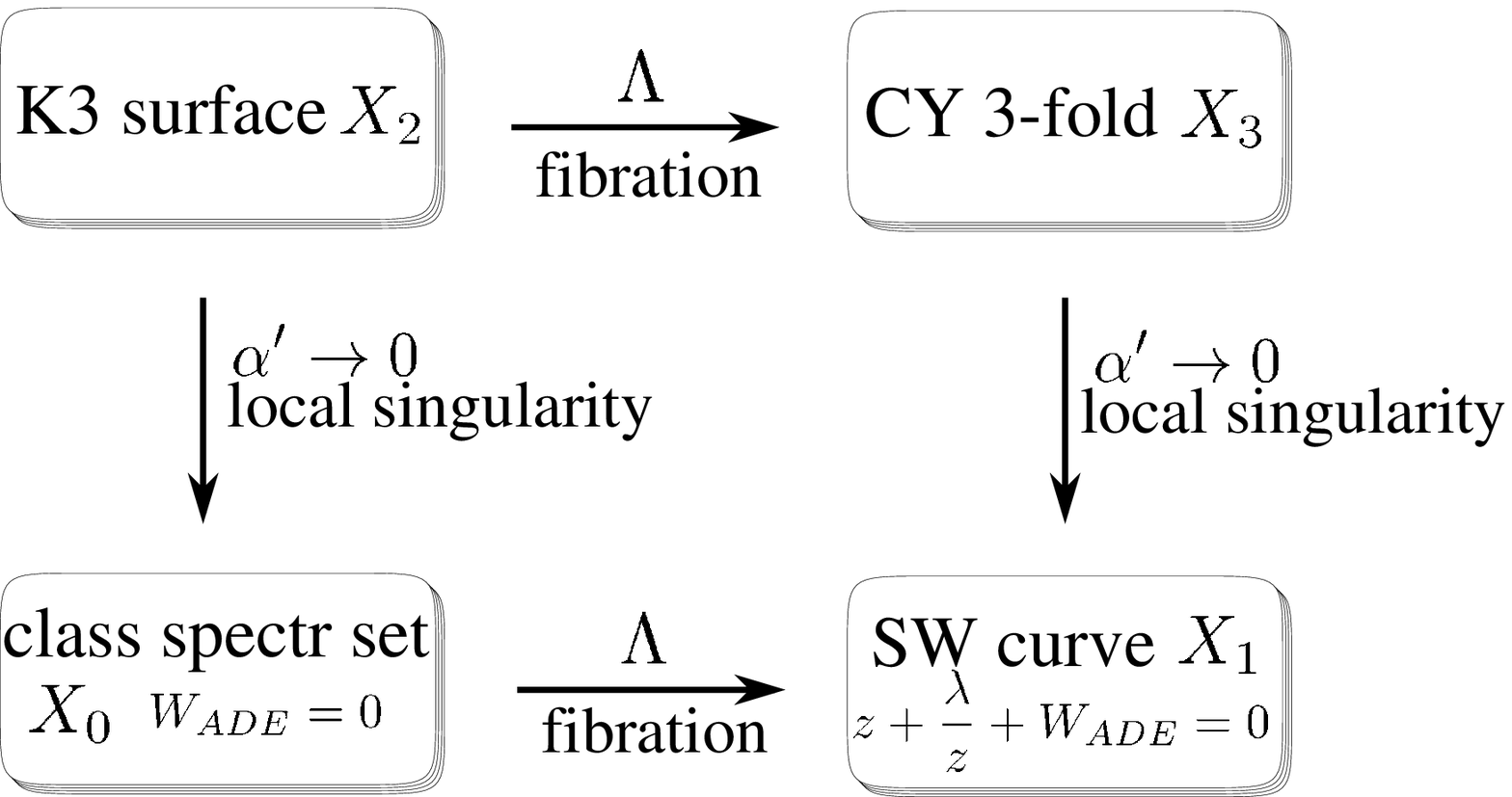}

More specifically, what we have in mind is a structure roughly as
depicted in \figref{CYfibr}. That is, starting from the spectral set
$X_0$, which describes classical Yang-Mills theory, one can go to the
quantum version via fibering $X_0$ over $\IP^1$ with size $1/\La$ --
this was described in section 3.4. However, one may also go
vertically and view $X_0$ (and thus classical, $N=4$ Yang-Mills
theory itself) as arising from a particular multiple scaling limit of
a $K3$ compactification (which includes the limit $\al'\to 0$). This
$K3$ surface $X_2$ may then itself be taken as a fiber over $\IP^1$,
to yield a Calabi-Yau threefold $X_3$. One may thus expect to close
the circle in \figref{CYfibr} by performing a suitable limit of
$X_3$, in order to get back to the SW curve. How this exactly works
will be explained in the next couple of sections.

\subsection{ALE Spaces and Heterotic-Type II
           String Duality in Six Dimensions}

\figref{CYfibr} may look suggestive, but so far we did not specify
the precise physical context to which it is supposed to apply. The
correct framework is the duality \cite{HT,KV,FHSV} between the
heterotic string and the type II string that appears when the strings
are compactified on suitable manifolds. See \cite{aspreview} for a
detailed review on this subject.

Let us first review the original Hull-Townsend hypothesis, which
states the non-perturbative equivalence of the heterotic string
(compactified on $T_4$) with the type IIA string (compactified on
$K3$). As is well-known, gauge symmetry arises in the six-dimensional
heterotic string from the Narain lattice, $\Gamma_{20,4}$. More
precisely, $ADE$ type of gauge symmetries arise if the background
moduli are such that the Narain lattice becomes $ADE$ symmetric,
i.e., if there are lattice vectors with (length)$^2=2$. These are
just the root vectors associated with the gauge group, and give rise
to space-time gauge fields via the Frenkel-Kac mechanism (see eg.,
\cite{LSW}).

On the type IIA string side, gauge symmetries arise from $ADE$ type
of singularities of $K3$ \cite{HT,Wi,Asp,BSV,DM}. More specifically,
if
the $K3$ moduli are tuned appropriately, $K3$ can locally (near the
singularity and in some suitable coordinate patch) be written as:
\beq
W_{K3}\ =\ \epsilon\,\big[\, {\cal W}_{ADE}^{ALE}\,\big]+
{\cal O}(\epsilon^2)\ =\ 0\ ,
\eel{K3limit}
where, in physical terms, $\epsilon\!\sim\! (\al')^\sigma\to 0$ for
some power $\sigma$. Moreover, $S_{ADE}:\,{\cal
W}_{ADE}^{ALE}(x_i)=0$ is the {\bf A}symptotically {\bf L}ocal {\bf
E}uclidean space \cite{EH} with $ADE$ singularity at the origin; it
is a non-compact space obtained by excising a small neighborhood
around the singularity on $K3$. This kind of spaces is essentially
given by the $ADE$ simple singularities \equ{WAnSing}, \equ{WDE}, up
to ``morsification'' by extra quadratic pieces:
\bea
{\cal W}_{A_{n-1}}^{ALE} &=& {W}_{A_{n-1}}(x_1,u_k)+{x_2}^2+{x_3}^2
\non\\ &=& {x_1}^n+{x_2}^2+{x_3}^2 + \dots\label{ALEdef}\\
{\cal W}_{D,E}^{ALE} &=& {W}_{D,E}(x_1,x_2,u_k)+{x_3}^2\ .\non
\eeal{ALEdef1}
In fact, the $ADE$ singularities can be characterized in a most
uniform and natural way if they are written, exactly as above, as ALE
space singularities in terms of three variables. By definition,
$S_{ADE}=\IC^3/[{\cal W}_{ADE}=0]$, but one may also write
$S_{ADE}=\IC^2/\Gamma$, where $\Gamma={\cal C}_n, {\cal D}_n, {\cal
T}, {\cal O}, {\cal I}\subset SL(2,\IC)$ are the discrete isometry
groups
of the sphere that are canonically associated with the $ADE$ groups
\cite{Arn,Slodo}.

The variables $u_k$ provide a minimal resolution of these
singularities, obtained by iterated blow up's of points. That is, the
$ADE$ singularity at the origin in $\IC^3$ is blown up into a
connected union of 2-spheres, with self-intersections equal to $-2$.
Pairwise intersections are either null or transverse, and one may
encode this information in a Dynkin diagram -- surprisingly, these
Dynkin diagrams agree precisely with those of the corresponding $ADE$
groups; see \figref{ALE} for an example. In fact, the second homology
group is isomorphic to the root lattice,
\beq
H_2(S_{ADE},\ZZ)\ \cong\ \Gamma_R^{ADE}\ ,
\eel{H2rel}
since it is equipped with a geometric intersection form given by (the
negative of) the Cartan matrix. Indeed the vanishing 2-cycles of the
ALE space behave exactly like the root vectors of the corresponding
$ADE$ group. Since the addition of quadratic pieces does not change
the singularity type, the relevant features of the classical spectral
surfaces $X_0$ thus apply here as well -- this is what is meant by
the left vertical arrow in \figref{CYfibr}. Eq.~\equ{H2rel} is
precisely the two-dimensional version of \equ{H0rel} for $A_{n-1}$,
and we may thus use \figref{claslevel} to visualize the ALE space
homology for $A_{n-1}$ as well, if we simply view the dashed lines as
2-cycles and not as 0-cycles.

\vskip .3cm
\figinsert{ALE}
{The vanishing 2-cycles of the $D_4$-type of ALE space
intersect in a Dynkin diagram pattern.
At the top we have drawn only
a real slice of the degeneration process.}
{1.3in}{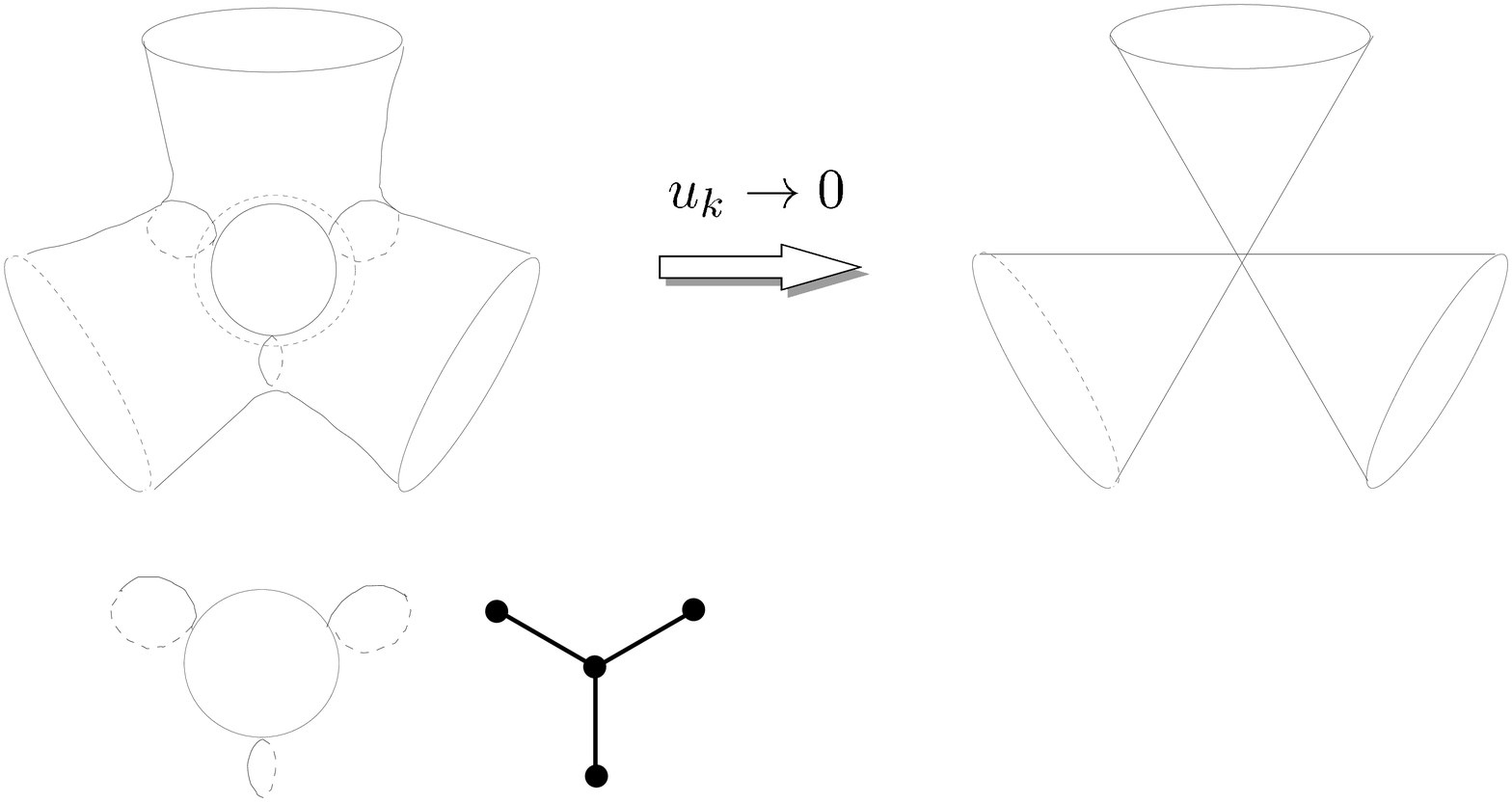}

That we meet here again the classical spectral set $X_0$
is no surprise in physics terms, since type IIA string
compactification on $K3$ does give classical, unrenormalized \nex2
gauge theory in $d=6$ (or \nex4 in $d=4$ after additional toroidal
compactification). The concrete physical mechanism that underlies the
appearance of gauge fields in space-time is the wrapping of type IIA
$2$-branes on the vanishing 2-cycles of the ALE space
\cite{HT,Wi,Asp,BSV,DM} -- obviously, if the volumina of the 2-cycles
shrink to zero, the corresponding BPS masses will vanish.

Summarizing, we see some sort of universality at work: for low-energy
physics ($\al'\to 0$) and small symmetry breaking VEVs, only the
local neighborhood of a singularity is relevant. That is, the local
geometry of an ALE space, sitting somewhere on $K3$, singles out a
subset of 2-cycles, namely those which tend to vanish near the
singularity. The embedding of these vanishing cycles in the full
2-homology of $K3$ exactly mirrors the embedding of a singled-out
$ADE$ root lattice into the Narain lattice on the heterotic side,
$H_2(S_{ADE},\ZZ) \subset H_2(K3,\ZZ)\cong \Gamma_{20,4}$. The
information that comes from more distant parts of $K3$, like effects
of 2-branes wrapping around the other 2-cycles, is suppressed by
powers of $\al'$. This mirrors the effect of massive winding states
on the heterotic side.

Note that in $d=6$ the duality is between the heterotic string on
$T_4$ and the type IIA string on $K3$, and not the type IIB string.
Since the type IIB string does not have any 2-branes, clearly
no gauge fields can arise from the vanishing 2-homology of $K3$.
Rather, since type IIB strings have 3-branes, one needs to
consider 3-branes wrapped around the vanishing 2-cycles. But what
this gives is not massless particles, but {\it tensionless strings}
in six dimensions \cite{witcom,selfd}.

Such kind of strings have not much to do with the perturbative
ten-dimensional superstrings that we started with. Rather, these
strings are {\it non-critical} and therefore do not involve gravity
at all -- indeed we have already taken the limit $\al'\to0$. They
couple to 2-form fields $B_{\mu\nu}^i$, in analogy to particles that
couple to gauge fields $A_\mu^i$. The antisymmetric tensor field
$B_{\mu\nu}^i$ forms together with 5 scalars (plus some fermions) a
tensor multiplet of $(0,2)$ supersymmetry in six dimensions; there is
one such multiplet for each 2-cycle in the ALE space (labelled by the
index $i$).

The non-critical string represents a novel quantum theory on its own,
but is hard to study directly \cite{witcom,selfd}. That is, since the
field strength $H^{(3)}=dB$ to which it couples arises from a
self-dual five-form in ten dimensions, that is (anti-)self-dual too,
$*H^{(3)}=-H^{(3)}$. This implies that the coupling must be equal to
one and thus that this string cannot be accessed in terms of usual
(conformal field theoretic) perturbation theory.

As we will see later in section 5, it is these anti-self-dual
strings that provide the natural dual, geometric representation
of the SW theory that we are looking for.

\subsection{$K3$-Fibrations and SW Geometry}

In the previous section we discussed \nex2 supersymmetric
compactifications in six dimensions, which gives rise to
``classical'' ($N=4$) Yang-Mills theory in $d=4$. We now like to
understand how \nex2 supersymmetric gauge theories emerge in four
dimensions. For our purposes, the appropriate framework\foot {The are
in fact various ways to obtain \nex2 $d=4$ YM theories, see, for
example, \cite{probes,MDML}.} to describe such theories is the
duality \cite{FHSV,KV} between the heterotic string, compactified on
$K3\times T_2$, and the type IIA (IIB) string, compactified on a
particular class of Calabi-Yau threefolds, $X_3$ (its mirror\foot{For
a review on mirror symmetry, see e.g., \cite{yau}.} $\tilde X_3$).
(See ref.\ \cite{NexTwoDual,Ftheory} for some selection of further
work related to this duality.) The particular choice of $X_3$
corresponds to the choice of gauge bundle data on the heterotic side.

Crucial is the insight that that the CY moduli space factors
into two (generically) decoupled pieces:
\beq
\cM_{X_3}\ =\ \cM_V(s,t_i)\,\otimes\, \cM_H(d,h_k)\ ,
\eel{totalM}
where the vector multiplet moduli space $\cM_V$ has a complex,
and the hyper\-multiplet moduli space $\cM_H$ a
quaternionic structure -- this follows directly from \nex2
supersymmetry \cite{nonre}. It is well-known that
\bea
{\rm dim}_C\cM_V(X_3) &=& h_{11}(X_3)\non\\
{\rm dim}_C\cM_H(X_3) &=& h_{21}(X_3)+1\ ,
\eeal{dimM}
where, of course, the Hodge numbers $h_{11}$ and $h_{22}$ exchange
under mirror symmetry, $X_3\leftrightarrow \tilde X_3$. The shift
``$+1$'' accounts for the type IIA dilaton $d$.

The point is that the heterotic dilaton $s$ does not enter in $\cM_H$
and the type IIA dilaton $d$ does not enter in $\cM_V$. Therefore, a
tree-level computation in $\cM_H$ in the heterotic string will give
the exact quantum result, while a tree-level computation in the type
II string will give the exact quantum corrected vector multiplet
moduli space, $\cM_V$. More precisely, $\cM_V$ will not get any
contributions from type IIA space-time quantum effects, but it will
still get corrections from world-sheet instantons. But one can invoke
mirror symmetry and obtain these corrections from a completely
classical computation on the type IIB string side, see
\figref{mirror}. As is well-known \cite{mirrorsymm}, this
boils down to evaluating period integrals or solving Picard-Fuchs
equations, much like we did in previous sections for rigid
Yang-Mills theories.

It is certainly $\cM_V$ that we are presently interested in, since we
expect it to contain the SW moduli spaces $\cM_q$ for a variety
of gauge groups. As explained before, these moduli
spaces do get complicated corrections, which correspond, in heterotic
language, to space-time instanton effects. But from the above we see
that in the string framework, the functional complexity of the SW
theory can equally well be attributed to type IIA world-sheet
instanton corrections \cite{KLTY,FHSV}. So what this means is that
the \nex2 supersymmetric heterotic-type II duality implies a map
between non-perturbative space-time instanton effects and world-sheet
instanton effects~!

\figinsert{mirror}
{Heterotic-type IIA duality plus mirror symmetry allows to compute
non-perturbatively exact quantities in the heterotic string
from tree-level computations in the type IIB string. In practice, in
the vector multiplet sector indicated above, this amounts to solving
Picard-Fuchs equations for the IIB moduli $y_i,y_s$ in terms of the
IIA (and heterotic) variables $t_i,s$. For the notation see below.}
{2.in}{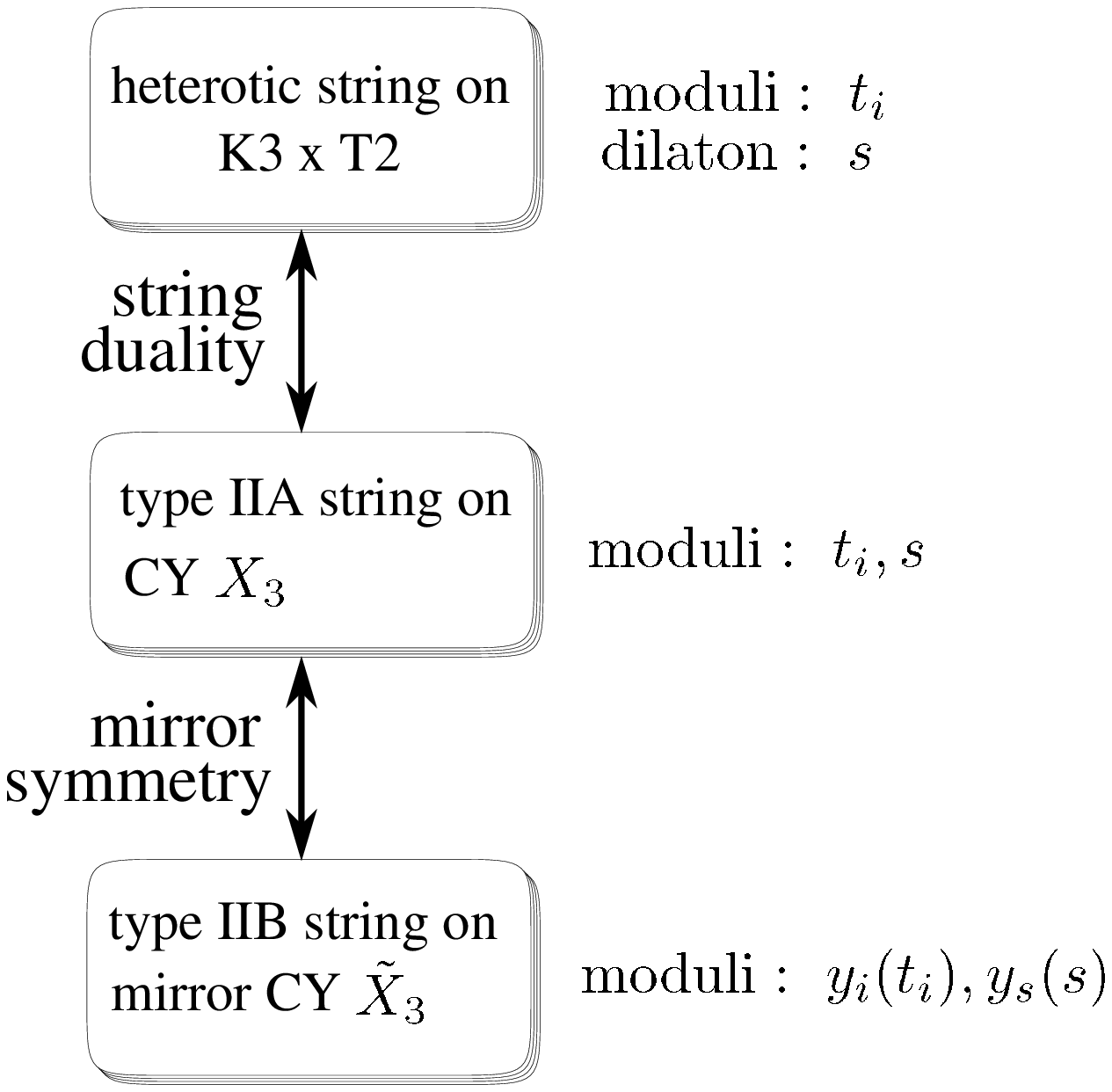}

To recover the SW physics from string theory, it is thus most natural
to start with the type IIB formulation. As pointed out in
\cite{strom,andyrev}, in IIB language the role of the vanishing
1-cycles of the SW curve is played by vanishing 3-cycles of the
Calabi-Yau threefold (corresponding to ``conifold'' singularities).
More
specifically, the relevant periods are those of the canonical
holomorphic
3-form $\Omega$ on $\tilde X_3$ \cite{specialG}:
\beq
X^I\ =\ \int_{\Gamma_{\al_I}}\!\Omega\ ,
\qquad\ \
F_J\ =\ \int_{\Gamma_{\be^J}}\!\Omega\ ,
\eel{CYperiods}
where ${\Gamma_{\al_I}},\ ,{\Gamma_{\be^J}}$, $I,J=1,\dots,
h_{11}(\tilde X_3)+1$ span a integral symplectic basis of $H_3$,
with the $\al$-type of cycles being dual to the $\be$-type of cycles:
${\Gamma_{\al_I}}\circ {\Gamma_{\be^J}}=\delta^I_J$, etc.
Concretely, assuming that the mirror threefold can be represented by
the vanishing of a polynomial in some weighted projective space,
$W_{\tilde X_3}=0$, the holomorphic 3-form on $\tilde X_3$ can
be written as:
\bea
\Omega &=& \int_\gamma\!{1\over W_{\tilde X_3}}\,\omega\ ,\ \ \ {\rm
with} \label{Omdef}\\ \omega &\equiv&
\sum_{A=1}^{5}(-1)^Ax^A\,dx^1\!\wedge\!...\wedge\!{\widehat{dx^A}}
\!\wedge\!...\!\wedge\!dx^{5}\ ,\non\cr
\eeal{Omdef1}
where $\gamma$ is a small, one-dimensional curve winding around the
hypersurface $W_{\tilde X_3}=0$, and where the hat means that the
indicated term is to be left out.

The periods \eq{CYperiods} are defined up to symplectic basis
transformations which belong to $Sp(2h_{11}(\tilde X_3)+2,\ZZ)$. The
duality group, which represents the exact quantum symmetries of the
theory and which is analogous the SW monodromy group $\Gamma_0(4)$ in
section 2.4, must be a subgroup of this symplectic group. The precise
form of this duality group depends, of course, on the choice of
threefold. See refs.\ \cite{dualities} for further discussion of
duality groups in this context.

In terms of the periods \eq{CYperiods}, the \nex2 central charge,
which enters in the BPS mass formula, then
looks very similar to \eq{zdef},
\beq
Z\ \sim\ M_I X^I\,+\,N^JF_J\ ,
\eel{ZXF}
so that again vanishing cycles will lead to massless BPS states. More
precisely, if some 3-cycle $\nu = M_I
{\Gamma_{\al_I}}+N^J{\Gamma_{\be^J}}$ vanishes in some region of the
moduli space, then $Z\equiv \int_{\nu}\Omega=0$ and we get, just like
in the rigid SW theory, a massless hypermultiplet, with quantum
numbers $(N^J,M_I)$. Physically, it arises from a type IIB 3-brane
wrapped around $\nu$ \cite{strom}.

Furthermore, by the general properties of vanishing cycles, governed
by the Picard-Lefshetz formula \eq{PicLef}, one finds a logarithmic
behavior for the dual period $F_Z\equiv {\del \cF(Z) /\del
Z}={1\over2\pi i} Z\log[Z/\Lambda]+\dots$, so that the effective
action $\cF(Z)$ near the conifold singularity $Z=0$ looks very
similar to the SW effective action ${\cal F}_D(a_D)$ near the
monopole singularity.

However, to really recover the SW geometry, we should
not look just to a single conifold singularity (of local form
$\sum_{i=1}^4 {x_i}^2= u$), since it does not carry enough
information. Indeed, by analogy just looking at the local
singularity ${x_1}^2+{x_2}^2=u$ of the SW curve (or its $SU(n)$
extension, which is an Argyres-Douglas singularity), we cannot learn
much
about the global SW geometry. So what we need to look at is an
appropriate ``semi-local'' neighborhood of the conifold singularity
in the CY, in order to capture both SW type singularities (or both AD
singularities, that is) at once. As we will see below, it is indeed
not a local singularity, but rather a fibration of a local
singularity, that is the right thing to consider.

Now, the CY manifolds that are relevant here have a very special
structure: namely they must be $K3$-fibrations \cite{KLM,VaWi,AspJl}.
Before we will briefly explain what this means mathematically, we
first point out why such threefolds are physically important. That
is, if and only if a threefold is a $K3$-fibration, the effective
prepotential (in the large radius limit) has this particular
asymptotic form:
\beq
\cF(s,t)  = \  {1\over2}s\,Q_{ij}t^it^j +
{1\over6}C_{ijk}t^it^jt^k + \dots.
\eel{Flarget}
Here, $Q$ and $C$ directly reflect the classical intersection
properties of 2-cycles, and $s,t^i$ are K\"ahler moduli of the
CY, where $s$ is singled out in that it couples only linearly. This
means that $s$ can naturally be identified (to leading order) with
the semi-classical dilaton of the heterotic string, since the
dilaton couples exactly in this way.

The $K3$ fibration structure turns out to be crucial for our
purposes. To see this, let us assume, for simplicity, that the
(mirror) threefold $\tilde X_3$ can be represented by some polynomial
\beq
\tilde X_3:\ \ W_{\tilde X_3}(x_1,x_2,x_3, x_4,x_5; y_i,y_s)\ =\ 0
\eel{kthreefib}
in weighted projective space $W\IP^{2d}_{ 1, 1, 2k_1 ,2k_2, 2k_3 }$,
with overall degree $2d \equiv 2(1 + k_1 + k_2 + k_3)$. Above, $y_i$
denote the moduli, and, in particular, $y_s$ denotes the special
distinguished modulus that is related to the heterotic dilaton,
$y_s\sim e^{-s}$. A list of this kind of $K3$ fibered
threefolds was given in \cite{KLM}.
(For more general classes of fibrations, see \cite{KthreeFib}.
Considerations similar to those below apply to these cases as well.)

The statement that $\tilde X_3$ is a $K3$ fibration of this
particular type means that it can be written as
\bea
W_{\tilde X_3}\!(x_j; y_i,y_s) &=& \!\!\!{1\over {2d}}\!\Big(
\!{x_1}^{2d}\!+\!{x_2}^{2d}\!+\!
{2\over \sqrt{y_s}}(x_1x_2)^{d}\!\Big) \non\\
&+& \!\!\widehat W({x_1x_2\over {y_s}^{1/d}},x_k;y_i)\ .\
\eeal{Kfib}
Upon the variable substitution
\bea
x_1 &=& \sqrt{x_0}\,\zeta^{1/2d}\non\\
x_2 &=& \sqrt{x_0}\,\zeta^{-1/2d}{y_s}^{1/2d}
\eeal{xirep}
this gives
$$
W_{\tilde X_3}(x_j;\zeta, y_i)\ =\
{1\over {2d}}\Big(
\zeta+{y_s\over \zeta}+2\Big)x_0^{d} + \widehat W(x_0,x_k;y_i).
$$
That is, if we now alter\-na\-tive\-ly view $\zeta$ as a mo\-du\-lus,
and not as a coordinate, then $W_{\tilde X_3} (\zeta, x_2, x_3, x_4,
x_5; y_i, y_s)$$\equiv$$W_{K3} (x_2, x_3, x_4, x_5; \\ \zeta + {y_s
\over \zeta}, y_i) = 0$ describes a $K3$ -- this is precisely what is
meant by fibration (of course, if we continue to view $\zeta$ as a
coordinate, then this equation still describes the Calabi-Yau
threefold). More precisely, $\zeta$ is the coordinate of the base
$\IP^1$, and $y_s\to0$ corresponds to the large base limit --
obviously, the fibration looks in this limit locally trivial,
and one expects then the theory to be dominated by the
``classical'' physics of the $K3$ (cf., the top horizontal step in
\figref{CYfibr}). This is the ``adiabatic limit'' \cite{VaWi}, in
which the
$K3$ fibers vary only slowly over the base and where one can
apply the original Hull-Townsend duality fiber-wise.

The left-over piece $\widehat W$ is precisely such that ${1\over
{d}}x_0^{d}+\widehat W(x_0,x_k;y_i)=0$ describes a $K3$ in canonical
parametrization in $W\IP^{d}_{1, k_1 ,k_2,k_3}$. Now assuming that
the $K3$ is singular of type $ADE$ in some region of the
moduli space, we can expand it around the critical point and thereby
replace it locally by the ALE normal form \eq{ALEdef} of the
singularity, ${1\over d}x_0^{d}+\widehat W \sim \epsilon\,{\cal
W}^{ALE}_{ADE}(x_i,u_k)$. Going to the patch $x_0=1$ and
rescaling\foot{ This fixes $\epsilon=(\al')^{h/2}$.}
$y_s=\epsilon^2\La^{2h}$ and $\zeta=\epsilon\,z$, we then obtain the
following fibration of the ALE space:
\bea
&&\!\!\!\!\!\!\!\!\!\!\!\!W_{\tilde X_3}(x_j,z;u_k)\ =\
\label{ALEfibr}\\ &&\!\!\!\!\!
\epsilon\Big(z+{\Lambda^{2h}\over z}+2 {\cal W}^{ALE}_{ADE}(x_j,u_k)
\Big)+\cO(\epsilon^2) = 0.\non
\eeal{ALEfibr1}
This is not totally surprising: since the CY was a $K3$
fibration, considering a region in moduli space where the $K3$ can be
approximated by an ALE space simply produces locally a corresponding
fibration of the ALE space.

Now, focusing on $G=SU(n)\sim A_{n-1}$ and remembering the definition
of the ALE space \eq{ALEdef}, we see that \eq{ALEfibr} is exactly the
same as the fibered form of the SW curve \eq{nuCu}, apart from the
extra quadratic pieces in $x_2,x_3$ ! Since quadratic pieces do not
change the local singularity type, this means that the local geometry
of the threefold in the SW regime of the moduli space is indeed
equivalent to the one of the Seiberg-Witten curve. However, just
because of these extra quadratic pieces, the SW curve itself is,
strictly speaking, {\it not} geometrically embedded in the threefold,
though this distinction is not very important.

One may in fact explicitly integrate out the quadratic pieces in
\eq{ALEfibr}, and verify \cite{KLMVW} that the {\it holo}morphic
three-form $\Omega$ of the threefold then collapses precisely to the
{\it mero}morphic one-form $\la_{SW}$ \eq{lswII} that is
associated with the SW curve:
\beq
\Omega = {d\zeta\over\zeta}\!\wedge\!
\Big[{d x_1 \wedge d x_2 \over {\del W\over \del x_3}}
\Big]\ \mathop{\longrightarrow}^{\epsilon\to0}\ x_1
{dz\over z}\ \equiv\ \lambda_{SW}.
\eel{omred}
This implies that the periods $a_i, a_{D,i}$ are indeed among the
periods of the threefold,
\bea
(X^I;F_J)&\equiv&\int_{({\Gamma_{\al_I}};{\Gamma_{\be^J}})}
\!\!\!\!\Omega \non\\
\ \ \ \ \ {\mathop{\supset}^{\epsilon\to0}}\quad
\int_{({{\al_i}};{{\be^j}})}\!\!\!\la_{SW}
&\equiv&  (a_i;a_{D,j})\ ,
\eeal{pereduce}
and thus that the string effective action, $\cF\equiv {1\over2}X^I
F_I$ \cite{specialG}, contains the SW effective action.

We should note there that we have tacitly assumed that the mirror
$\tilde X_3$ is a $K3$ fibration. However, our starting point was
really that the original threefold $X_3$ is a $K3$ fibration, since a
priori it is type IIA strings on $X_3$ that are dual to the heterotic
string. But in general, if $X_3$ is a $K3$ fibration, the mirror
$\tilde X_3$ is not necessarily a $K3$ fibration as well. However,
our above arguments are nevertheless correct, because all what counts
is that {\it locally} near the relevant singularity, the mirror is a
fibration of an ALE space. One can indeed show, using ``local''
mirror symmetry \cite{KKV}, that whenever we have an asymptotically
free gauge theory on the type IIA side, the mirror $\tilde X_3$ has
locally the required form.

As for the other simply laced groups, we face a complication similar
to the one of section 3.2: namely, the CY geometry implies fibrations
of ALE spaces, whereas the corresponding SW curves \eq{nuCu1} involve
the characteristic polynomials $R^{{\cal R}}_{ADE}$, which coincide
with the simple singularities only for of $SU(n)$ (for the
fundamental representation). For the other groups, the expressions
\eq{ALEfibr} are quite different as compared to the corresponding
Riemann surfaces, and are not Riemann surfaces even if we drop the
extra quadratic terms. But it can be shown that the independent
periods of these spaces indeed do coincide with those of the curves
\eq{nuCu1} (see \cite{ALESW} for details on how this works for
$E_6$). This represents a good test of the string duality, because
that {\it predicts} that the fibered ALE spaces describe the rigid YM
theories.

Moreover, note that these ideas carry over to gauge theories with
extra matter, though we will very brief here. In the $F$-theoretical
\cite{Ftheory} formulation of gauge symmetry enhancement
\cite{Fmatter}, there is a very systematic way to construct \nex2 YM
theory on the type IIA side, for almost any matter content. Via local
mirror symmetry \cite{KKV}, this maps over to the type IIB side and
directly produces the relevant SW curves, which generically exhibit
extra matter fields. Similar to \eq{ALEfibr}, one still has $ADE$
singularities fibered over some base $\IP^1$, but in general the
dependence of $z$ will be more complicated.

Summarizing, not only pure \nex2 gauge theory of $ADE$ (and
non-simply laced) type, but also gauge theories with extra matter
fields, are geometrized in string theory and the corresponding
Riemann surfaces can be constructed in a systematic fashion from
appropriate threefolds.

There is a very simple generalization of the above,\foot {The
rest of this section refers to unpublished work
\cite{sadlyenough}.} which however has no easy interpretation in
terms of Yang-Mills theory. Remember that we
focused above on $K3$ fibrations associated with weighted projected
spaces of type $W\IP^{2d}_{ 1, 1, 2k_1 ,2k_2, 2k_3 }$. There are many
other types of $K3$ fibrations \cite{KthreeFib}, for example related
to $W\IP^{(\ell+1)d}_{ 1, \ell, (\ell+1)k_1 ,(\ell+1)k_2,
(\ell+1)k_3}$. These have the generic form
\bea
W_{\tilde X_3}\ &=& {1\over2d}\Big({x_1}^{(\ell+1)d} +
\sum_{k}
 \tilde y_k\, {x_1}^{k}{x_2}^{(\ell+1)d-k} + \non\\
&&\!\!\!\!\!\!\!{x_2}^{(\ell+1)d/\ell}
\Big) +
\widehat W(x_k;\tilde y_k,y_i) = 0,
\eeal{otherfib1}
where the sum runs over appropriate values of $k$. Note that before,
in eq.~\eq{Kfib}, we had just one variable $y_s$ in the bracket. The
difference is that we have now $\ell$ moduli of this sort, $\tilde
y_k$. Still, one of the $\tilde y_k$ is distinguished by the linear
coupling property and thus is related to the heterotic dilaton; we
will continue to denote it by $y_s\sim e^{-s}$. Also note that
before, when we tuned $y_s\to1$, the term in the bracket became a
perfect square, and hence the threefold singular (of type $A_1$).
This singularity leads in fact to an IR free $SU(2)$ gauge symmetry
that is non-perturbative from the heterotic point of view (since
$y_s=1$ corresponds to strong coupling).

For the $K3$ fibrations above, we have now $\ell$ such parameters,
and tuning them appropriately, there arises an analogous $SU(\ell+1)$
``strong coupling'' gauge symmetry \cite{strongcoup}. But this is not
the gauge symmetry that we are interested in here, we are rather
still interested in the perturbative gauge symmetries that come from
the $K3$ singularities. By taking now a singular limit similar
to before, we arrive at the following modified fibration of the ALE
spaces:
\bea
&&\!\!\!\!\!\!\!\!\!\!\!\!W_{\tilde X_3}(x_j,z;u_k) \sim
\label{ALEfibr2}\\ &&\!\!\!\!\!\!\!\!\!\!
\epsilon\Big(z^\ell+\!\! \sum _{m=1}^{\ell-1}\Lambda^{(m)}\,z^m\!+
\!\!{\Lambda^{2h}\over z}\!+\!
2 {\cal W}^{ALE}_{ADE}(x_j,u_k)
\Big) = 0.\non
\eeal{ALEfibr3}
This kind of fibrations has obviously a more complicated structure in
the base $\IP^1$, and indeed we have now a whole series of extra
moduli, $\Lambda^{(m)}$, that move the new singular points on the
base around. If we send the extra $\Lambda$'s to infinity, we
obviously recover the SW theories as explained before.

The interesting point is here that the physics associated to the
extra parameters has nothing to do with the $K3$ fibers, and thus is
intrinsically non-perturbative from the heterotic point of view. So
what we have here is a marriage of the SW physics with physics
inherited from the strong-coupling singularity. It results in SW type
of curves that do {\it not} simply describe ordinary Yang-Mills
physics. More concretely, in the classical limit, $\Lambda\to 0$, the
discriminant of \eq{ALEfibr2} looks
\bea
\Delta(u_k,\Lambda^{(m)},\Lambda=0)\ &=&\
\big(\Delta_0(u_k)\big)^2
\tilde \Delta(u_k,\Lambda^{(m)})\non\\
\tilde \Delta(u_k,\Lambda^{(m)}=0) &\equiv&
\big(\Delta_0(u_k)\big)^{\ell-1},
\eeal{deltalambda}
where $\Delta_0$ is the classical discriminant \eq{cdisc} associated
with the corresponding $ADE$ gauge group. The discriminant
$\Delta(\Lambda=0)$ describes a classical gauge
symmetry $G=G_{ADE}\times U(1)^\ell$, with extra matter fields
charged under the various group factors; in general this theory is
non-asymptotically free.

However, for $\ell>1$ the dependence of $\Delta$ on the
$\Lambda^{(m)}$ is such that it does not describe any classical
gauge theory (an exception is for $G=SU(2)$ with $\ell=2$, where the
theory is identical to the $SU(2)$ gauge theory with $N_f=1$, with
$\Lambda^{(1)}$ playing the role of the bare mass parameter). In
other
words, even though we can identify gauge symmetries and matter
representations, the dependence of the theory on the extra parameters
is in general not like the dependence on any mass parameters, or
VEV's (because $\tilde \Delta\not=\prod_{{\rm weights}\
\lambda}(a\cdot\lambda+m)$).

Rather, the moduli $\Lambda^{(m)}$ are novel, dilaton-like parameters
that reflect non-perturbative physics of the heterotic string. Their
effects persist even for weak coupling, $\Lambda\to0$, and are thus
like ``small instanton'' effects \cite{smallinst}.

The lesson we draw from this is that ``rigid'' limits in string
theory will in general not only reproduce known field theories like
Yang-Mills theories and their curves, but also new kinds of SW like
theories that do not have the interpretation in terms of conventional
physics. Indeed, since the variety of possible CY singularities is
quite large, one may expect to find a whole zoo of supersymmetric
effective theories, and there is no reason why such theories should
always be interpretable in terms of conventional field theories like
gauge theories.

\section{Anti-Self-Dual Strings on Riemann Surfaces}

\subsection{Geometry of Wrapped 3-Branes}

In the previous section we have mentioned that the r\^ole of the SW
monopole singularity is played in the CY threefold by a conifold
singularity \cite{strom}. This corresponds in the type IIB
formulation to a vanishing 3-cycle in the mirror, $\tilde X_3$. The
relevant BPS states are given by wrapping type IIB 3-branes around
such 3-cycles.

What we are currently interested in is however not the full string
theory (which includes gravity), but only the $\alpha'\to0$ limit,
where the local geometry is given by an ALE space fibered over
$\IP^1$; cf., \eq{ALEfibr}. Thus we need to understand the properties
of wrapped 3-branes in this local geometry in some more detail.

For this, reconsider the SW curve
as fibration of a weight diagram over $\IP^1$; cf., \figref{SWfib}.
The difference to the present situation is that now the vanishing
0-cycles of the weight diagram are replaced by the vanishing
2-cycles of the ALE space. These 2-cycles, when fibered in this
manner,
produce 3-cycles in the CY in exactly the same way
as the 0-cycles in \figref{SWfib} produce 1-cycles on the SW curve;
this is sketched in \figref{threecyc}.

\figinsert{threecyc}
{Geometry of a 3-cycle in the CY, given by the fibration of an ALE
2-cycle. The 3-cycle arises by dragging the 2-cycle between branch
points $z^\pm$ on the base $\IP^1$. The 3-cycle thus projects
in the base to an open line $\cal L$ that joins the branch points --
it represents a non-critical string with variable local tension.
}{1.6in}{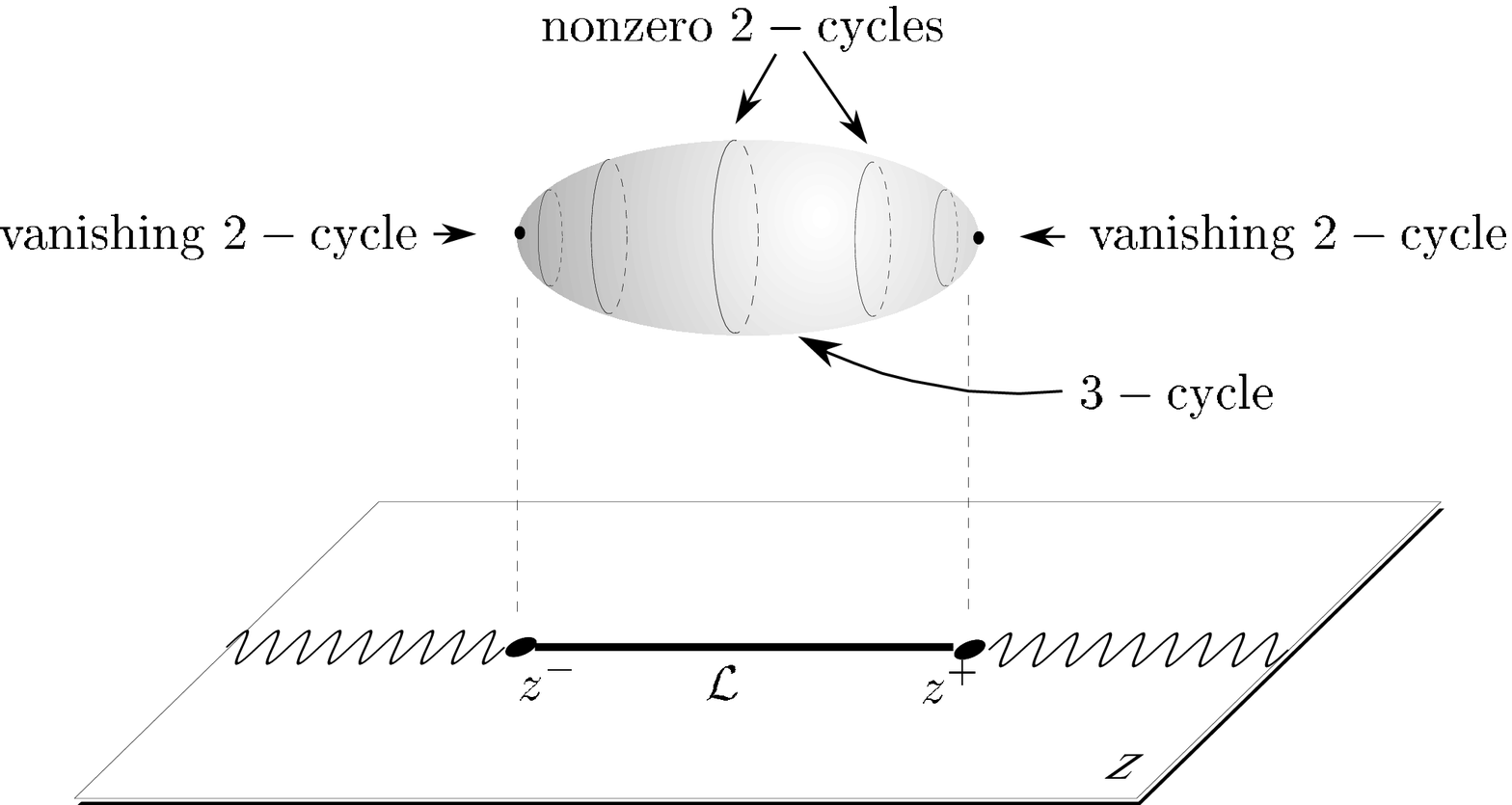}

{}From this picture we can see that a 3-brane that is wrapped around
the 3-cycle can be viewed as a fibration of a wrapped 2-brane over an
open line segment $\cal L$ in the base. This open line in the base is
thus a ``1-brane left-over'', obtained by wrapping two dimensions of
the 3-brane around the ALE 2-cycle. But, according to what we said at
the end of section 4.2, this corresponds precisely to a non-critical,
anti-self-dual IIB string on the base; its local tension varies along
$\cal L$, depending on the size of the 2-cycle over it. Obviously, if
the branch points $z^\pm$ coincide in some region of the moduli
space, the volume of the 3-brane and, in particular, the 1-brane
$\cal L$ vanishes, producing a massless ``monopole'' hypermultiplet
in four dimensions.\foot{We consider here only $\be$-type of
3-cycles, which give rise to (potentially massless) hypermultiplets.
In fact, $\al$-type of cycles, which describe electrically charged
fields like the massive gauge multiplets $W^\pm$, project on closed
lines in the base that wrap around the branch cuts -- precisely as it
is for rigid SW curves.}

Now, there is a partner of the open line $\cal L$ on another sheet of
the $z$-plane, which runs in the opposite direction. In effect,
taking the sheets together and forgetting about the ALE fibers, these
two lines corresponds to a {\it closed} anti-self dual string,
wrapping around a $\be$-cycle of a Riemann surface. This is indeed
precisely how the SW curves were obtained in section 3.4~! More
generally, for $G=SU(n)$ there are $n$ sheets, as well as $n-1$
fundamental vanishing 2-cycles in the ALE space (which are associated
with the simple roots of $SU(n)$). These gives rise to $n-1$
different types of anti-self dual strings that run on the various
sheets. It is now a simple mathematical fact of the representation of
Riemann surfaces in terms of branched coverings, that the monodromy
properties of these various strings are precisely such that they
correspond to a {\it single} type of string that winds around the
genus $g=n-1$ Riemann surface given in \eq{nuCu}.

Thus, what this discussion boils down to is that, effectively,
3-branes wrapped around the cycles of the threefold \eq{ALEfibr} are
equivalent to 1-branes, or anti-self dual strings, wrapped around the
Seiberg-Witten Riemann surface \cite{KLMVW}~! This gives then finally
a physical interpretation of the SW curves: just like the classical
spectral surfaces $X_0$ \eq{levsurf} represent ``$D$-manifolds''
\cite{BSV} in a dual formulation of classical gauge symmetry
breaking, the curves $X_1$ represent, in a dual formulation
of the \nex2 gauge theory, compactification manifolds on which the
six-dimensional, non-critical $(0,2)$ supersymmetric string lives;
see \figref{rigidual}. This is similar in spirit to ideas in refs.\
\cite{erik,witcom,claim}.

\figinsert{rigidual}
{The duality between conventional \nex2 Yang-Mills
theory and anti-self-dual strings winding
around the SW curve is just the {\it rigid remnant}
of the non-perturbative duality between heterotic and type II
strings.}
{1.4in}{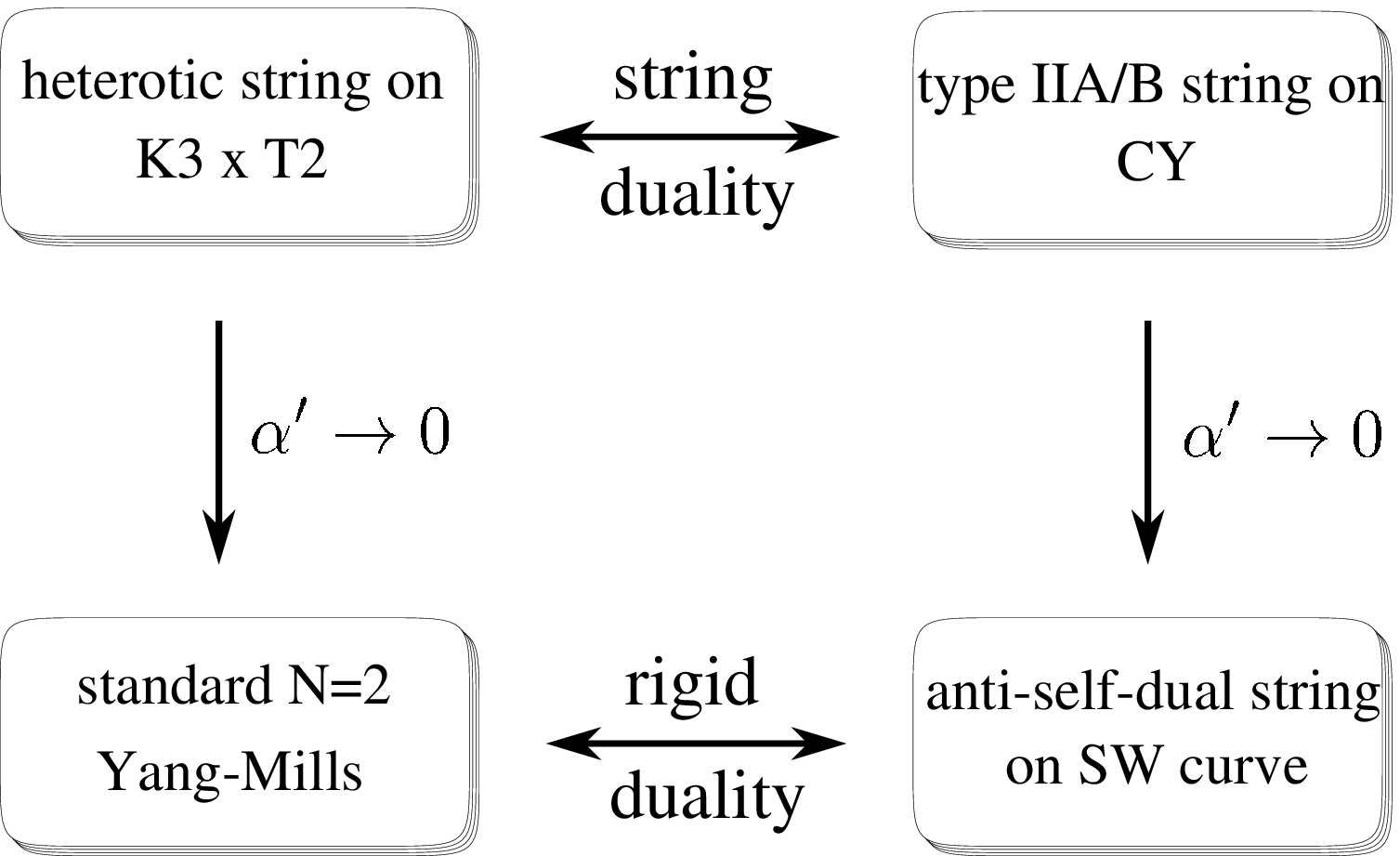}

\subsection{Geodesics and BPS States}

We now focus on some of the properties of the anti-self dual string
when wrapped around a SW curve; as we will see, this gives valuable
insight into the \nex2 YM theory itself. There is a crucial point to
make in this context, which can be easier appreciated if we ask
first: if we compactify the six dimensional anti-self dual string on
an ordinary flat torus $T_2$, it is known \cite{witcom} that this
gives an \nex4 supersymmetric gauge theory in four dimensions (this
theory has BPS states for all coprime $(g,q)$; see \figref{traj1}).
Why are we then supposed to get only an \nex2 gauge theory, if we
wrap the string on the SW curve, which is also a torus (for
$SU(2)$)~?

\figinsert{traj1}
{On a torus with standard flat metric, there are string geodesics for
all homology classes $(g,q)$ where $g$ and $q$ are coprime. This
reflects the BPS spectrum of \nex4 supersymmetric Yang-Mills theory.
}{1.3in}{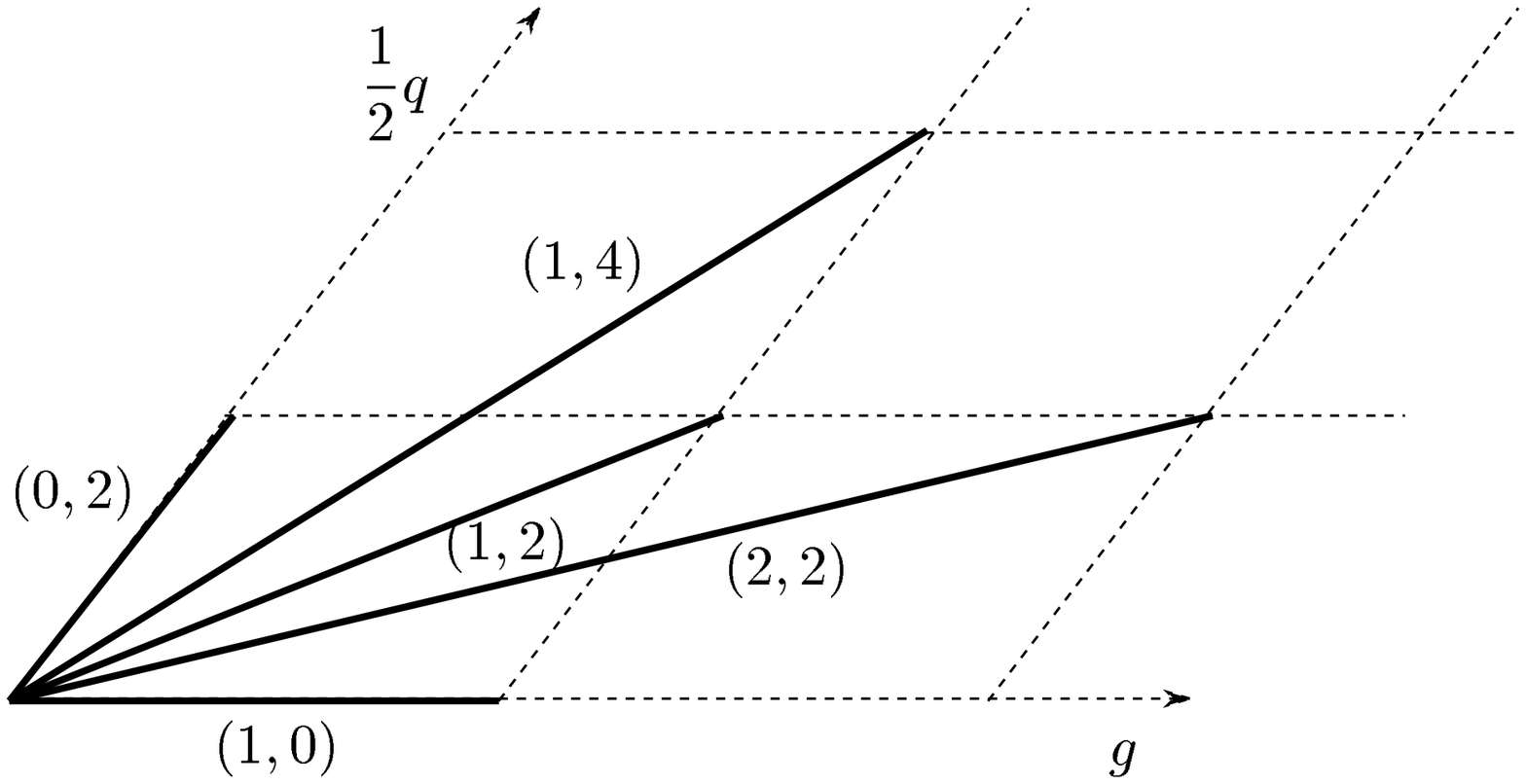}

The answer is that the non-critical string cannot wrap the curve in
an arbitrary fashion, rather it must wrap it in a particular
way. Indeed, a BPS state always corresponds to a minimal volume,
or ``supersymmetric'' cycle -- otherwise, it does not have the lowest
energy in a given homology class. This mean that the string
trajectory (or space-part of the string world-sheet) must be a {\it
geodesic} on the curve, with respect to a suitable metric. But, what
is then here the right metric ? To see this, recall the condition
for a ``supersymmetric 3-cycle'' \cite{BBS} in the threefold:
\beq
\del_\al X^\mu \del_\be X^\nu \del_\gamma X^\rho\,
\Omega_{\mu\nu\rho}e^{-\alpha' K}\ =\ {\rm
const.}\epsilon_{\al\be\gamma}
\eel{susycycle}
Here, $K$ is the K\"ahler potential (in which we have at present no
further interest), $\Omega$ the holomorphic 3-form on the CY, and
$\del_\al X^\mu$ the pull-back from the compactified space-time to
the 3-brane world volume. Now remember that in the limit $\al'\to0$,
the three-form $\Omega$ can be reduced by integration to give the
meromorphic SW differential $\lambda_{SW}=x(z){dz\over z}$.
Therefore, in the rigid limit the supersymmetric cycle condition
\eq{susycycle} reduces to:
\beq
{\del\over\del t}\lambda_{z}(z(t)) = {\rm const.}\ ,\ \ \ {\rm
where}\
\lambda_{SW}\equiv\lambda_{z}(z)dz\ ,
\eel{geodes}
and where $t$ parametrizes the space part of the string world-sheet.
This is precisely the geodesic differential equation for $z(t)$,
associated with the flat metric\foot {That this is the metric that
governs the shape of the self-dual strings on the SW curve, can also
be seen by some more direct geometrical reasoning, see \cite{KLMVW}.
Note that $\lambda_{SW}$ is multi-valued on the $z$-plane, but is
single valued on the curve $X_1$. This means that $g$ really is a
metric on the SW curve, and not on the base space of the fibration.}
\cite{KLMVW}
\beq\
g_{z\bar z}= \lambda_z \bar\lambda_{\bar z}\ .
\eel{metric}
Now, in contrast to the usual flat metric on the torus
used in the \nex4 supersymmetric compactification,
this metric has {\it poles} since $\lambda_{SW}$ is meromorphic.
Thus, metrically the SW curves have ``spikes'' sticking out,
which severely influence the form and kind of the possible
geodesics.

The main effect from these singularities on the world-sheet metric
is, for generic windings of type $(g,q)$, that the shortest
trajectory in the homology class $(g,q)$ may not be the ``direct''
one. Rather, the shortest trajectory may be one that is just a
composition of fundamental trajectories; see \figref{traj2} for an
example. In other words, the lightest state in a given charge class
$(g,q)$ may not be a single-particle, but a multi-particle state~!
When this happens, the state with charges $(g,q)$ cannot be counted
as a stable BPS state.

\figinsert{traj2}
{\nex2 Yang-Mills theory corresponds to a compactification torus with
a metric given by $|\lambda_z|^2$. The poles in the differential
$\lambda$ deflect the string trajectories, with the effect that the
shortest trajectories for $g>1$ are composite of several fundamental
ones. Specifically, the straight dashed line in charge class $(2,2)$
is not the shortest trajectory in this class, but the composite one
shown above is. On the other hand, we see from this picture that the
states of type $(1,2\ell)$ are good single-particle BPS states (in
the semi-classical region).
}{1.3in}{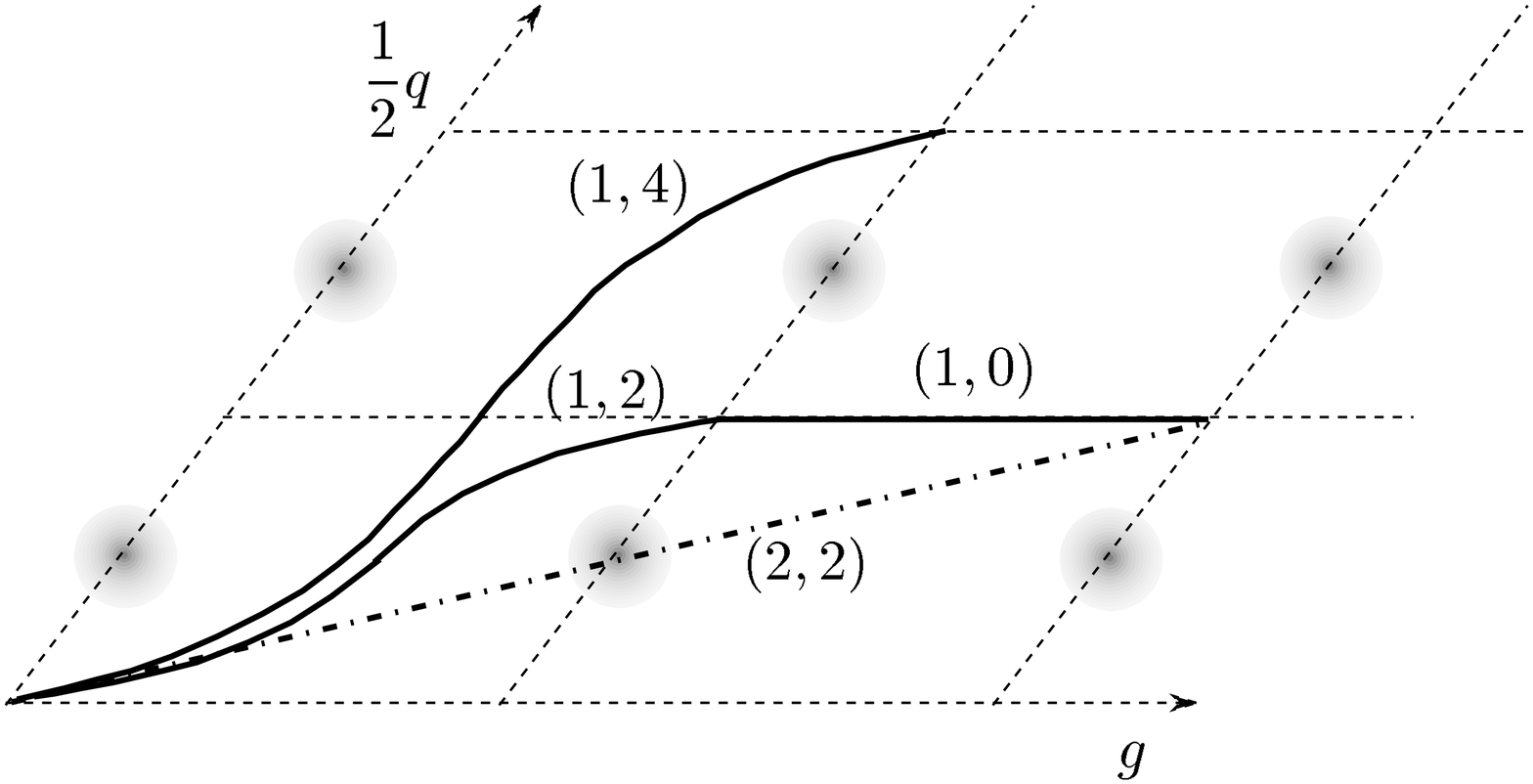}

We thus see that this dual formulation of \nex2 SYM theory gives a
new method for determining quantum BPS spectra \cite{KLMVW}. With
conventional field theoretic methods, without the use of SW geometry,
it is really hard to make statements about stable BPS spectra, see,
for example \cite{BPSspectra}. The geodesic string method has been
successfully applied \cite{KLMVW} to re-derive the BPS spectrum
\cite{SW,BiFe} of the $SU(2)$ SYM theory that we had exhibited in
section 2.5. More recently, it has been extended to discuss BPS
spectra when extra matter is added \cite{ABSS}. It has also been
studied \cite{nicketal} what happens if one starts with an \nex2
gauge theory with adjoint matter (which in total has \nex4
supersymmetry), and breaks \nex4$\to$\nex2 supersymmetry by giving
the adjoint matter field a mass. Then can explicitly see how most of
the \nex4 BPS states decay into the allowed semi-classical states of
the \nex2 Yang-Mills theory.

It is most interesting to study from this viewpoint what happens on
the line of marginal stability $\cal C$ discussed in section 2.5.
Here, $a_D/a\in\IR$, and thus the period lattice spanned by $(a_D,a)$
degenerates to a real line. That is, {\it all} possible string
trajectories lie on top of each other. In particular, the string
trajectory of the gauge field (and similarly for all the other
semi-classical BPS states with $(g,q)=(1,2\ell)$), is
indistinguishable from the composite trajectory made out of the
monopole $(1,0)$ and the dyon $(1,2)$; see \figref{degen}. This is
very similar to the considerations of section 2.5, where we mentioned
that for kinematical reasons the gauge field might decay into a
monopole-dyon pair.

\figinsert{degen}
{On the line $\cal C$ of marginal stability the string representation
degenerates, and the only indecomposable geodesics are the ones of
the monopole and the dyon. Shown is here the trajectory $(0,2)$ for
the gauge boson, which cannot be distinguished from the one of the
monopole-dyon pair.}
{1.in}{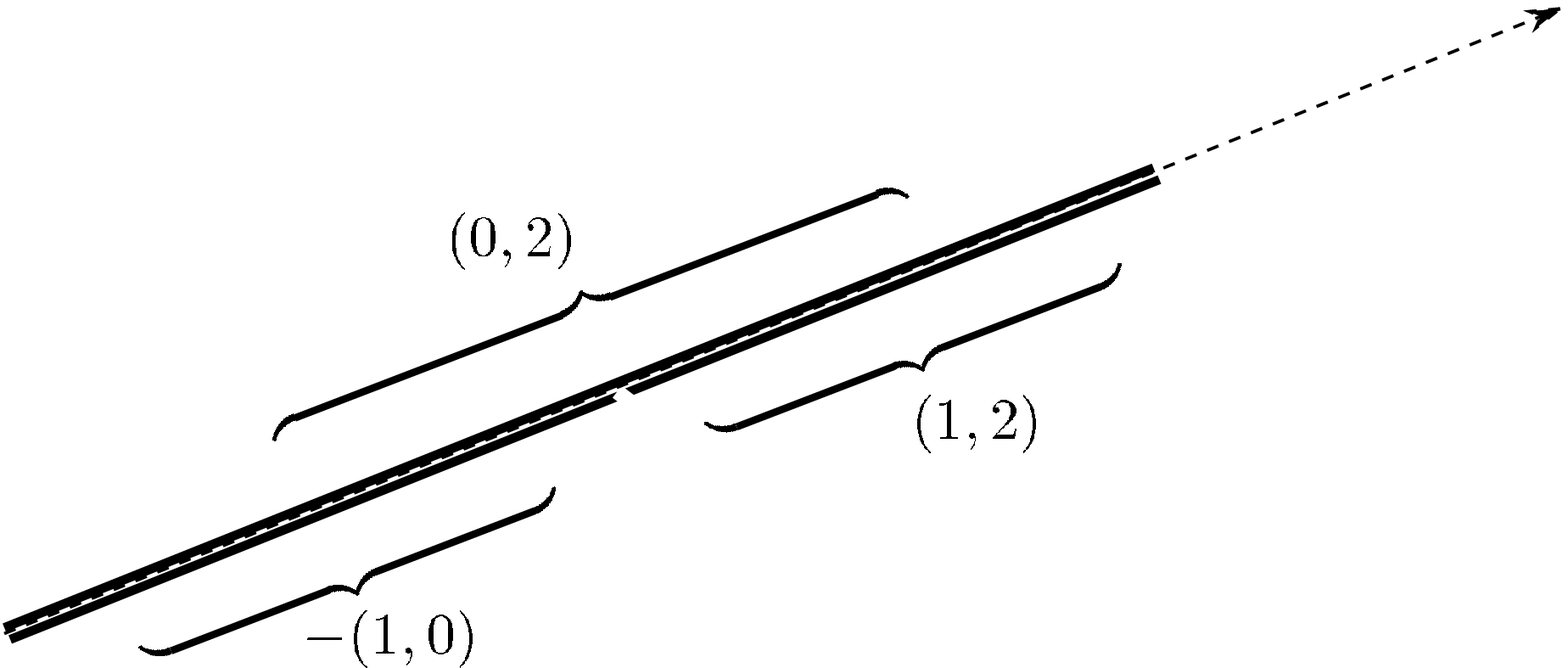}

However, the situation is quite different here, because we do not
merely talk about kinematics, but about a dual representation of the
BPS states. Thus, while \figref{degen} may simultaneously also
represent some simple kinematics, in the present context it shows
that the {\it string representation itself} degenerates. Remember
that even though we use classical physics (geometry) in the IIB
formulation, by rigid string duality we supposedly capture the exact
non-perturbative quantum behavior of the Yang-Mills theory.
Therefore, what we see here is, roughly speaking, that on $\cal C$
the (single particle) non-perturbative quantum state of the gauge
field degenerates into a two-particle state made from the monopole
and the dyon.

This clearly shows the conceptional power of the anti-self-dual
string representation. Insights like the one above are extremely hard
to obtain in ordinary quantum field theory, where the
gauge field is elementary and the monopole/dyon solitonic (or vice
versa). In contrast, in terms of non-critical strings, these fields
are treated on equal footing \cite{witcom}. Other applications of
this dual representation of \nex2 gauge theory include the
description of non-abelian gauge flux tubes and confinement in terms
of non-critical strings \cite{GMV}.

\section{Conclusions}

Note that the above construction of the SW theory is deductive -- the
SW theory in its full glory can be systematically derived from string
theory, once one takes the heterotic-type II string duality \cite{HT}
for granted and works out its consequences. This convergent evolution
of {\it a priori} disparate physical ideas seems to indicate that we
are really on the right track for understanding non-perturbative
phenomena in supersymmetric field and string theory.


\vskip 1.cm
\centerline{{\bf Acknowledgements}}

I like to thank S.\ Kachru, A.\ Klemm, P.\ Mayr, S.\ Theisen, C.\
Vafa, N.\ Warner and S.\ Yankielowicz for pleasant collaboration, and
K.\ Pilch for comments on the manuscript. I also thank the organizers
of the school for their hard work.

\goodbreak
\newcommand\nil[1]{{}}
\newcommand\nihil[1]{{\sl #1}}
\newcommand\ex[1]{}
\lref\SW{N.\ Seiberg and E.\ Witten, \nup426(1994) 19, hep-th/9407087;
\nup431(1994) 484, hep-th/9408099.} \lref\HT{C.\ Hull and P.\ Townsend,
\nup438 (1995) 109, hep-th/9410167.} \lref\suN{ A. Klemm, W. Lerche, S.
Theisen and S. Yankielowicz, \plt344 (1995) 169, hep-th/9411048; \\ P.\
Argyres and A.\ Faraggi, {Phys.\ Rev.\ Lett.} {\bf 74} (1995) 3931,
hep-th/9411057; \\ M. Douglas and S. Shenker, \nup447 (1995) 271,
hep-th/9503163;\\ P.\ Argyres and M.\ Douglas, \nup448 (1995) 93,
hep-th/9505062; \\ {A.\ Klemm, W.\ Lerche and S.\ Theisen,
\ex{Nonperturbative effective actions of $N=2$ supersymmetric gauge
theories,} Int.\ J.\ Mod.\ Phys.\ {\bf A11} (1996) 1929-1974,
hep-th/9505150;}\\ {C.\ Fraser and T.\ Hollowood, \nihil{On
the Weak Coupling Spectrum of $N=2$ Supersymmetric $SU(n)$ Gauge
theory,} hep-th/9610142.}} \lref\FHSV{ S.\ Ferrara, J.\
Harvey, A.\ Strominger and C.\ Vafa, \plt361 (1995) 59,
hep-th/9505162.}
\lref\KV{ S.\ Kachru and C.\ Vafa, \nup450 (1995) 69, hep-th/9506024.}
\lref\KLM{A.\ Klemm, W.\ Lerche and P.\ Mayr, \ex{K3 Fibrations and
Heterotic-Type II String Duality,} Phys.\ Lett.\ {\bf B357} (1995)
313-322, hep-th/9506112.} \lref\VaWi{C.\ Vafa and E.\ Witten,
\nihil{Dual String Pairs With $N=1$ And $N=2$ Supersymmetry In Four
Dimensions}, hep-th/9507050.} \lref\strom{ A. Strominger,
\nup451 (1995) 96, hep-th/9504090;\\ B.\ Greene, D.\ Morrison and A.\
Strominger, \nup451 (1995) 109, hep-th/9504145.} \lref\KKLMV{S.\
Kachru, A.\ Klemm, W.\ Lerche, P.\ Mayr and C.\ Vafa, \nup459 (1996)
537, hep-th/9508155.} \lref\KLMVW{A.\ Klemm, W.\ Lerche, P.\
Mayr, C.\ Vafa and N.\ Warner, \nihil{Self-Dual Strings and N=2
Supersymmetric Field Theory,} hep-th/9604034.}
\lref\pbranes{See eg.,\\ E.\ Witten, as in \cite{witcom};\\ {P.\ K.\
Townsend, \nihil{P-brane Democracy,} hep-th/9507048;}\\ {K.\
\&\ M.\ Becker and A.\ Strominger,\\ Nucl.\ Phys.\ {\bf B456} (1995)
130-152, hep-th/9507158;}\\ {J.\ Polchinski,
\ex{Dirichlet-Branes and Ramond-Ramond Charges,} Phys.\ Rev.\ Lett.\
{\bf 75} (1995) 4724-4727, hep-th/9510017;}\\ {Cumrun Vafa,
\ex{Evidence for F-Theory,} Nucl.\ Phys.\ {\bf B469} (1996) 403-418,
hep-th/9602022.}\\ {D.\ Kutasov and E.\ Martinec, \nihil{New
Principles for String/Membrane Unification,}
hep-th/9602049;}\\ {M.\ Green, \nihil{World-volumes and string
target spaces,} hep-th/9602061.} }
\lref\NSbeta{N.\ Seiberg, \plt206(1988) 75.} \lref\SCHIF{ V.\
Novikov, M.\ Schifman, A.\ Vainstein, M\ Voloshin and V.\ Zakharov,
\nup229(1983) 394;\\ V.\ Novikov, M.\ Schifman, A.\ Vainstein and V.\
Zakharov, \nup229(1983) 381, 407;\\ M.\ Schifman, A.\ Vainstein and
V.\ Zakharov, \plt166(1986) 329.} \lref\MO{C.\ Montonen and D.\
Olive, \plt72 (1977) 117.} \lref\thooft{G.\ `t Hooft, \nup190(1981)
455.} \lref\OW{D.\ Olive and E.\ Witten, \plt78 (1978) 97.}
\lref\AD{P.\ Argyres and M.\ Douglas, as in ref. \cite{suN}.}
\lref\AspJl{ {P.\ Aspinwall and J.\ Louis, \ex{On the Ubiquity of K3
Fibrations in String Duality,} Phys.\ Lett.\ {\bf B369} (1996)
233-242, hep-th/9510234;}\\ {P.\ Aspinwall and M.\ Gross,
\ex{Heterotic-Heterotic String Duality and Multiple K3 Fibrations,}
Phys.\ Lett.\ {\bf B382} (1996) 81-88, hep-th/9602118.}}
\lref\Asp{P.\ Aspinwall, \ex{Enhanced Gauge Symmetries and K3
Surfaces,} Phys.\ Lett.\ {\bf B357} (1995) 329-334,
hep-th/9507012.} \lref\andyrev{A.\ Strominger, \nihil{Black
Hole Condensation and Duality in String Theory,}
hep-th/9510207.} \lref\nahm{W.~Nahm, {\it On the
Seiberg-Witten Approach to electric-magnetic Duality,}
hep-th/9608121.} \lref\Arn{See e.g., V.\ Arnold, A.\
Gusein-Zade and A.\ Varchenko, {\it Singularities of Differentiable
Maps I, II}, Birkh\"auser 1985;\\ V.\ Arnold, {\it Dynamical Systems
VI}, Singularity Theory I, Enc.\ of Math.\ Sci.\ Vol.\ 6, Springer
1993. } \lref\MW{E.\ Martinec and N.P.\ Warner, \ex{Integrable
Systems and Supersymmetric Gauge Theory,} Nucl.\ Phys.\ {\bf B459}
(1996) 97-112, hep-th/9509161.} \lref\LSW{W.\ Lerche, A.\ N.\
Schellekens and N.\ P.\ Warner, \nihil{Lattices and Strings,} Phys.\
Rept.\ {\bf 177} (1989) 1.}
 \lref\Wi{E.\ Witten, \nup443 (1995) 85, hep-th/9503124.} \lref\BSV{M.\
Bershadsky, V.\ Sadov and C.\ Vafa,\\ \ex{D-Strings on D-Manifolds,}
Nucl.\ Phys.\ {\bf B463} (1996) 398-414, hep-th/9510225.}
\lref\Slodo{see eg., P.\ Slodowy, {\it Platonic Solids, Kleinian
Singularities and Lie Groups}, in: Algebraic Geometry, {\sl Lecture
Notes in Mathematics} {\bf 1008}, Springer 1983.} \lref\reviews{ {A.\
Bilal, \nihil{Duality in N=2 SUSY SU(2) Yang-Mills Theory: A
pedagogical introduction to the work of Seiberg and Witten,}
hep-th/9601007};\\ {C.\ Gomez and R.\ Hernandez,
\nihil{Electric-Magnetic Duality and Effective Field Theories,}
hep-th/9510023}; {C.\ Gomez, R.\ Hernandez and E.\ Lopez,
\nihil{Integrability, Duality and Strings,} hep-th/9604043},
\nihil{Integrability, Jacobians and Calabi-Yau Threefolds,}
hep-th/9604057;\\ {L.\ Alvarez-Gaum\'e and M.\ Marino, to
appear.} } \lref\BiFe{ {A.\ Bilal and F.\ Ferrari, \nihil{The
Strong-Coupling Spectrum of the Seiberg-Witten Theory,} Nucl.\ Phys.\
{\bf B469} (1996) 387-402, hep-th/9602082.}} \lref\HKP{E.\
D'Hoker, I.\ Krichever and D.\ Phong, \nihil{The Effective
Prepotential of $N=2$ Supersymmetric $SU(N_c)$ Gauge Theories,}
hep-th/9609041; \nihil{The Effective Prepotential of $N=2$
Supersymmetric $SO(N_c)$ and $Sp(N_c)$ Gauge Theories,}
hep-th/9609145.} \lref\conformal{ {P.\ Argyres, M.\ Plesser
and N.\ Seiberg and E.\ Witten, \ex{New N=2 Superconformal Field
Theories in Four Dimensions,} Nucl.\ Phys.\ {\bf B461} (1996) 71-84,
hep-th/9511154;}\\ {T.\ Eguchi, K.\ Hori, K.\ Ito and S.K.\
Yang, Nucl.\ Phys.\ {\bf B471} (1996) 430-444,
hep-th/9603002.}}
 \lref\withmatter{ {N.\ Seiberg and E.\ Witten, as in \cite{SW};}\\
{A.\ Hanany and Y.\ Oz, \ex{On the quantum moduli space of vacua of
N=2 supersymmetric SU(N(c)) gauge theories,} Nucl.\ Phys.\ {\bf B452}
(1995) 283-312, hep-th/9505075;}\\ {P.\ Argyres, M.\ Plesser
and A.\ Shapere, \ex{The Coulomb Phase of N=2 Supersymmetric QCD,}
Phys.\ Rev.\ Lett.\ {\bf 75} (1995) 1699-1702,
hep-th/9505100;}\\ {J.\ Minahan and D.\ Neme\-schans\-ky,
\ex{Hyper\-elliptic Curves for Super\-symmetric Yang-Mills,} Nucl.\
Phys.\ {\bf B464} (1996) 3-17, hep-th/9507032;\\ \ex{$N=2$
Super Yang-Mills and Subgroups of $SL(2,Z)$,} Nucl.\ Phys.\ {\bf
B468} (1996) 72-84, hep-th/9601059;}\\ {R.\ Donagi and E.\
Witten, \ex{Supersymmetric Yang-Mills Theory And Integrable Systems,}
Nucl.\ Phys.\ {\bf B460} (1996) 299-334, hep-th/9510101;}\\
{P.\ Argyres and A.\ Shapere, Nucl.\ Phys.\ {\bf B461} (1996)
437-459, hep-th/9509175;}\\ {A.\ Hanany, \ex{On the Quantum
Moduli Space of N=2 Supersymmetric Gauge Theories,} Nucl.\ Phys.\
{\bf B466} (1996) 85-100, hep-th/9509176;}\\ {P.\ Argyres, M.\
Plesser and N.\ Seiberg, \ex{The Moduli Space of N=2 SUSY QCD and
Duality in $N=1$ SUSY QCD,} Nucl.\ Phys.\ {\bf B471} (1996) 159-194,
hep-th/9603042;}\\ {K.\ Ito and S.K.\ Yang,
\nihil{Picard-Fuchs Equations and Prepotentials in $N=2$
Supersymmetric QCD,} hep-th/9603073;}\\ {A.\ Bilal and F.\
Ferrari, \nihil{Exact Multiplets of Spontaneously Broken Discrete
Global Symmetries: The Example of $N=2$ SUSY QCD,}
hep-th/9606111;}\\ {F.\ Ferrari, \nihil{Charge fractionization
in $N=2$ supersymmetric QCD,} hep-th/9609101;}\\ {A.\
Brandhuber and S.\ Stieberger, \nihil{Periods, Coupling Constants and
Modular Functions in $N=2$ $SU(2)$ SYM with Massive Matter,}
hep-th/9609130;}\\ {N.\ Dorey, V.\ Khoze and M.\ Mattis,
\nihil{On $N=2$ Supersymmetric QCD with 4 Flavors,}
hep-th/9611016.} } \lref\othergrps{ U.\ Danielsson and B.\
Sundborg, \plt358 (1995) 273, hep-th/9504102; \\ A.\ Brandhuber and K.\
Landsteiner, \plt358 (1995) 73, hep-th/9507008;\\ {U.\ Danielsson and
B.\ Sundborg, \ex{Exceptional Equivalences in $N=2$ Supersymmetric
Yang-Mills Theory,} Phys.\ Lett.\ {\bf B370} (1996) 83-94,
hep-th/9511180;}\\ {M.\ Alishahiha, F.\ Ardalan and F.\
Mansouri, Phys.\ Lett.\ { \bf B381} (1996) 446-450,
hep-th/9512005; M.\ R.\ Abolhasani, M.\ Alishahiha and A.\ M.\
Ghezelbash, \nihil{The Moduli Space and Monodromies of the $N=2$
Supersymmetric Yang-Mills Theory with any Lie Gauge Groups,}
hep-th/9606043;}\\ {W.\ Lerche and N.\ Warner,
\nihil{Exceptional SW Geometry from ALE Fibrations,}
hep-th/9608183;}\\ {K.\ Landsteiner, J.\ Pierre and S.\
Giddings, \nihil{On the Moduli Space of N = 2 Supersymmetric $G_2$
Gauge Theory,} hep-th/9609059.} }
 \lref\instaComp{ {D.\ Finnell and P.\ Pouliot, \ex{Instanton
Calculations versus exact Results in Four-Dimensional SUSY Gauge
Theories,} Nucl.\ Phys.\ {\bf B453} (1995) 225-239,
hep-th/9503115;}\\ {A.\ Yung, \nihil{Instanton-induced
Effective Lagrangian in the Seiberg-Witten Model,}
hep-th/9605096;}\\ {K.\ Ito and N.\ Sasakura,
\ex{One-Instanton Calculations in $N=2$ Supersymmetric $SU(N_c)$
Yang-Mills Theory,} Phys.\ Lett.\ {\bf B382} (1996) 95-103,
hep-th/9602073;}\\ {F.\ Fucito and G.\ Travaglini,
\nihil{Instanton Calculus and Nonperturbative Relations in $N = 2$
Supersymmetric Gauge Theories,} hep-th/9605215;}\\ {N.\ Dorey,
V.A.\ \ Khoze and M.\ Mattis, \nihil{Multi-Instanton Check of the
Relation Between the Prepotential $F$ and the Modulus $u$ in $N=2$
SUSY Yang-Mills Theory,} hep-th/9606199;\\ \ex{Multi-Instanton
Calculus in $N=2$ Supersymmetric Gauge Theory,} Phys.\ Rev.\ {\bf
D54} (1996) 2921-2943, hep-th/9603136; \nihil{A Two-Instanton
Test of the Exact Solution of $N=2$ Supersymmetric QCD,}
hep-th/9607066; \nihil{Multi-Instanton Calculus in $N=2$
Supersymmetric Gauge Theory II: Coupling to Matter,}
hep-th/9607202;}\\ {T.\ Harano and M.\ Sato,
\nihil{Multi-Instanton Calculus versus Exact Results in $N=2$
Supersymmetric QCD,} hep-th/9608060;\\} {H.\ Aoyama, T.\
Harano, M.\ Sato and S.\ Wada, \nihil{Multi-Instanton Calculus in
$N=2$ Supersymmetric QCD,} hep-th/9607076.} }
 \lref\ALESW{W.\ Lerche and N.\ P.\ Warner, as in \cite{othergrps}.}
\lref\AF{P.\ Argyres and A.\ Faraggi, as in \cite{suN}.}
\lref\KLT{A.\ Klemm, W.\ Lerche and S.\ Theisen, last reference in
\cite{suN}.} \lref\KLTY{A. Klemm, W. Lerche, S. Theisen and S.
Yankielowicz, first reference in \cite{suN}.} \lref\WDo{E.\ Witten,
\nihil{Monopoles and Four Manifolds,} hep-th/9411102.}
\lref\luisetal{ {L.\ Alvarez-Gaum\'e, J.\ Distler, C.\ Kounnas and
M.\ Marino, \nihil{Large Softly broken $N=2$ QCD,}
hep-th/9604004;}\\ {L.\ Alvarez-Gaume and M.\ Marino,
\nihil{Softly broken $N=2$ QCD,} hep-th/9606168; \nihil{More
on softly broken $N=2$ QCD,} hep-th/9606191.} } \lref\PF{ {A.\
Ceresole, R.\ D'Auria, S.\ Ferrara, \ex{On the Geometry of Moduli
Space of Vacua in $N=2$ Supersymmetric Yang-Mills Theory,} Phys.\
Lett.\ {\bf B339} (1994) 71-76, hep-th/9408036;}\\ {A.\ Klemm,
W.\ Lerche and S.\ Theisen, as in ref.\ \cite{suN};}\\ {S.\ Ryang,
\ex{The Picard-Fuchs Equations, Monodromies and Instantons in the
$N=2$ SUSY Gauge Theories,} Phys.\ Lett.\ {\bf B365} (1996) 113-118,
hep-th/9508163;}\\ {J. Isidro, A.\ Mukherjee, J.\ Nunes and
H.\ Schnitzer, \nihil{A New Derivation of the Picard-Fuchs Equations
for Effective ¢N=2¢ Super Yang-Mills Theories,}
hep-th/9609116;}\\ {M.\ Alishahiha, \nihil{On the Picard-Fuchs
equations of the SW models,} hep-th/9609157;\\ {H.\ Ewen and
K.\ Foerger, \nihil{Simple Calculation of Instanton Corrections in
Massive $N=2$ $SU(3) $SYM,} hep-th/9610049.} } } \lref\prepot{
{M.\ Matone, \ex{Instantons and recursion relations in N=2 SUSY gauge
theory,} Phys.\ Lett.\ {\bf B357} (1995) 342-348,
hep-th/9506102;}\\ {J.\ Sonnenschein, S.\ Theisen and S.\
Yankie\-lowicz, \ex{On the Relation Between the Holomorphic
Prepotential and the Quantum Moduli in SUSY Gauge Theories,} Phys.\
Lett.\ {\bf B367} (1996) 145-150, hep-th/9510129;}\\ {T.\
Eguchi and S.K.\ Yang, \nihil{Prepotentials of $N=2$ Supersymmetric
Gauge Theories and Soliton Equations,} hep-th/9510183;}\\ {H.\
Itoyama and A.\ Morozov, \nihil{Prepotential and the Seiberg-Witten
Theory,} hep-th/9512161;}\\ {G.\ Bonelli and M.\ Matone,
\ex{Nonperturbative Renormalization Group Equation and Beta Function
in $N=2$ SUSY Yang-Mills,} Phys.\ Rev.\ Lett.\ {\bf 76} (1996)
4107-4110, hep-th/9602174; \nihil{Nonperturbative Relations in
$N=2$ Susy Yang-Mills and WDVV Equation,} hep-th/9605090.}}
\lref\massprep{ {K.\ Ito and S.K.\ Yang, \nihil{Picard-Fuchs
Equations and Prepotentials in $N=2$ Supersymmetric QCD,}
hep-th/9603073;}\\ {Y.\ Ohta, \nihil{Prepotential of $N=2$
$SU(2)$ Yang-Mills Gauge Theory Coupled with a Massive Matter
Multiplet,} hep-th/9604051; \nihil{Prepotential of $N=2$
$SU(2)$ Yang-Mills Gauge Theory Coupled with Massive Matter
Multiplets,} hep-th/9604059;}\\ {T.\ Masuda and H.\ Suzuki,
\nihil{Prepotential of $N=2$ Supersymmetric Yang-Mills Theories in
the Weak Coupling Region,} hep-th/9609065; \nihil{Periods and
Prepotential of N=2 SU(2) Supersymmetric Yang- Mills Theory with
Massive Hypermultiplets,} hep-th/9609066;}\\ {H.\ Ewen, K.\
Forger and S.\ Theisen, \nihil{Prepotentials in $N=2$ Supersymmetric
$SU(3)$ YM-Theory with Massless Hypermultiplets,}
hep-th/9609062.} } \lref\witcom{E.\ Witten, \nihil{Some
Comments on String Dynamics,} hep-th/9507121.}
\lref\unique{R.\ Flume, M.\ Magro, L.\ O'Raifeartaigh, I.\ Sachs and
O.\ Schnetz, \nihil{Uniqueness of the Seiberg-Witten Effective
Lagrangian}, hep-th/9611123.}
 \lref\DM{M.\ Douglas and G.\ Moore, \nihil{D-branes, Quivers, and
ALE Instantons,} hep-th/9603167.} \lref\strongcoup{ {A.\ Klemm
and P.\ Mayr, \nihil{Strong Coupling Singularities and Non-abelian
Gauge Symmetries in $N=2$ String Theory,} hep-th/9601014;}\\
{S.\ Katz, D.\ Morrison and M. Plesser, \ex{Enhanced Gauge Symmetry
in Type II String Theory,} Nucl.~ Phys.~{\bf B477} (1996) 105-140,
hep-th/9601108.} } \lref\selfd{ {M.\ Duff and J.\ Lu, \nup416
(1994) 301, hep-th/9306052;}\\ {N.\ Seiberg and E.\ Witten,
\ex{Comments on String Dynamics in Six Dimensions,} Nucl.\ Phys.\
{\bf B471} (1996) 121-134, hep-th/9603003;}\\ {M.\ J.\ Duff,
H.\ Lu and C.\ N.\ Pope, \ex{Heterotic phase transitions and
singularities of the gauge dyonic string,} Phys.\ Lett.\ {\bf B378}
(1996) 101-106, hep-th/9603037;}\\ {E.\ Witten, \ex{Phase
Transitions In M-Theory And F-Theory,} Nucl.\ Phys.\ {\bf B471}
(1996) 195-216, hep-th/9603150;}\\ {P.\ Argyres and K.\
Dienes, \ex{On the worldsheet formulation of the six-dimensional
self- dual string.,} Phys.~ Lett.~{\bf B387} (1996) 727-734,
hep-th/9607190;}\\ {M.\ Perry and J.\ Schwarz,
\nihil{Interacting Chiral Gauge Fields in Six Dimensions and
Born-Infeld Theory,} hep-th/9611065;}\\ {J.\ Distler and A.\
Hanany, \nihil{(0,2) Noncritical Strings in Six Dimensions,}
hep-th/9611104.} } \lref\NexTwoDual{ {V.\ Kaplunovsky, J.\
Louis, and S.\ Theisen, \ex{Aspects of duality in N=2 string vacua,}
Phys.\ Lett.\ {\bf B357} (1995) 71-75, hep-th/9506110;}\\
A.Klemm, W. Lerche and P.\ Mayr, as in \cite{KLM};\\ C.\ Vafa and E.\
Witten,as in \cite{VaWi};\\ G.\ Cardoso, D.\ L\"ust and T.\ Mohaupt,
hep-th/9507113;\\ I.\ Antoniadis, E.\ Gava, K.\ Narain and T.\
Taylor, \nup455 (1995) 109, hep-th/9507115;\\ S.\ Kachru, A.\
Klemm, W.\ Lerche, P.\ Mayr and C.\ Vafa, as in \cite{KKLMV};\\ {A.\
Sen and C.\ Vafa, \ex{Dual pairs of type II string compactification,}
Nucl.\ Phys.\ {\bf B455} (1995) 165-187, hep-th/9508064.}\\
I.\ Antoniadis and H.\ Partouche, \nup460 (1996) 470,
hep-th/9509009;\\ G.\ Curio, \plt366 (1996) 131,
hep-th/9509042; \plt368 (1996) 78, hep-th/9509146; \\
G.\ Aldazabal, A.\ Font, L.E.\ Ibanez and F.\ Quevedo,
hep-th/9510093;\\ {I.\ Antoniadis, S.\ Ferrara, E.\ Gava, K.\
Narain and T.\ Taylor, \ex{Duality Symmetries in $N=2$ Heterotic
Superstring,} Nucl.\ Phys.\ Proc.\ Suppl.\ {\bf 45BC} (1996) 177-187,
hep-th/9510079;}\\ {P.\ Aspinwall, \ex{An $N=2$ Dual Pair and
a Phase Transition,} Nucl.\ Phys.\ {\bf B460} (1996) 57-76,
hep-th/9510142;}\\ {J.\ Harvey, G.\ Moore, \nihil{Algebras,
BPS States, and Strings,} Nucl.~ Phys.~{\bf B463} (1996) 315-368,
hep-th/9510182;}\\ P.\ Aspinwall and J.\ Louis, as in
\cite{AspJl};\\ I.\ Antoniadis, S.\ Ferrara and T.\ Taylor,\nup 460
(1996) 489, hep-th/9511108;\\ {M.\ Henningson and G.\ Moore,
\ex{Counting Curves with Modular Forms,} Nucl.\ Phys.\ {\bf B472}
(1996) 518-528, hep-th/9602154;}\\ {G.\ Cardoso, G.\ Curio,
D.\ L\"ust and T\ Mohaupt, \nihil{Instanton Numbers and Exchange
Symmetries in $N=2$ Dual String Pairs,} hep-th/9603108;}\\
{P.\ Candelas and A.\ Font, \nihil{Duality Between Webs of Heterotic
and Type II Vacua,} hep-th/9603170;}\\ {I.\ Antoniadis, E.\
Gava, K.\ Narain and T.\ Taylor, \nihil{Topological Amplitudes in
Heterotic Superstring Theory,} hep-th/9604077;}\\ {J.\ Louis,
J.\ Sonnenschein, S.\ Theisen and S.\ Yankielowicz,
\nihil{Non-Perturbative Properties of Heterotic String Vacua
Compactified on ${K3\times T^2}$,} hep-th/9606049;}\\
{I.\ Antoniadis and B.\ Pioline, \nihil{Higgs branch, HyperKaehler
Quotient and Duality in SUSY $N = 2$ Yang-Mills Theories,}
hep-th/9607058;}\\ {C.\ Gomez, R.\ Hernandez and E.\ Lopez,
{\it $K3$-Fibrations and Softly Broken $N=4$ Super-\\symmetric Gauge
Theories,} hep-th/9608104;}\\ {G.\ Cardoso, G.\ Curio and D.\
L\"ust, \nihil{Perturbative Couplings and Modular Forms in $N=2$
String Models with a Wilson Line,} hep-th/9608154;}\\ {J.\
Harvey and G.\ Moore, \nihil{Fivebrane Instantons and $R^2$ couplings
in $N=4$ String Theory,} hep-th/9610237.} } \lref\KthreeFib{
{B.\ Hunt and R.\ Schimmrigk, \ex{Heterotic Gauge Structure of Type
II K3 Fibrations,} Phys.\ Lett.\ {\bf B381} (1996) 427-436,
hep-th/9512138;}\\ {S.\ Hosono, B.\ Lian and S.\ T.\ Yau,
\nihil{Calabi-Yau Varieties and Pencils of K3 Surfaces,}
alg-geom/9603020;}\\ {A.\ C.\ Avram, M.\ Kreuzer, M.\
Mandelberg and H.\ Skarke, \nihil{Searching for K3 Fibrations,}
hep-th/9610154.} } \lref\GKMMM{A.\ Gorskii, I.\ Krichever, A.\
Marshakov, A.\ Mironov and A.\ Morozov, \ex{Integrability and
Seiberg-Witten exact solution,} Phys.\ Lett.\ {\bf B355} (1995)
466-474, hep-th/9505035.} \lref\integ{ {T.\ Nakatsu and K.\
Takasaki, \ex{Whitham-Toda hierarchy and $N=2$ supersymmetric
Yang-Mills theory,} Mod.\ Phys.\ Lett.\ {\bf A11} (1996) 157-168,
hep-th/9509162;}\\ {E.\ Martinec, \ex{Integrable Structures in
Supersymmetric Gauge and String Theory,} Phys.\ Lett.\ {\bf B367}
(1996) 91-96, hep-th/9510204;}\\ {H.\ Itoyama and A.\ Morozov,
\nihil{Integrability and Seiberg-Witten Theory: Curves and Periods,}
hep-th/9511126; \nihil{Prepotential and the Seiberg-Witten
Theory,} hep-th/9512161; \nihil{Integrability and
Seiberg-Witten theory,} hep-th/9601168;}\\ {I.\ Krichever and
D.\ Phong, \nihil{On the Integrable Geometry of Soliton Equations and
$N=2$ Supersymmetric Gauge Theories,} hep-th/9604199;}\\ {A.\
Gorskii, \nihil{Peierls model and Vacuum Structure in $N=2$
Supersymmetric Gauge Theories,} hep-th/9605135;}\\ {A.\
Marshakov, A.\ Mironov and A.\ Morozov, \nihil{WDVV-like equations in
$N=2$ SUSY Yang-Mills Theory,} hep-th/9607109;}\\ {A.\
Marshakov, \nihil{From Nonperturbative SUSY Gauge Theories to
Integrable Systems,} hep-th/9607159; \nihil{Integrability as
Effective Principle of Nonperturbative Field and String Theories,}
hep-th/9608161; \nihil{Non-perturbative Quantum Theories and
Integrable Equations,} hep-th/9610242;} }
 \lref\nonre{B.\ de Wit, P.\ Lauwers, R.\ Philippe, S.\ Su and A.\
van Proeyen, \plt134(1984)37;\\ B.\ de Wit, A.\ van Proeyen,
\nup245(1984)89; \\ J.P.\ Derendinger, S.\ Ferrara, A.\ Masiero and
A.\ van Proeyen, \plt140(1984)307; \\ B.\ de Wit, P.G.\ Lauwers, A.\
van Proeyen, \nup255(1985)569;\\ E.\ Cremmer, C.\ Kounnas, A.\ van
Proeyen, J.P.\ Derendinger, S.\ Ferrara, B.\ de Wit and L.\
Girardello, \nup250(1985)385;\\ N.\ Berkovits and W.\ Siegel,
\nup462(1996)213. } \lref\smallinst{ {E.\ Witten, \ex{Small
Instantons in String Theory,} Nucl.~ Phys.~{\bf B460} (1996) 541-559,
hep-th/9511030.} } \lref\sadlyenough{ A.\ Klemm, W.\ Lerche,
P.\ Mayr and N.\ Warner, unpublished and perhaps to appear.}
\lref\marg{ {A.\ Fayyazuddin, \nihil{Some comments on $N=2$
supersymmetric Yang-Mills,} hep-th/9504120;}\\ {U.\ Lindstrom
and M.\ Rocek, \ex{A Note on the Seiberg-Witten solution of N=2
superYang- Mills theory,} Phys.~ Lett.~{\bf B355} (1995) 492-493,
hep-th/9503012;}\\ {P.\ Argyres, A.\ Faraggi and A.\ Shapere,
\nihil{Curves of marginal stability in $N=2$ super QCD,}
hep-th/9505190;}\\ {M.\ Matone, \nihil{Koebe 1/4 theorem and
inequalities in $N=2$ super QCD,} Phys.~ Rev.~{\bf D53} (1996)
7354-7358, hep-th/9506181;}\\ {M.\ Henningson,
\ex{Discontinuous BPS spectra in $N = 2$ gauge theory,} Nucl.~
Phys.~{\bf B461} (1996) 101-108, hep-th/9510138;}\\ A.\ Bilal
and F.\ Ferrari, as in \cite{BiFe};\\ {\nihil{Curves of Marginal
Stability and Weak and Strong-Coupling BPS Spectra in $N2$
Supersymmetric QCD,} hep-th/9605101;} {\nihil{Exact Multiplets
of Spontaneously Broken Discrete Global Symmetries: The Example of
$N=2$ SUSY QCD,} hep-th/9606111;}\\ {A.\ Bilal,
\nihil{Discontinuous BPS spectra in $N = 2$ SUSY QCD,}
hep-th/9606192.}\\ {F.\ Ferrari, \nihil{Duality and BPS
Spectra in $N=2$ Supersymmetric QCD,} hep-th/9611012.} }
\lref\probes{ {A.\ Sen, \nihil{F-theory and Orientifolds,}
hep-th/9605150;}\\ {T.\ Banks, M.\ Douglas and N.\ Seiberg,
\nihil{Probing F-theory with Branes,} hep-th/9605199;}\\ {K.\
Dasgupta and S.\ Mukhi, \nihil{F-theory at Constant Coupling,}
hep-th/9606044.} } \lref\MDML{M.\ Douglas and M.\ Li,
\nihil{D-Brane Realization of $N=2$ Super Yang-Mills Theory in Four
Dimensions,} hep-th/9604041.} \lref\Ftheory{ {C.\ Vafa,
\ex{Evidence for F-Theory,} Nucl.\ Phys.\ {\bf B469} (1996) 403-418,
hep-th/9602022;}\\ {D.\ Morrison and C.\ Vafa,
\ex{Compactifications of F-Theory on Calabi--Yau Threefolds -- I,}
Nucl.\ Phys.\ {\bf B473} (1996) 74-92, hep-th/9602114;
\nihil{Compactifications of F-Theory on Calabi--Yau Threefolds --
II,} hep-th/9603161.} } \lref\Neqone{See for reviews eg., {N.\
Seiberg, \nihil{The Power of holomorphy: Exact results in 4-D SUSY
field theories,} hep-th/9408013;}\\ {K.\ Intriligator and N.\
Seiberg, \nihil{Lectures on Supersymmetric Gauge Theories and
Electric-Magnetic Duality,} Nucl.\ Phys.\ Proc.\ Suppl.\ {\bf 45BC}
(1996) 1-28, hep-th/9509066;}\\ {S.\ Elitzur, A.\ Forge, A.\
Giveon and E.\ Rabinovici, \nihil{Summary of Results in $N=1$
Supersymmetric $SU(2)$ Gauge Theories,} hep-th/9512140.} }
\lref\PAMG{P.\ Aspinwall and M.\ Gross, \nihil{The $SO(32)$ Heterotic
String on a $K3$ Surface,} hep-th/9605131.} \lref\Fmatter{
{S.\ Katz and C.\ Vafa, \nihil{Matter from Geometry,}
hep-th/9606086;}\\ {M.\ Bershadsky et. al., \nihil{Geometric
Singularities and Enhanced Gauge Symmetries,} hep-th/9605200.}
} \lref\KKV{S.\ Katz, A.\ Klemm and C.\ Vafa, \nihil{Geometric
Engineering of Quantum Field Theories,} hep-th/9609239.}
\lref\higherDer{ {Mans Henningson, \ex{Extended Superspace, higher
Derivatives and SL(2,Z) Duality,} Nucl.~ Phys.~{\bf B458} (1996)
445-455, hep-th/9507135;}\\ {M.\ Matone, \nihil{Modular
Invariance and Exact Wilsonian Action of $N=2$ SYM,}
hep-th/9610204.} } \lref\yau{S.\ Yau (ed.), {\it Essays on
Mirror Manifolds}, International Press 1992.} \lref\specialG{ S.\
Ferrara and A.\ Strominger, {\it $N=2$ Space-Time Supersymmetry and
Calabi-Yau Moduli Space}, in {\it Strings '89}, eds.\ R.\ Arnowitt et
al.\ (World Scientific, 1989) p.245;\\ A.\ Strominger, {Comm.\ Math.
Phys} {\bf 133} (1990) 163-180;\\ L.\ Dixon, V.\ Kaplunovsky and J.
Louis, {Nucl.\ Phys.\ } {\bf B329}, (1993) 27;\\ L.\ Castellani, R.\
D'Auria and S.\ Ferrara, {Phys.\ Lett.} {\bf B241},(1990) 57.\ }
\lref\erik{E.\ Verlinde, \nup 455 (1995) 211, hep-th/9506011.}
\lref\claim{ {A.\ Ceresole, R.\ D'Auria, S.\ Ferrara and A.\ Van
Proeyen, as in ref.\ \cite{dualities};}\\ {S.\ Ferrara, R.\ Minasian
and A.\ Sagnotti, \ex{Low-Energy Analysis of $M$ and $F$ Theories on
Calabi-Yau Threefolds,} Nucl.~ Phys.~{\bf B474} (1996) 323-342,
hep-th/9604097.} } \lref\BBS{K.\ \&\ M.\ Becker and A.\
Strominger, as in ref.\ \cite{pbranes}.} \lref\EH{T.\ Eguchi and A.\
Hanson, Phys.\ Lett.\ {\bf B74} (1978) 249.} \lref\witdyon{E.\
Witten, Phys.\ Lett.\ {\bf B86} (1979) 283.} \lref\ABSS{A.\
Brandhuber and S.\ Stieberger, \nihil{Self-Dual Strings and Stability
of BPS States in $N=2$ $SU(2)$ Gauge Theories,}
hep-th/9610053.} \lref\mirrorsymm{ {P.\ Candelas, X.\ De La
Ossa, P.\ Green and L.\ Parkes, \ex{A Pair of Calabi-Yau manifolds as
an exactly soluble superconformal theory,} Nucl.~ Phys.~{\bf B359}
(1991) 21-74;}\\ {S.\ Hosono, A.\ Klemm and S.\ Theisen,
\nihil{Lectures on Mirror Symmetry,} hep-th/9403096.} }
 \lref\BPSspectra{A.\ Sen, \plt329 (1994) 217, hep-th/9402032;\\ S.\
Sethi, M.\ Stern, and E.\ Zaslow, \nup457 (1995) 484,
hep-th/9508117;\\ J.\ Gauntlett and J. Harvey, preprint EFI-95-56,
hep-th/9508156.} \lref\GMV{B.\ Greene, D.\ Morrison and C.\ Vafa,
\nihil{A Geometric Realization of Confinement,}
hep-th/9608039.} \lref\nicketal{J.\ Schulze and N.\ Warner, to
appear.} \lref\dualities{ {A.\ Ceresole, R.\ D'Auria, S.\ Ferrara and
A.\ Van Proeyen, \nup444 (1995) 92, hep-th/9502072;}\\ {B.\ de Wit,
V.\ Kaplunovsky, J.\ Louis and D.\ L\"{u}st, {\it Perturbative
Couplings of Vector Multiplets in $N=2$ Heterotic String Vacua},
hep-th/9504006;}\\ {I.\ Antoniadis, S.\ Ferrara, E.\ Gava, K.\
Narain and T.\ Taylor, \ex{Perturbative Prepotential and Monodromies
in $N=2$ Heterotic Superstring,} Nucl.\ Phys.\ {\bf B447} (1995)
35-61, hep-th/9504034;}\\ {M.\ Billo, A.\ Ceresole, R.\
D'Auria, S.\ Ferrara, P.\ Fr\'e, T.\ Regge, P.\ Soriani, and A. Van
Proeyen, \ex{A Search for nonperturbative dualities of local N=2
Yang- Mills theories from Calabi-Yau threefolds,} Class.\ Quant.\
Grav.\ {\bf 13} (1996) 831-864, hep-th/9506075;}\\ {G.\
Cardoso, D.\ L\"ust and T.\ Mohaupt, \ex{Nonperturbative Monodromies
in N=2 Heterotic String Vacua,} Nucl.~ Phys.~{\bf B455} (1995)
131-164, hep-th/9507113;}\\ {S.\ Kachru, A.\ Klemm, W.\
Lerche, P.\ Mayr and C.\ Vafa, as in \cite{KKLMV}.} }
\lref\aspreview{P.\ Aspinwall, \nihil{K3 Surfaces and String
Duality,} hep-th/9611137.} 


\begin{thebibliography}{99}
\bibitem{SW}{\mref{\SW}} \bibitem{HT}{\mref{\HT}}
\bibitem{pbranes}{\mref{\pbranes}} \bibitem{reviews}{\mref{\reviews}}
\bibitem{Arn}{\mref{\Arn}} \bibitem{KLMVW}{\mref{\KLMVW}}
\bibitem{GKMMM}{\mref{\GKMMM}} \bibitem{NSbeta}{\mref{\NSbeta}}
\bibitem{SCHIF}{\mref{\SCHIF}} \bibitem{WDo}{\mref{\WDo}}
\bibitem{instaComp}{\mref{\instaComp}}
\bibitem{luisetal}{\mref{\luisetal}} \bibitem{OW}{\mref{\OW}}
\bibitem{Neqone}{\mref{\Neqone}}
\bibitem{higherDer}{\mref{\higherDer}}
\bibitem{unique}{\mref{\unique}} \bibitem{thooft}{\mref{\thooft}}
\bibitem{AD}{\mref{\AD}} \bibitem{witdyon}{\mref{\witdyon}}
\bibitem{BiFe}{\mref{\BiFe}} \bibitem{PF}{\mref{\PF}}
\bibitem{KLT}{\mref{\KLT}} \bibitem{withmatter}{\mref{\withmatter}}
\bibitem{massprep}{\mref{\massprep}} \bibitem{suN}{\mref{\suN}}
\bibitem{othergrps}{\mref{\othergrps}} \bibitem{MW}{\mref{\MW}}
\bibitem{Slodo}{\mref{\Slodo}} \bibitem{ALESW}{\mref{\ALESW}}
\bibitem{conformal}{\mref{\conformal}}
\bibitem{prepot}{\mref{\prepot}} \bibitem{KLTY}{\mref{\KLTY}}
\bibitem{AF}{\mref{\AF}} \bibitem{HKP}{\mref{\HKP}}
\bibitem{integ}{\mref{\integ}} \bibitem{PAMG}{\mref{\PAMG}}
\bibitem{nahm}{\mref{\nahm}} \bibitem{BSV}{\mref{\BSV}}
\bibitem{witcom}{\mref{\witcom}} \bibitem{selfd}{\mref{\selfd}}
\bibitem{KV}{\mref{\KV}} \bibitem{FHSV}{\mref{\FHSV}}
\bibitem{aspreview}{\mref{\aspreview}} \bibitem{LSW}{\mref{\LSW}}
\bibitem{Wi}{\mref{\Wi}} \bibitem{Asp}{\mref{\Asp}}
\bibitem{DM}{\mref{\DM}} \bibitem{EH}{\mref{\EH}}
\bibitem{probes}{\mref{\probes}} \bibitem{MDML}{\mref{\MDML}}
\bibitem{yau}{\mref{\yau}} \bibitem{NexTwoDual}{\mref{\NexTwoDual}}
\bibitem{Ftheory}{\mref{\Ftheory}} \bibitem{nonre}{\mref{\nonre}}
\bibitem{mirrorsymm}{\mref{\mirrorsymm}}
\bibitem{strom}{\mref{\strom}} \bibitem{andyrev}{\mref{\andyrev}}
\bibitem{specialG}{\mref{\specialG}}
\bibitem{dualities}{\mref{\dualities}} \bibitem{KLM}{\mref{\KLM}}
\bibitem{VaWi}{\mref{\VaWi}} \bibitem{AspJl}{\mref{\AspJl}}
\bibitem{KthreeFib}{\mref{\KthreeFib}} \bibitem{KKV}{\mref{\KKV}}
\bibitem{Fmatter}{\mref{\Fmatter}}
\bibitem{sadlyenough}{\mref{\sadlyenough}}
\bibitem{strongcoup}{\mref{\strongcoup}}
\bibitem{smallinst}{\mref{\smallinst}} \bibitem{erik}{\mref{\erik}}
\bibitem{claim}{\mref{\claim}} \bibitem{BBS}{\mref{\BBS}}
\bibitem{BPSspectra}{\mref{\BPSspectra}} \bibitem{ABSS}{\mref{\ABSS}}
\bibitem{nicketal}{\mref{\nicketal}} \bibitem{GMV}{\mref{\GMV}}
\bibitem{KKLMV}{\mref{\KKLMV}}
\end{thebibliography}
\end{document}